\documentclass[12pt]{article}
\usepackage{amsmath,amssymb,graphicx,bbm}
\usepackage[utf8]{inputenc}
\numberwithin{equation}{section}
\usepackage{color}
\usepackage[dvipsnames]{xcolor}

\usepackage[linktocpage=true]{hyperref}
\hypersetup{
  colorlinks=true,
  linkcolor=blue,
  citecolor=blue,
  pdfpagemode=FullScreen,
  }

\edef\restoreparindent{\parindent=\the\parindent\relax}
\usepackage{parskip}
\restoreparindent

\usepackage{tikz}
\usetikzlibrary{arrows,arrows.meta,intersections, calc,positioning,decorations.pathreplacing,decorations.pathmorphing,shapes}
\usetikzlibrary{patterns}
\usetikzlibrary{decorations.markings}
\usetikzlibrary{knots}

\usepackage{here}
\usepackage{caption}
\begin{document}
\begin{titlepage}

\renewcommand{\thefootnote}{\fnsymbol{footnote}}
\begin{flushright}
\begin{tabular}{l}
\end{tabular}
\end{flushright}

\vfill
\begin{center}

\noindent{\Large \textbf{Bulk-cone singularities and echoes} }

\smallskip

\noindent{\Large \textbf{from AdS exotic compact objects}}



\vspace{1.5cm}

\noindent{Heng-Yu Chen,$^{a}$\footnote{heng.yu.chen@phys.ntu.edu.tw} Yasuaki Hikida$^b$\footnote{yasuaki.hikida@oit.ac.jp} and Yasutaka Koga$^b$\footnote{y.koga.tj@gmail.com}}

\bigskip

\vskip .6 truecm

\centerline{\it $^a$Department of Physics, National Taiwan University, Taipei 10617, Taiwan }

\medskip

\centerline{\it $^b$Department of Information and Computer Science, Osaka Institute of Technology, }
\centerline{\it Kitayama, Hirakata,  Osaka 573-0196, Japan}

\end{center}

\vfill
\vskip 0.5 truecm
\begin{abstract}

The region near a black hole horizon may be modified by quantum gravity effects that resolve the singularity. Such geometry may be represented by an exotic compact object. Because the horizon is enclosed by a photon sphere, it is difficult to probe this region directly. In this paper, we develop a method to study the region inside the photon sphere by applying the AdS/CFT correspondence. We extract signatures of the modified geometry from the retarded Green functions of the dual conformal field theory. The retarded Green functions can be computed from bulk wave functions of scalar field. We show that exotic compact objects leave two characteristic imprints: bulk-cone singularities and echoes. The bulk-cone singularities correspond to null geodesics in the bulk, allowing us to detect null trajectories that are specific to exotic compact objects. The echoes arise from wave modes trapped inside the photon sphere, and thus signal the absence of a horizon. As concrete examples, we study AdS gravastar and AdS wormhole. We compute the corresponding bulk wave functions both via the WKB approximation and through numerical analysis and observe the bulk-cone singularities and echoes explicitly.

\end{abstract}
\vfill
\vskip 0.5 truecm

\setcounter{footnote}{0}
\renewcommand{\thefootnote}{\arabic{footnote}}
\end{titlepage}

\newpage

\hrule
\tableofcontents

\bigskip
\hrule
\bigskip


\section{Introduction}

A black hole has a singularity inside its horizon and a particularly interesting question is whether the singularity is resolved or not in (quantum) gravity theory. There are plenty of works on the subject, such as, AMPS firewall \cite{Almheiri:2012rt}, Fuzzballs \cite{Lunin:2001jy,Lunin:2002qf} and so on. Even at the classical level, we can construct a black hole alternative which smoothly replaces the entire region inside the horizon without any singularity. 
The regions near but outside the horizon can be modified due to its internal geometry. However, such a region is surrounded by a photon sphere thus it is difficult to be observed.
Such alternatives to black holes are given by exotic compact objects (ECOs) like compact boson stars, wormhole, gravastar and so on, see, e.g., \cite{Cardoso:2019rvt}. 
In this paper, we develop a method to probe the region inside the photon sphere by making use of AdS/CFT correspondence and apply it to several concrete examples of ECOs. 
In \cite{Chen:2025cee}, we constructed AdS gravastar in four dimensions and numerically analyzed the correlation functions in dual conformal field theory (CFT). In this paper, we consider more generic ECOs, such as, AdS gravastar and AdS wormhole in generic dimensions. We analyze the CFT correlation functions by semi-analytical approaches along with numerical computations. A part of our results were already presented in a short letter \cite{Chen:2025ywj}.

AdS/CFT correspondence provides a powerful way to probe the detailed structures of geometries that asymptote to the Lorentzian anti–de Sitter (AdS) spacetime in terms of the lower-dimensional CFT on its conformal boundary. We compute the retarded Green functions of scalar operator in dual CFT from gravity theory and examine how the effects around black hole horizon are encoded in the correlation functions. Among others, we focus on the following two types of signals. The first one is the singularity in the retarded Green function which is related to a null trajectory of the scalar field in the bulk geometry. See, e.g., \cite{Hubeny:2006yu,Maldacena:2015iua,Hashimoto:2018okj,Hashimoto:2019jmw,Dodelson:2020lal,Kinoshita:2023hgc,Terashima:2023mcr,Hashimoto:2023buz,Dodelson:2023nnr,Caron-Huot:2025she,Chen:2025cee} for previous works.
The other is the bump in the retarded Green function which is related to the echoes. 
These echoes indicate the tunneling effects of wave functions localized at the region inside the photon sphere. 
See, e.g., \cite{Cardoso:2016rao,Cardoso:2016oxy,Oshita:2018fqu,Oshita:2020dox,Terashima:2025tct} for the case of asymptotic flat spacetime.

The two-point functions of scalar operators in a Lorentzian CFT develop singularities when the two operator insertion points become light-like separated. This type of singularity is called a light-cone singularity. If the CFT admits a dual gravitational description, then its two-point functions are expected to exhibit another type of singularity associated with a null trajectory of a bulk scalar particle. 
Let us consider a scalar field $\phi$ with mass $m$ and denote its dual CFT operator by $\mathcal{O}(x)$. For large $m$, the two-point function of $\mathcal{O}(x)$ can be approximated as: 
\begin{align} \label{eq:semiclassical}
\langle \mathcal{O}(x_1) \mathcal{O} (x_2) \rangle \sim e^{- m \, d(x_1,x_2)},
\end{align}
where $d(x_1,x_2)$ is the proper distance along a bulk space-like geodesic between the boundary points $x_1$ and $x_2$. When the geodesic approaches to light-like one, the two-point function develops a singularity.
This type of singularity is called a bulk-cone singularity. Conversely, from the bulk-cone singularities, we can read off the geometric structures of the bulk gravity theory. 
In particular, we shall show that it is possible to distinguish the black hole and its alternatives from the information of the bulk-cone singularities.

More concretely, we consider a CFT on $\mathbb{R}_t \times S^{d-1}$, where $\mathbb{R}_t$ denotes the time direction and $S^{d-1}$ is $(d-1)$-dimensional sphere described by the metric
\begin{align}
\label{q:metricS}
    d\Omega_{d-1}^2 = d \theta^2 + \sin ^2 \theta d \Omega_{d-2}^2 . 
\end{align}
The dual geometry is given by an asymptotic AdS spacetime, which is written in the global coordinates with the Lorentzian signature. We examine the retarded Green functions
\begin{align} \label{eq:defretarded}
    G_\text{R} (t , \theta) = i H(t) \langle [\mathcal{O} (t , \theta) , \mathcal{O} (0,0)] \rangle .
\end{align}
Here $H(t)$ is the Heaviside step function and the points in $S^{d-2}$ are set to be the same.
Following \cite{Son:2002sd}, the retarded Green functions in a CFT can be computed from the wave functions of scalar field in the dual bulk geometry. The wave functions are solutions to the Klein-Gordon equations of scalar field, which can be put into the form of Schr{\"o}dinger equations. We solve the Schr{\"o}dinger equations by applying the WKB approximation. We then develop a way to read off the bulk-cone singularities from the wave functions obtained in the WKB approximation. 
The results are confirmed by comparing with those obtained by numerical computations.
This is one of the main results of this paper. See, e.g., \cite{Dodelson:2023nnr,Chen:2025cee} for previous works.

As argued above, the retarded Green functions of dual CFT can be computed from solutions to Schr{\"o}dinger-type equations. The bulk-cone singularities are associated with null trajectories, which correspond to the wave functions localized in classically allowed region. For trajectories with large angular momentum $\ell$, the Planck constant is given by $\ell$. For relatively small $\ell$, we may observe tunneling effects, which would create extra bumps in the retarded Green functions. These bumps are called the echoes.
In the case of AdS-Schwarzschild black hole, the potential for the Schr{\"o}dinger equation has only a maximum and there would be no tunneling effects. The existence of minima in the potential is an important characteristic of an ECO with a photon sphere but not a black hole horizon, see, e.g., \cite{Cardoso:2014sna}. Therefore, if we can observe echoes, then we can clearly see that the object should be not a black hole but its alternative. 
Analyzing the wave function of scalar field semi-analytically and numerically, we show that echoes can be observed also for asymptotic AdS spacetime.

Applying the methods developed, we examine two concrete examples of ECOs. This is another main result of this paper.
We believe that the two examples capture the generic features of these black hole alternatives.
The first example is a gravastar solution as an asymptotic AdS spacetime. 
In the gravastar solution, the outer region is AdS-Schwarzschild black hole and the inner region is de Sitter (dS) space. We consider a simple model where the two regions are connected by a thin-shell. The thin-shell is located inside the photon sphere and it is assumed that there is no surface energy. We construct such a geometry by generalizing the gravastar solution as an asymptotic flat spacetime in \cite{Mazur:2001fv,Visser:2003ge,Pani:2009ss,Cardoso:2014sna}. 
We have already constructed four-dimensional AdS gravastar in our previous work \cite{Chen:2025cee} and here we extend it to the arbitrary dimensional solution. We compute the retarded Green function
from the wave function of bulk scalar field both by the WKB analysis and the numerical computations.
We find that the Green function exhibits bulk-cone singularities, which correspond to null trajectories propagating through the region inside the photon sphere as well as those that orbit near and reflect around the photon sphere.
We also see that the Green function would have echoes just after the bumps corresponding to null trajectories around the photon sphere. From these signals we can distinguish the black hole and the gravastar geometry.

The potential in the Schr{\"o}dinger equation for the AdS gravastar takes a form typical of ECOs. In particular, the region inside the photon sphere is replaced by a non-singular one, and a high potential barrier would appear near the center due to the centrifugal force.  As the second example, we consider a wormhole solution, which connects the two asymptotic AdS boundaries.
This example provides a different type of potential. The region near the center is connected to another outer region and there is no high potential barrier due to the centrifugal force. With these two explicit examples, we believe that the representative cases of ECOs are covered.
We construct AdS wormhole solution by connecting two AdS-Schwarzschild geometries via a thin-shell at a radius 
inside the photon sphere. This is an extension of thin-shell wormhole as an asymptotic flat spacetime in \cite{Visser:1989kg,Morris:1988tu}.
AdS wormhole solutions have been constructed in the context of superstring theory \cite{Maldacena:2004rf}.
See, e.g.,  \cite{Kokubu:2014vwa,Kokubu:2015spa,Kokubu:2020lxs,Huang:2020qmn,Nozawa:2020gzz,Nozawa:2023aep} for other works.
Recently, there have been plenty of works on the applications to AdS/CFT correspondence, see, e.g., \cite{Gao:2016bin,Maldacena:2017axo,Maldacena:2018lmt}.
In particular, it has been studied how the two CFTs living on the opposite AdS boundaries communicate with each other.
In this paper, we treat the AdS thin-shell wormhole as an example of ECOs and examine the correlation functions of operators on the same AdS boundary. 
As in the case of AdS gravastar, we compute the retarded Green function from the wave function of bulk scalar field both by the WKB analysis and numerical computations.
In the AdS wormhole, there are null trajectories which bounce at the opposite AdS boundary and the retarded Green function is shown to possess bulk-cone singularities corresponding to such null trajectories along with those going around the photon sphere. We also observe the echoes which follow the signals corresponding to the null trajectories around the photon sphere.

The organization of this paper is as follows. In the next section, we show that the retarded Green functions \eqref{eq:defretarded} have bulk-cone singularities. 
As a concrete example, we focus on the case of AdS-Schwarzschild black hole.
We first use the geodesic approximation as in \eqref{eq:semiclassical} and then adopt the recipe of \cite{Son:2002sd} by solving the bulk wave equations in the WKB approximation.
In section \ref{sec:AdSECOs}, we apply the method to the case of ECOs. We first show that the retarded Green functions have bulk-cone singularities in these case as well. We further see that there are bumps called echoes which originate from the bulk wave functions localized inside the photon sphere. As concrete examples of ECOs, we explicitly compute the retarded Green functions from the wave functions of bulk scalar operators for AdS gravastar and AdS wormhole. In section \ref{sec:numerical}, we compute the retarded Green functions by solving the bulk wave equations numerically. We again consider AdS gravastar and AdS wormhole as concrete examples. We observe the bulk-cone singularities, which are consistent with null geodesics in the geometry of ECOs. We also show that the retarded Green functions exhibit echoes when the angular momentum $\ell$ is small. Section \ref{sec:conclusion} is devoted to conclusion and discussions. In appendix \ref{app:regge}, we use a method of \cite{Dodelson:2023nnr} with some modifications to derive the bulk-cone singularities corresponding to null geodesics bouncing at the AdS boundary, which was not covered in \cite{Dodelson:2023nnr}.
In appendix \ref{sec:bc-gravastar}, we summarize the details on the condition for the wave function at the junction of AdS gravastar.

\section{Bulk-cone singularities}
\label{sec:BCsingularities}

In this section, we evaluate the two-point functions or the retarded Greens functions of scalar operator from dual gravity theory. We first evaluate the two-point functions by the geodesic approximation of bulk scalar particle. We show that bulk-cone singularities appear when the space-like geodesics become null. We then compute the retarded Green functions from the wave function of bulk scalar field. The retarded Green function has bulk-cone singularities, which is consistent with the geodesic approach. 
As a concrete example, we consider the AdS-Schwarzschild black hole within this section. However, the analysis works for a large class of geometry including ECOs, which will be analyzed in the next section.

\subsection{Null geodesics}
\label{sec:nullgeodesics}

We would like to deal with a large class of geometry including AdS-Schwarzschild black hole and ECOs. As ECOs, we consider the geometry which approaches to the AdS spacetime near the boundary and have no singular behavior near the center. 
For the purpose, we assume that $(d+1)$-dimensional geometry is parameterized in the form:
\begin{align} \label{eq:metric}
    ds^2 = - f(r) dt^2 + f (r)^{-1} dr^2 + r^2 d \Omega_{d-1}^2 .
\end{align}
Here $f(r)$ is the warped factor and the metric of $(d-1)$-dimensional sphere is as in \eqref{q:metricS}.
In the case of AdS-Schwarzschild black hole, the metric function is given by 
\begin{align} \label{eq:AdSS}
        f(r) = r^2 + 1 - \frac{\mu}{r^{d-2}} .
\end{align}
The asymptotic AdS condition requires that $f(r) = r^2 + \mathcal{O} (r^0)$ for $r \to \infty$. The regularity condition means $f(r)$ is finite except for the large $r$ regime. Note that the metric function for the black hole in \eqref{eq:AdSS} has a divergence at $r \to 0$, but the region is removed as it is surrounded by the black hole horizon located at $r=r_h$ satisfying
\begin{align} \label{eq:rh}
f(r_h) = r_h^2 + 1 - \frac{\mu}{r_h^{d-2}} = 0 .
\end{align}

As discussed in the introduction, the two-point functions of scalar operator can be approximated as in \eqref{eq:semiclassical}.
The expression would diverge if the proper length $d(x_1 , x_2)$ between two boundary points $x_1,x_2$ vanishes as there would be no suppression factor.
This is realized when the two boundary points are connected by a bulk null geodesic.
The equation for the null geodesic  can be written as
\begin{align}
g_{\mu \nu} k^\mu k^\nu = - f(r ) \dot t^2 + f(r)^{-1} \dot r^2 + r^2 \dot \theta^2 = 0 .
\end{align}
Here $k^\mu = dx^\mu / d \lambda$ is the geodesic tangent with the affine parameter $\lambda$. The dot represents the derivative with respect to the affine parameter $\lambda$, i.e., $\dot A = d A/d \lambda$.
The conserved quantities are the energy $E$ and the angular momentum $L$. They are given respectively by:
\begin{align} \label{eq:EL}
E = - k_t = f(r) \dot t , \quad L = k_\theta = r^2 \dot \theta .
\end{align}
In terms of $E$ and $L$,
the geodesic equation becomes
\begin{align} 
\dot r^2 + L^2 f(r) r^{-2}  =  E^2 .
\end{align}
Rescaling the affine parameter $\lambda$, we can rewrite the equation as
\begin{align} \label{eq:Schcla}
\dot r^2 + V_\text{null} (r) = u^2 , \quad V_\text{null} (r) =   f(r) r^{-2} .
\end{align}
Here we have defined $u = E/L$ and $L \neq 0$ is assumed.

\begin{figure}
 \centering
 \includegraphics[height=4cm]{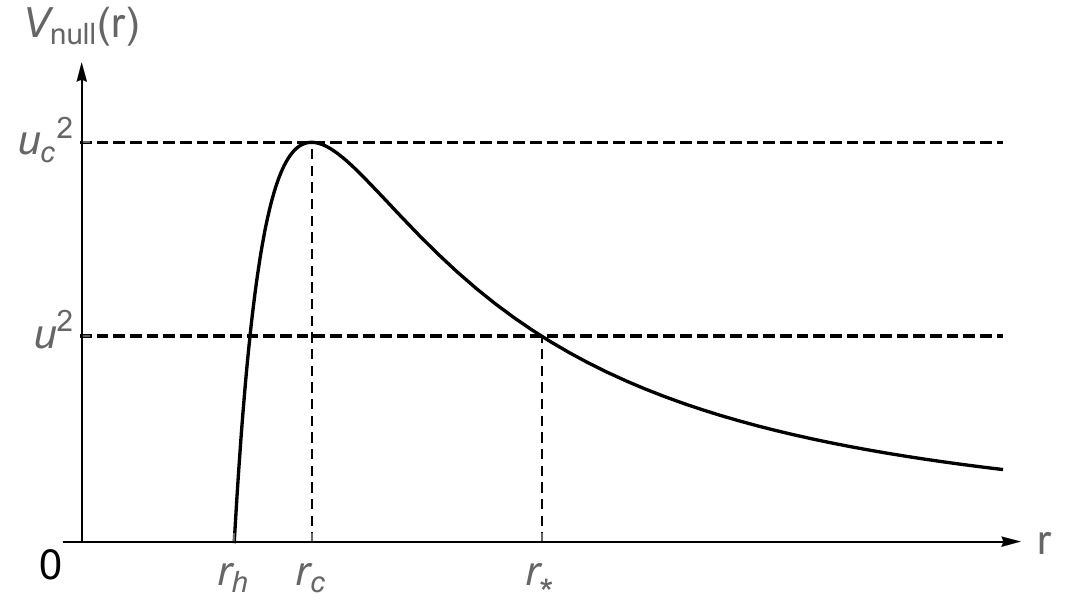}
 \caption{The shape of potential for the null geodesic in AdS-Schwarzschild black hole. 
 }
 \label{fig:AdSSnull}
\end{figure}

We would like to see the features of null trajectory by examining the case of AdS-Schwarzschild black hole. 
In this case, the potential $V_\text{null} (r)$ is given by (see fig.\,\ref{fig:AdSSnull})
\begin{align}
V_\text{null} (r) = 1 + \frac{1}{r^2} - \frac{\mu}{r^{d}} .
\end{align}
The AdS boundary is located at $r \to \infty$, where the potential approaches to $V_\text{null} (r) \to 1$. 
The potential vanishes at the black hole horizon, $r = r_h$, see \eqref{eq:rh}.
The maximum of the potential is realized at $r =r_c$, where
\begin{align}
V_\text{null} (r_c) = u_c^2, \quad \left. \partial_r V_\text{null} (r) \right|_{r = r_c} = 0. 
\end{align}
Solving these equations,  we find 
\begin{align} \label{eq:rc}
     r_c = \left(  \frac{d \mu}{2} \right)^{\frac{1}{d-2}}
\end{align}
and 
\begin{align} \label{eq:Vmax}
V_\text{null} (r_c)= 1 + \left( 1 - \frac{2}{d}\right) \left( \frac{2}{d \mu}\right)^{\frac{2}{d-2}}  \, (\equiv u_c) .
\end{align}
There is an unstable null trajectory at $r = r_c$, where a photon can stay there for a long time. Such a surface is called the photon sphere. We consider a null trajectory starting at $r \to \infty$ and moving to the center with small $r$. For a fixed $u$ with $1 < u < u_c$, there is a turning point $r= r_*$ satisfying $V_\text{null} (r_*) = u^2$. The null trajectory then comes back to the AdS boundary at $r \to \infty$. The AdS boundary may reflect the trajectory, and it may move toward the center of the bulk geometry again.

\begin{figure}
 \centering
 \includegraphics[height=4.5cm]{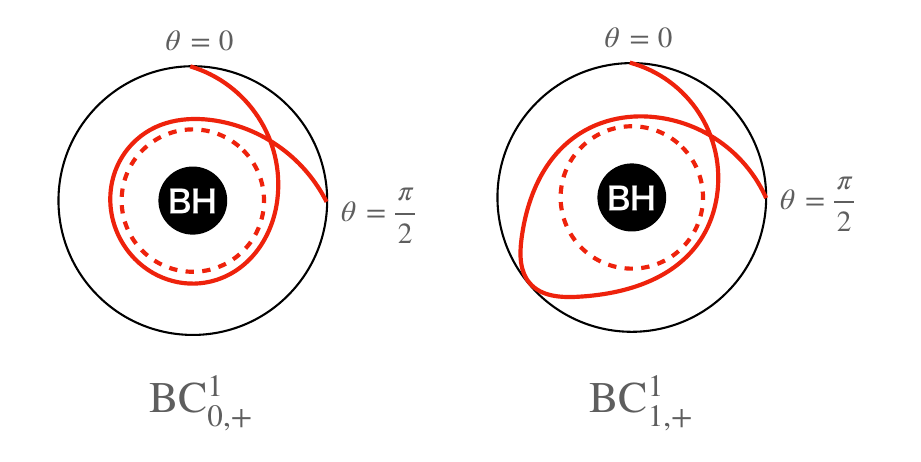}
 \caption{Illustration of null geodesics connecting two points on the boundary of AdS Schwarzschild black hole. The left trajectory $\mathrm{BC}_{0+}^1$ with $\theta=\pi/2$ has the angular displacement $2\pi j+\theta=5\pi/2$ with no bounce ($n-1=0$) at the boundary. The right trajectory $\mathrm{BC}_{1+}^1$ has the same displacement $2\pi j+\theta=5\pi/2$ with $n-1=1$ bounce. They wind around the photon sphere radius (the red circle).}
 \label{fig:obt_bh}
 \end{figure}

\begin{figure}
 \centering
 \includegraphics[height=7cm]{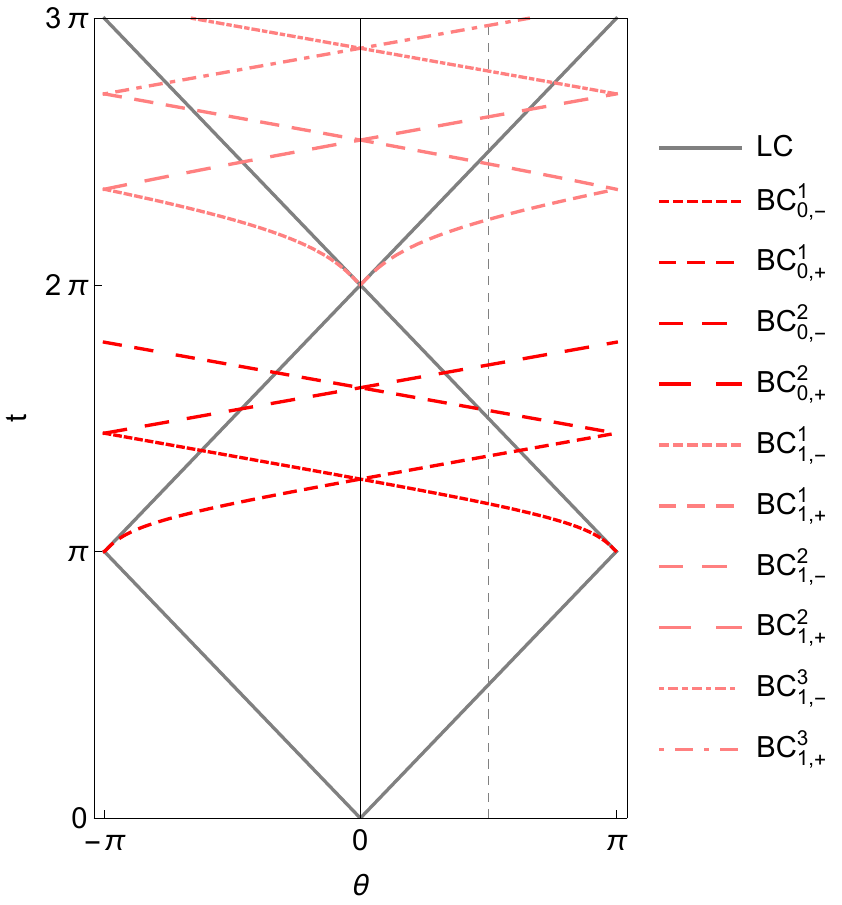}
 \includegraphics[height=7cm]{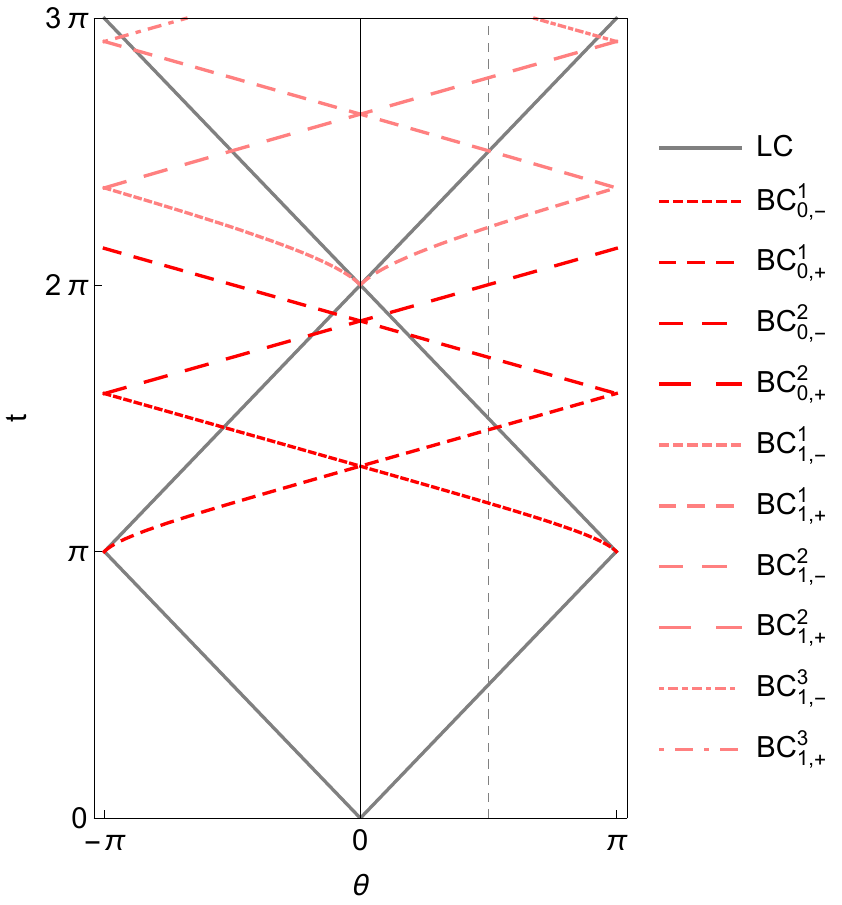}
 \caption{The bulk-cone structure for AdS-Schwarzschild black hole with $d=3$, $\mu=1/15$ (left) and $d=4$, $\mu=1/50$ (right).
 The retarded Green function $G_\mathrm{R}(t,\theta)$ is divergent along the red curves.
 It is also divergent along the gray lines corresponding to the light-cone singularity.
 }
 \label{fig:BCstructure_BH_4d}
 \end{figure}

In order to obtain the physical picture of null geodesics, we  rewrite eqs.\,\eqref{eq:Schcla} with \eqref{eq:EL} by removing the explicit dependence on the affine parameter $\lambda$. 
In terms of the derivative with respect to $r$, we have
\begin{align}
\frac{dt}{dr} = \pm \frac{1}{f(r)} \frac{u}{\sqrt{u^2 - V_\text{null} (r)}} , \quad 
\frac{d\theta}{dr} = \pm \frac{1}{r^2} \frac{1}{\sqrt{u^2 - V_\text{null} (r)}} .
\end{align}
Integrating over $r$, the arrival time and angle of null geodesic are given by
\begin{align} \label{eq:arrival}
    t = n T_\text{null} (u) , \quad 2  \pi j \pm \theta =  n \Theta_\text{null} (u) .
\end{align}
Here we have introduced
\begin{align} \label{eq:TTheta}
\begin{aligned}
&T_\text{null}(u) = 2 u \int_{r_*}^\infty \frac{dr}{f(r)} \frac{1}{\sqrt{u^2 -V_\text{null} (r) }} , \\
&\Theta_\text{null}(u) = 2  \int_{r_*}^\infty \frac{dr}{r^2} \frac{1}{\sqrt{u^2 -V_\text{null} (r) }} .
\end{aligned}
\end{align}
The null geodesics are labeled by two parameters $j,n$ $(n=1,2,\ldots , j = 0 ,1,\dots)$. The parameter $j$ counts how many times the geodesic goes around the photon sphere and the sign $\pm$ represents the direction of the geodesic. The number $(n-1)$ counts how many times the geodesic bounces at the AdS boundary.%
\footnote{The arrival time of a null trajectory that starts from the AdS boundary, turns around at the photon sphere, and returns to the AdS boundary is $t = T_{\text{null}}(u)$.
As illustrated in the right panel of fig.\,\ref{fig:obt_bh}, the trajectory may further bounce off the AdS boundary. In that case, the arrival time of the corresponding trajectory is $ t = 2 T_{\text{null}}(u)$.
More generally, if the trajectory bounces $(n-1)$ times at the AdS boundary, the arrival time is $t = n T_{\text{null}}(u)$.}
The trajectories are illustrated in fig.~\ref{fig:obt_bh}.
The structure of null geodesics for $d =4$ can be found in fig.\,\ref{fig:BCstructure_BH_4d}, where the null trajectories are labeled by $j,n,\pm$ as in $\text{BC}^{j}_{n-1,\pm}$.
For more details, see \cite{Dodelson:2023nnr,Chen:2025cee}.

\subsection{Geodesic approximation}
\label{sec:geodesic}

Applying the geodesic approximation, the two-point function can be expressed as (see eq.\,\eqref{eq:semiclassical})
\begin{align} \label{eq:semiclassical2}
\langle  \mathcal{O} (t,\theta) \mathcal{O}(0,0) \rangle \sim e^{- m \, d(t , \theta)} .
\end{align}
Here $ d(t , \theta)$ is the proper length of space-like geodesic between the two boundary points $(0,0)$ and $(t , \theta)$.
At high energy, the space-like geodesics approach null, and singularities are developed in the two-point function. In this subsection, we examine the space-like geodesics and examine how the two-point function \eqref{eq:semiclassical2} behaves near the singularities.
A space-like geodesic can be obtained by solving
\begin{align}
g_{\mu \nu} k^\mu k^\nu = - f(r ) \dot t^2 + f(r)^{-1} \dot r^2 + r^2 \dot \theta^2 = 1 .
\end{align}
Using the conserved quantities such as the energy $E$ and the angular momentum $L$ introduced in \eqref{eq:EL}, the geodesic equation can be reduced to 
\begin{align} \label{eq:space-like}
\dot r^2 + L^2 f(r) r^{-2} - f(r) =  E^2 .
\end{align}
The equation may be written as
\begin{align} \label{eq:Veff}
\dot r^2 + V_\text{eff} (r) = E^2 , \quad V_\text{eff} (r) = (L^2 - r^2) f(r) r^{-2} .
\end{align}
For large $L$ and small $r \ll L$, the effective potential $V_\text{eff} (r)$ reduces to $L^2 V_\text{null} (r)$
with $V_\text{null}(r) = f(r) r^{-2}$, see \eqref{eq:Schcla}.

As in the case of null geodesic, it is convenient to rewrite the geodesic equations without using the affine parameter $\lambda$.
In terms of the derivative with respect to $r$, the eqs.\,\eqref{eq:EL} and \eqref{eq:space-like} reduce to
\begin{align}
\frac{dt}{dr} = \pm \frac{1}{f(r)} \frac{E}{\sqrt{E^2 - V_\text{eff} (r)}} , \quad 
\frac{d\theta}{dr} = \pm \frac{1}{r^2} \frac{L}{\sqrt{E^2 - V_\text{eff} (r)}} .
\end{align}
We represent the turning point by $r=r_*$, which satisfies the equation $E^2 - V_\text{eff} (r_*) = 0$. 
For the trajectory without any bounce at the AdS boundary,%
\footnote{For the space-like geodesic, the potential behaves as $V_\text{eff} (z) \sim - r^2$ for large $r$. This makes the analysis on the bounce at AdS boundary subtle, so we focus on the case without any bounce here. In order to deal with bouncing trajectories, it might be necessary to use the analysis of wave function  as below.}
the arrival time and angle of the space-like geodesic are given by
\begin{align} \label{eq:tthetasp}
    t =   T(E,L)  , \quad 2  \pi j \pm \theta =  \Theta(E,L) .
\end{align}
Here we set
\begin{align} 
\label{eq:TThetanull}
\begin{aligned}
&T(E,L) = 2 E \int_{r_*}^\infty \frac{dr}{f(r)} \frac{1}{\sqrt{E^2 -V_\text{eff} (r) }} , \\
&\Theta(E,L) = 2 L \int_{r_*}^\infty \frac{dr}{r^2} \frac{1}{\sqrt{E^2 -V_\text{eff} (r) }} .
\end{aligned}
\end{align}
Not that these functions are different from those defined in \eqref{eq:TTheta}. In particular, the potential $V_\text{eff} (r)$ significantly differs from $L^2 V_\text{null} (r)$ for large $r$ region.

With the above preparations, we study the two-point function by geodesic approximation as in \eqref{eq:semiclassical2}, see \cite{Dodelson:2020lal} for closely related analysis.
Using the notations introduced above, the proper length can be expressed as
\begin{align}
    d (t , \theta) = 2 \int_{r_*}^{r_\text{max}} \frac{dr}{\dot r} 
     = 2 \int_{r_*}^{r_\text{max}} \frac{dr}{\sqrt{E^2 - V_\text{eff} (r)}} .
\end{align}
We have introduced a cut off at  $r = r_\text{max}$ in order to regularize the logarithmic divergence, which arises as $E^2 - V_\text{eff}(r) \sim r^2$ for $r \to \infty$.
Here we assume that the difference $E^2 - L^2$ is large. We separate the integral domain into 
\begin{align} \label{eq:rregion}
(i)\ r_* < r < R , \quad (ii)\ R < r < r_\text{max} ,
\end{align}
where we set $R$ to satisfy
\begin{align} \label{eq:intdomain}
r_\text{max} \gg \sqrt{E^2 - L^2} \gg R \gg r_*. 
\end{align}
Here we have assumed that $L$ is large and $E ,\sqrt{E^2 - L^2}$ are of order $\mathcal{O}(L)$.
The integral for the region $(i) \ r_* < r < R$ can be neglected as we assumed that $\sqrt{E^2  - L^2} \gg R$. The integral for the region $(ii)\ R < r < r_\text{max}$ can be approximated as
\begin{align}
    2 \int_{R}^{r_\text{max}} \frac{dr}{\sqrt{r^2 + E^2 - L^2}} \sim 2 \log \left(\frac{r_\text{max}}{\sqrt{E^2 - L^2}} \right) .
\end{align}
We use a regularized geodesic length $d (t , \theta)_\text{ren}$, where the term proportional to $\log r_\text{max}$ is dropped.
We conclude that  the two-point function behaves as
\begin{align} \label{eq:Dren}
    e^{- m \, d (t , \theta)_\text{ren}} = (E^2 - L^2)^m 
\end{align}
with $m$ as the mass of bulk scalar field.

The right-hand side of \eqref{eq:Dren} is written in terms of $E, L$. Since the arrival time and angle of space-like geodesic are functions of $E , L$, as in \eqref{eq:TThetanull}, they can be re-expressed in terms of $t , \theta$ by solving \eqref{eq:tthetasp}.
We would like to obtain approximated expressions as in the case of the proper length $d(t ,\theta)$. 
We separate the integral region into two as in \eqref{eq:rregion}. Here we can set $r_\text{max} \to \infty$ as the integrals \eqref{eq:TThetanull} converge for large $r$. 
Let us first consider the region $(ii)\ R < r < \infty$. In the region, we can use $f(r) r^{-2} \simeq 1$ and hence $V_\text{eff}(r) \simeq L^2 - r^2$. 
The contributions to $T(E,L)$ and $\Theta (E,L)$ from the region are given by
\begin{align}
\begin{aligned}
   & 2 E \int_{R}^\infty \frac{dr}{r^2} \frac{1}{\sqrt{E^2 - L^2 + r^2}} = \Delta T - \frac{2E}{E^2 - L^2} + \mathcal{O}(L^{-2}), \\  
   & 2 L \int_{R}^\infty \frac{dr}{r^2} \frac{1}{\sqrt{E^2 - L^2 + r^2}} = \Delta \Theta - \frac{2L}{E^2 - L^2}  + \mathcal{O}(L^{-2}),
    \end{aligned}
\end{align}
respectively. Here we set
\begin{align} \label{eq:Deltas}
    \Delta T  = 2 E \int_{R}^\infty \frac{dr}{r^2} \frac{1}{\sqrt{E^2 - L^2}} , \quad \Delta \Theta = 2 L \int_{R}^\infty \frac{dr}{r^2} \frac{1}{\sqrt{E^2 - L^2}} .
\end{align}
We then focus on another region $(i)\ r_* < r < R$. In this region, the term with $-r^2$ in \eqref{eq:Veff} can be neglected and $V_\text{eff}(r)$ is replaced by $L^2 V_\text{null}(r)$. Thus, the contributions to $T(E,L)$ and $\Theta (E,L)$ from the region plus $\Delta T$ and $ \Delta \Theta$ in \eqref{eq:Deltas} are given by $T_\text{null}(u)$ and $\Theta_\text{null}(u)$, respectively.
Combining the both contributions, we have
\begin{align} \label{eq:Tnull}
\begin{aligned}
&T (E,L) = T_\text{null} (u) - \frac{2 E}{E^2 - L^2} + \mathcal{O} (L^{-2})= T_\text{null} (u) - \frac{2 u}{(u^2 - 1) L} + \mathcal{O} (L^{-2}), \\
&\Theta (E,L) = \Theta_\text{null} (u) - \frac{2 L}{E^2 - L^2} + \mathcal{O} (L^{-2})= \Theta_\text{null} (u) - \frac{2}{(u^2 - 1) L} + \mathcal{O} (L^{-2}).  
\end{aligned}
\end{align}
With the expressions, the two-point function \eqref{eq:Dren} can be rewritten as
\begin{align} \label{eq:2ptsemi}
    e^{- m \, d(t,\theta)_\text{ren}} \sim \left(\frac{4  u^2}{u^2 - 1} \right)^m \frac{1}{(T_\text{null} (u) - t)^{2m}},
\end{align}
where we have set $t = T(E,L)$ as in \eqref{eq:tthetasp}. The result is consistent with the one obtained in \cite{Dodelson:2020lal} for the geodesic going around the photon sphere of AdS-Schwarzschild black hole. We have extended their analysis such as to hold also for more generic geometry and shown that the conclusion does not depend on the details of the geometry.

Note that the power of the singularity is given by $2m$. The conformal dimension of the dual scalar operator is 
\begin{align} \label{eq:Delta}
    \Delta = \frac{d}{2} + \sqrt{\left( \frac{d}{2} \right)^2 + m^2  } \, \left(\equiv \frac{d}{2} + \nu \right),
\end{align}
where the AdS radius is set to be the unity.
The usual light-cone singularity for $t - \theta\sim 0$ is
\begin{align}
    \langle \mathcal{O} (t ,\theta) \mathcal{O} (0,0) \rangle \propto \frac{1}{(t - \theta)^\Delta} \sim \frac{1}{(t - \theta)^m} ,
\end{align}
where we set $m \gg d/2$. Thus the power of singularity for the bulk-cone singularity is different from the one for the light-cone singularity. 
An exception is given for the case of pure AdS, where the bulk-cone singularity coincides with the light-cone one. In the above arguments, we assumed that $L$ is large and $E,\sqrt{E^2 - L^2}$ are of order $\mathcal{O}(L)$, which is used to separate the integral region as in \eqref{eq:rregion} with \eqref{eq:intdomain}.
The assumption does not hold for the pure AdS,
where $\sqrt{E^2 - L^2}$ is of order $\mathcal{O}(L^0)$, see, e.g., \cite{Hubeny:2006yu}.
For geometry whose space-like geodesics are given with large $E,L$ and $E^2 - L^2$, the expression \eqref{eq:2ptsemi} with the power $2m$ holds. In the case of AdS Schwarzschild black hole, it was shown in \cite{Dodelson:2023nnr} that the power is corrected to be $2 \Delta - (d-1 )/ 2 $, which reduces to $2m$ for large $m$. We will see below that the power will also be corrected to be $2 \Delta - (d-1 ) /2$ for generic geometries and the power $2 \Delta - (d-1 / )2$ is not specific to the case with the photon sphere and appears generically.

\subsection{Retarded Green functions}
\label{sec:mometum}

In the previous subsection, we have examined the two-point function of scalar operator from the bulk gravity by applying the geodesic approximation. This approximation is valid when the mass of bulk scalar field, denoted by $m$, is large. For cases where $m$ is not so large, we evaluate the two-point function or the retarded Green function \eqref{eq:defretarded} by applying the recipe of \cite{Son:2002sd}. Namely, we first solve the Klein-Gordon equation for the bulk scalar field on the geometry. We then obtain the retarded Green function from the behavior of the solution near the AdS boundary. In this subsection, we explain this strategy and apply it to the case of AdS-Schwarzschild black hole.

We study the propagation of the bulk scalar field in the $(d+1)$-dimensional geometry of the form \eqref{eq:metric}.
We consider the mode expansion of the scalar field given by
\begin{align}
 \phi (t , \Omega_{d -1} ,r) = e^{- i \omega t} Y_{\ell \vec m} (\Omega_{d-1}) r^{-\frac{d-1}{2}} \psi_{\omega \ell} (r) . 
\end{align}
Here the spherical harmonics on $S^{d-1}$ are denoted by $Y_{\ell \vec m}$.
The wave equation for the radial part can be put into the form
\begin{align}
\label{eq:zwaveeq}
    (- \partial_z^2 + V(z) - \omega^2) \psi_{\omega \ell} (z) = 0  .
\end{align}
The radial coordinate $z$ is introduced via
\begin{align} \label{eq:zcoordinate}
    dz = - \frac{dr}{f(r)} 
\end{align}
and the potential is given by
\begin{align} \label{eq:potential}
\begin{aligned}
    V(z) = f(r)&  \Bigg[\frac{(\ell+\alpha)^2-\frac{1}{4}}{r^2}+ \nu^2 - (\alpha + 1)^2 \\
    & \quad +\left(\alpha^2 - \frac14\right)\frac{f(r)-1}{r^2}+\left(\alpha + \frac12\right)\frac{\frac{d}{dr}f(r)}{r}\Bigg] ,
    \end{aligned}
\end{align}
where $\nu$ was introduced in \eqref{eq:Delta} and $\alpha = (d-2)/2$.
In terms of  $z$ in \eqref{eq:zcoordinate}, 
the AdS boundary is located at $z \to 0$. 
For the AdS-Schwarzschild black hole,
the metric function $f(r)$ is given by \eqref{eq:AdSS} and the potential $V(z)$ becomes
\begin{align}
    V(z) = f(r) \left[ \frac{(\ell + \alpha)^2 - \frac14}{r^2} + \nu^2 - \frac14 + \frac{(d-1)^2 \mu}{4 r^d}\right]  .
\end{align}
The black hole horizon is at $z \to \infty$.
Near the AdS boundary $(z \to 0)$, the scalar field behaves as
 \begin{align} \label{eq:waveSon}
     \psi_{\omega \ell} (z) \sim \mathcal{A} (\omega , \ell) z^{\frac12 - \nu} + \mathcal{B} (\omega , \ell) z^{\frac12 + \nu} .
 \end{align}
The retarded Green function is then obtained as 
 \begin{align} \label{eq:GRSon}
     G_R(\omega , \ell) = \frac{\mathcal{B} (\omega , \ell)}{\mathcal{A} (\omega , \ell)}  .
 \end{align}

\begin{figure}
\centering
\includegraphics[height=4cm]{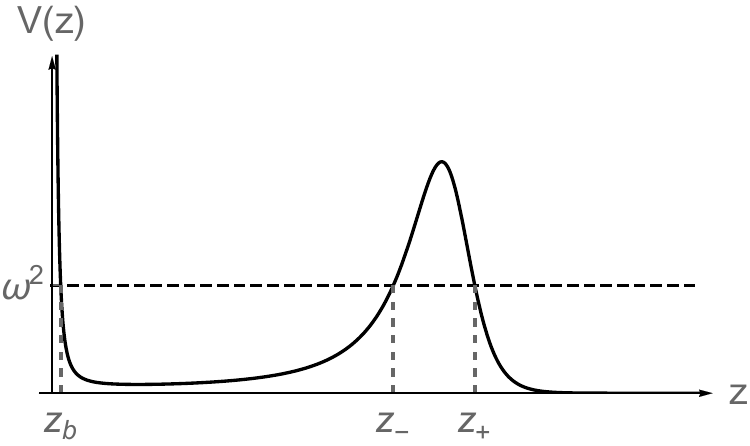}
\caption{The potential for AdS-Schwarzschild black hole.}
\label{fig:Vsch}
\end{figure}

Now the task is solving the wave equation \eqref{eq:zwaveeq} and reading off the behavior of the solution near the AdS boundary. However, the wave equation is generically difficult to solve and additional techniques have to be utilized.
Since the wave equation is written in the form of Sch{\"o}dinger equation, we can adopt the WKB approach. For the time-being, we use the semi-analytic method to study the wave equation. In section \ref{sec:numerical}, we will adopt the numerical method to compute the retarded Green function for comparison. We consider the case where the angular momentum $\ell \in \mathbb{Z}$ to be large as $\ell \to \infty$, thus regard $1/\ell$ as the Planck constant in the Sch{\"o}dinger equation.
For large $\ell$, the wave equation can be approximated by
\begin{align} \label{eq:wave}
    ( \partial_z^2 +  \kappa (z)^2  ) \psi (z) = 0
\end{align}
with 
\begin{align} \label{eq:kappa}
    \kappa (z) = \sqrt{\omega^2 - \tilde V (z)}  , \quad 
     \tilde V(z) =  \left [\ell^2 + \left (\nu^2 - \frac{1}{4} \right) r^2 \right] \frac{f(r)}{r^2} .
\end{align}
The potential is usually
approximated as $\tilde V(z) = \ell^2 f(r)r^{-2}$, see, e.g., \cite{Dodelson:2023nnr}. Here we rather keep the term proportional to $r^2$ as the term would be relevant when $r$ is large as $r \sim \mathcal{O}(\ell)$. We will see that the extra term play an important role to obtain the behavior of singularity in the retarded Green function.
In order to apply the WKB approximation, we separate the region by the points satisfying the equation $\kappa (z)^2 = 0$.
In the case of AdS-Schwarzschild black hole, the potential is of the form as in fig.\,\ref{fig:Vsch}.
We consider the case with $1 < u < u_c$, where
$u = \omega/\ell$ and $u_c$ is defined in \eqref{eq:Vmax}. In the case, there are three zeros in the equation $\kappa (z)^2 = 0$ and the zeroes are denoted as
 $ z_b , z_- , z_+$ with $ z_b < z_- < z_+ $. 
Note that the potential diverges for $z \to 0$ (or $r \to \infty$) when $\nu \neq 1/2$ as in fig.\,\ref{fig:Vsch}.
Thus there should be a zero for $z \lesssim 1/\ell$, which is represented by $z_b \,(\simeq 0)$.

We would like to solve the wave equation \eqref{eq:wave} in the WKB approximation for the explicit example of AdS-Schwarzschild black hole. 
We find the wave functions for each region bounded by the zeros of the equation $\kappa (z)^2 = 0$ and then apply the WKB connection formula to relate the wave functions in different regions.
We start from the wave function for $z_+ < z  < \infty$. The black hole horizon is located at $z \to \infty$ and the ingoing boundary condition is required at the horizon, i.e.,
 \begin{align}
 \label{eq:bc-ingoing}
     \psi (z)  \sim e^{i \omega z}  \quad (z \to \infty) .
 \end{align}
The wave function satisfying the condition is given by
\begin{align} \label{eq:0zzm}
    \psi (z) \sim \frac{1}{\sqrt{\kappa(z)}} e^{i  \int_{z_+}^z d z' \kappa (z')}  .
\end{align}
Applying the WKB connection formula, the  solution for $0 \simeq z_b < z < z_-$ can be found as
\begin{align} \label{eq:psito0}
    \psi (z) \sim \frac{C_+}{\sqrt{\kappa(z)}} e^{i \int_{0}^z d z' \kappa (z')} + \frac{C_-}{\sqrt{\kappa(z)}} e^{ - i \int_{0}^z d z' \kappa (z')} 
\end{align}
with
\begin{align} \label{eq:CmCp}
\begin{aligned}
&C_+ = \left (e^{S(z_- , z_+)} + \frac14e^{-S(z_- , z_+)} \right ) e^{-i S(0,z_-)}, \\ 
&C_-  = \left (e^{S(z_- , z_+)} - \frac14e^{-S(z_- , z_+)} \right )e^{ - \frac12 i \pi}  e^{i S(0,z_-)} .
\end{aligned}
\end{align}
Here we set 
\begin{align} \label{eq:S0m}
S(0,z_-)  =  \int_{0}^{z_-}  \kappa (z') d z'
\end{align}
and 
\begin{align} \label{eq:Smp}
S(z_-,z_+)  =  \int_{z_-}^{z_+}  q (z') d z' , \quad q (z) = \sqrt{\tilde V (z) - \omega^2 } . 
\end{align}
Notice that the asymptotic behaviors for $z \sim 1/\ell$ are given by
\begin{align} \label{eq:asym1overl}
    e^{ \pm i \int_{0}^z d z' \kappa (z')} \sim e^{\pm i  \sqrt{\omega^2-\ell^2} z } \, .
\end{align}
For $z < 1/\ell$, the geometry can be approximated by the AdS space, and the solution is written in terms of Hankel functions. 
We have found that the solution is given by \eqref{eq:0zzm} for the region $0 \simeq z_b < z < z_-$ and its behavior near $z \sim 1/\ell$ can be read off by applying \eqref{eq:asym1overl}. We realize the behavior near $z \sim 1/\ell$ from the linear combination of Hankel functions as
\begin{align}
\begin{aligned}
    \psi (z) & =  C_+ \sqrt{\frac{\pi \ell z}{2}} e^{\frac{i \pi \nu}{2} + \frac{i \pi}{4}
     }H^{(1)}_{\nu} (\sqrt{\omega^2-\ell^2} z)  + C_-  \sqrt{\frac{\pi \ell z}{2}} e^{-\frac{i \pi \nu}{2} - \frac{i \pi}{4} } H^{(2)}_\nu ( \sqrt{\omega^2-\ell^2}z).
    \end{aligned}
\end{align}
Here we have used the asymptotic behaviors of Hankel functions
\begin{align}
    H^{(1)}_\nu (x) \sim \sqrt{\frac{2 }{\pi x}} e^{i (x - \nu \frac{\pi}{2} - \frac{\pi}{4})}\, , \quad      H^{(2)}_\nu (x) \sim \sqrt{\frac{2 }{\pi x}} e^{- i (x - \nu \frac{\pi}{2} - \frac{\pi}{4})} 
\end{align}
for $x \gg 1$.

The retarded Green function of scalar operator can be obtained as in \eqref{eq:GRSon}. 
Applying the asymptotic behaviors of Hankel functions near $x \sim 0$ given by
\begin{align}
\begin{aligned}
    &H^{(1)}_\nu (x) \sim \frac{i}{\sin \pi \nu} \left[\frac{e^{- i \nu \pi}}{\Gamma (1+\nu)} \left( \frac{x}{2}\right)^\nu - \frac{1}{\Gamma (1-\nu)} \left( \frac{x}{2}\right)^{-\nu} \right], \\
    &H^{(2)}_\nu (x)  \sim -\frac{i}{\sin \pi \nu} \left[\frac{e^{ i \nu \pi}}{\Gamma (1+\nu)} \left( \frac{x}{2}\right)^\nu - \frac{1}{\Gamma (1-\nu)} \left( \frac{x}{2}\right)^{-\nu} \right] ,
\end{aligned}
\end{align}
we can read off the retarded Green function as
\begin{align} \label{eq:retardedAdSG}
    G_R (\omega ,\ell) &=  \frac{\Gamma (- \nu)}{\Gamma(\nu)} 
    \left( \frac{\omega^2 - \ell^2}{4}\right)^\nu \frac{C_+ e^{-\frac{i \pi \nu}{2} + \frac{i \pi}{4}} - C_- e^{\frac{i \pi \nu}{2} - \frac{i \pi}{4}}}{C_+ e^{\frac{i \pi \nu}{2} + \frac{i \pi}{4}} - C_- e^{-\frac{i \pi \nu}{2} - \frac{i \pi}{4}}}  .
\end{align}
The explicit form of coefficients $C_\pm$ can be found in \eqref{eq:CmCp} and in particular they are functions of $e^{i S(0,z_-)}$ and $e^{-S(z_- , z_+)}$. For large $\ell$, the functions $S(0,z_-)$ and $S(z_- , z_+)$ take large values and it is convenient to expand in terms of  $e^{i S(0,z_-)}$ and $e^{-S(z_- , z_+)}$. In the following, we analytically continue $\omega$ to take a complex value and assume that $\text{Im} \, S(0, z_-) > 0$.%
\footnote{We can see that $\text{Im} \, S(0, z_-) > 0$ for our choice of analytic continuation, but the analysis works also for  $\text{Im} \, S(0, z_-) < 0$ only with minor modifications.}
We expand the retarded Green function as
\begin{align} \label{eq:GRte}
    G_R (\omega ,\ell) &=  \frac{\Gamma (- \nu)}{\Gamma(\nu)} 
    \left( \frac{\omega^2 - \ell^2}{4}\right)^\nu (A + B e^{-2 S(z_- , z_+)} + \mathcal{O} (e^{-4 S(z_- , z_+)} )) ,
\end{align}
where we set 
\begin{align} 
A = \sum_{n=0}^\infty a_n e^{2 i n S(0,z_-)} , \quad  B =  \sum_{n=1}^\infty b_{n} e^{2 i n S(0,z_-)} .
\end{align}
From the expressions of $C_\pm$ in \eqref{eq:CmCp}, the coefficients $a_n$ and $b_n$ are obtained as 
\begin{align}  \label{eq:an}
a_0 =  e^{-i\pi \nu} , \quad a_{n} = - 2 i  (-1)^n e^{ -i n \pi \nu}  \sin ( \pi \nu ) 
\end{align}
and
\begin{align}
b_{n} = i (-1)^{n} n e^{- i n \pi \nu} \sin ( \pi \nu )
\end{align}
for $n=1,2,3,\ldots$.
The terms proportional to $e^{- 2 S(z_- ,z_+)}$ are suppressed and they are neglected for a while.
We will see shortly that the integer $n$ has the same meaning as in \eqref{eq:arrival} and each term with $a_n$ corresponds to a null trajectory that bounces off the AdS boundary $(n-1)$ times.

\subsection{Conversion to a coordinate basis}
\label{sec:coordinate}

In subsection \ref{sec:geodesic}, we evaluated the two-point function of a scalar operator by geodesic approximation. We found that the two-point function has singularities related to the null geodesics in the bulk geometry as in \eqref{eq:2ptsemi}. In the previous subsection, we evaluated the regarded Green function in the momentum basis with $\omega, \ell$ from the wave analysis. In order to relate the two expressions, we need to covert the expressions in the momentum basis with $\omega, \ell$ into those in the coordinate basis with $t ,\theta$ by performing Fourier transformations.
The map from the frequency basis with $\omega$ to the time basis with $t$ is given by
\begin{align}
G_R(\omega,\theta) = \int_{0}^\infty dt e^{ i \omega t} G_R(t,\theta) .
\end{align}
We set $\omega \in \mathbb{R} +  i \delta$ with $\delta > 0$, which makes the integral convergent.
The map from the angular momentum basis with $\ell$ to the angle coordinate basis with $\theta$ can be constructed by utilizing the orthogonality of spherical harmonics $Y_{\ell \vec{m}}$. 
We are interested in the case where the $S^{d-2}$ angular coordinates of the two operators coincide.
Then the expression of the map simplifies to:
\begin{align}
G_R (t , \theta) = \frac{1}{2 \pi}\int_{- \infty + i \delta}^{\infty + i \delta} d \omega e^{-i\omega t} \sum_{\ell =0}^\infty G_R(\omega , \ell) \frac{\ell + \alpha}{\alpha} C_{\ell}^{(\alpha)}(\cos \theta) ,
\end{align}
where $C_{\ell}^{(\alpha)}(\cos \theta)$ is the Gegenbauer polynomials.

It is convenient to rewrite the sum over $\ell = 0,1,2,\ldots$ in terms of integral over a parameter, say $k$, by applying the Cauchy theorem.
As explained in \cite{Dodelson:2023nnr}, the summation can be replaced by an integration as
\begin{align} \label{eq:Gegen}
G_R (t , \theta) = \frac{i}{4\pi \alpha} \int_{- \infty + i \delta}^{\infty + i \delta} d \omega e^{-i\omega t} \int _{- \infty + i \delta '}^{\infty + i \delta '}  \frac{k dk}{\sin (\pi(k-\alpha))}G_R(\omega , k - \alpha) C_{k-\alpha}^{(\alpha)}( - \cos \theta)
\end{align}
with $\delta ' > 0$.
For $0 < \theta < \pi$, we can expand the Gegenbauer polynomials as
\begin{align}
    C^{(\alpha)}_{k - \alpha} (-z) 
    = \frac{i (2 \sin \theta)^{1 - 2 \alpha} \Gamma (k+\alpha)}{\Gamma (\alpha) \Gamma (k+1)} \left[ e^{- i (1 + k - \alpha) (\pi - \theta)} f(k , \theta) - e^{i (1 + k - \alpha) (\pi - \theta)} f(k , \pi - \theta)\right]
\end{align}
with 
\begin{align}
    f(k ,\theta) = {}_2 F_1 (1 - \alpha , 1+ k - \alpha , 1 + k , e^{2 i \theta}) .
\end{align}
Thus, for $0 < \theta < \pi$, we can rewrite the expression \eqref{eq:Gegen} as
\begin{align} \label{eq:sumj}
G_R (t , \theta) = \sum_{j = 0}^\infty \left[ g_R (t , |\theta| + 2 \pi j) + (-1)^{2 \alpha} g_R (t , 2\pi - |\theta| + 2 \pi j) \right]
\end{align}
with
\begin{align} \label{eq:gr}
\begin{aligned}
g_R (t , \theta)  &=  \frac{4^{1-\alpha}}{(\sin \theta)^{2 \alpha -1}} \frac{1}{4 \pi i \alpha} \int_{- \infty + i \delta}^{\infty + i \delta}d \omega  e^{-i\omega t} \int_{- \infty + i \delta '}^{\infty + i \delta '} d k  G_R(\omega , k - \alpha)e^{i (1 + k - \alpha ) \theta } \\
&\quad \times \frac{\Gamma (k + \alpha)}{\Gamma(k)\Gamma(\alpha)} F(1 -\alpha , 1 + k - \alpha , 1 + k , e^{2 i \theta}) . 
\end{aligned}
\end{align}
We may use
\begin{align} \label{eq:Gauss}
    \frac{\Gamma (k + \alpha)}{\Gamma(k)\Gamma(\alpha)} F(1 -\alpha , 1 + k -\alpha , 1 + k , e^{2 i \theta}) = \frac{k^\alpha}{\Gamma(\alpha)} (1 - e^{2 i \theta})^{\alpha -1} + \mathcal{O} \left( \frac{1}{k} \right)
\end{align}
for large $k$.

Compared with the analysis in \cite{Dodelson:2023nnr}, here we modified several points in order for the method to be applicable  not only to AdS-Schwarzschild black hole but also to ECOs including AdS gravastar and AdS wormhole. In \cite{Dodelson:2023nnr}, the authors deformed the integral contour such that the integral reduces to the sum of residues at the poles of $G_R (\omega , k - \alpha)$. This is possible only when there is no contribution from the integral over the  contour at infinity. In the case of AdS-Schwarzschild black hole, such contributions can be neglected. The contribution from infinity comes from high energy modes, which are expected to be falling inside the black hole horizon. In the case of ECOs without horizon, such contributions are expected to survive. 
Instead of summing over the residues at the poles, we directly perform the integration over $k$ in \eqref{eq:gr}. 
In order to make this integration tractable, we expand $G_R (\omega , k - \alpha)$ in terms of  $e^{i S(0,z_-)}$ and $e^{-S(z_- , z_+)}$ as in \eqref{eq:GRte} and perform the integration for each term.
As another difference, we add only small imaginary parts for $\omega , \ell$, whereas the authors of \cite{Dodelson:2023nnr} set $\omega$ and $\ell + \alpha$ to take pure imaginary values. 
They showed that the analytic continuation works in the case of $n=1$, where the corresponding trajectory does not bounce off the AdS boundary.
As we will see below, the cases with $n > 1$ can be properly treated by working with $\omega, \ell \in \mathbb{R} + i \delta$ with small $\delta > 0$.
See also appendix \ref{app:regge}.

Now we have prepared for the Fourier transformations. Applying the asymptotic behavior of hypergeometric function in \eqref{eq:Gauss} and the expansion of $G_R (\omega , k - \alpha)$  in \eqref{eq:GRte}, the function \eqref{eq:gr} can be put into the form
\begin{align} 
\begin{aligned}\label{eq:gE2}
    &g_R (t , \theta)  = \sum_{n=0}^\infty g_R^{(n)} (t , \theta) , \\
    &g_R^{(n)} (t , \theta) \simeq \tilde a_n    \int^{\infty + i \delta}_{- \infty + i \delta} d \omega \int _{- \infty + i \delta '}^{\infty + i \delta '} d k  e^{- i \omega t + i k \theta} k^{\alpha}
 (\omega^2 -k^2)^\nu   e^{2 i n S (0,z_-)}  .
\end{aligned}
\end{align}
Here we set
\begin{align}
    \tilde a_n =  \frac{\Gamma (- \nu)}{\Gamma(\nu)} \frac{e^{- i\frac{\pi}{2 } \alpha + i \pi \alpha \lfloor \frac{\theta}{\pi} \rfloor}}{2^{\alpha +1 + 2 \nu} \pi |\sin \theta|^\alpha \Gamma(1+\alpha)}a_n .
\end{align}
We can show that the term with $n=0$ reproduces the light-cone singularity (see \cite{Dodelson:2023nnr}), so we focus on the case with $n \neq 0$.

We would like to evaluate the integral \eqref{eq:gE2} by the saddle point approximation. 
Before examining the integral, we  recall the saddle point approximation with two integral variables.
We consider an integral of the form
\begin{align}
I (\lambda) = \iint g(x,y) e^{- \lambda f(x,y)} dx dy .
\end{align}
We first find a saddle point $(x,y) = (x_0 , y_0)$ satisfying 
\begin{align}
\left. \frac{\partial}{\partial x} f(x , y) \right|_{(x,y) = (x_0,y_0)} = 0 , \quad 
\left. \frac{\partial}{\partial y} f(x , y) \right|_{(x,y) = (x_0,y_0)} = 0 .
\end{align}
In the case with several saddle points, we sum over these saddle points.
The functions in the integrand can be approximated by
\begin{align}
&f(x ,y) \simeq f(x_0 , y_0) + \frac12 (x - x_0 , y - y_0) \cdot H \cdot 
\begin{pmatrix} x - x_0 \\ y - y_0 \end{pmatrix}  , \\ 
&g(x,y) \simeq g(x_0 , y_0) .
\end{align}
Here $H$ is defined by
\begin{align}
H = \left. 
\begin{pmatrix}
\frac{\partial ^2 f(x,y)}{\partial x \partial x} & \frac{\partial ^2 f(x,y)}{\partial x \partial y} \\ 
\frac{\partial ^2 f(x,y)}{\partial y \partial x} & \frac{\partial ^2 f(x,y)}{\partial y \partial y}
\end{pmatrix}
\right|_{(x,y) = (x_0 , y_0)} .
\end{align}
The integral is then approximated by
\begin{align} \label{eq:SPA}
I(\lambda ) \simeq \left( \frac{2 \pi}{\lambda}\right ) \cdot \frac{g(x_0 ,y_0)}{\sqrt{\det H}} \cdot e^{-\lambda f(x_0 , y_0) } .
\end{align}

We apply the saddle point approximation to the integral \eqref{eq:gE2}.
For this, we define the function
\begin{align}
\Phi (\omega ,k)= - \omega t + k \theta + 2 n S(0,z_-) .
\end{align}
The saddle points of $\Phi(\omega ,k)$ can be obtained by $\omega = \omega_*, k = k_*$ satisfying
\begin{align} \label{eq:saddlept}
\left.
\frac{\partial \Phi (\omega , k)}{\partial \omega}
\right|_{(\omega ,k) = (\omega_*, k_*)} = 0 , \quad 
\left.
\frac{\partial \Phi (\omega , k)}{\partial k}
\right|_{(\omega ,k) = (\omega_*, k_*)} = 0 .
\end{align}
The function $\Phi (\omega , k)$ includes $S(0,z_-)$ defined by \eqref{eq:S0m} with \eqref{eq:kappa}.
The derivatives of the function $S(0,z_-)$ with respect to $\omega ,k$ can be rewritten as
\begin{align} \label{eq:DerS}
\begin{aligned}
&2 \frac{\partial S (0,z_-)}{\partial \omega} = 2 \int_{r (z_-)}^\infty \frac{\omega}{ \sqrt{\omega^2 - (k^2 - (\nu^2 - \frac14)r^2 ) \frac{f(r)}{r^2}}} \frac{dr}{f(r)} = T(\varrho  \omega , \varrho k  ) , \\
&2 \frac{\partial  S (0,z_-)}{\partial k} = - 2 \int_{r (z_-)}^\infty \frac{k}{ \sqrt{\omega^2 - (k^2 - (\nu^2 - \frac14)r^2 ) \frac{f(r)}{r^2}}} \frac{dr}{r^2} = - \Theta(\varrho  \omega  ,\varrho k  ) ,
\end{aligned}
\end{align}
where we set $\varrho = 1/\sqrt{\nu^2 - 1/4}$.
The functions $T(E , L)$ and $\Theta (E , L)$
with $E = \varrho \omega $ and $L = \varrho k $ are the arrival time and angle for space-like geodesic introduced in \eqref{eq:TThetanull}.
With \eqref{eq:DerS}, the equations \eqref{eq:saddlept} can be reduced to
\begin{align} \label{eq:geodesicwave}
t = n T(\varrho\omega_*  , \varrho k_*) , \quad \theta = n \Theta (\varrho \omega_*  , \varrho k_*) . 
\end{align}
From \eqref{eq:SPA}, the term with $n > 0$ can be approximated as
\begin{align} \label{eq:SPAm}
g_R^{(n)} (t , \theta) \simeq \tilde a_n k_*^\alpha ((\omega_*)^2 - (k_*)^2 )^\nu \frac{2 \pi i }{\sqrt{\det H}}  e^{ - i \omega_* t + i k _* \theta + 2 i n S_*(0,z_-) } ,
\end{align}
where we set 
\begin{align} \label{eq:H}
H = \left. 
\begin{pmatrix}
\frac{\partial ^2 \Phi(\omega,k)}{\partial \omega \partial \omega} & \frac{\partial^2 \Phi(\omega,k)}{\partial \omega \partial k} \\ 
\frac{\partial ^2 \Phi(\omega,k)}{\partial k \partial \omega} & \frac{\partial ^2 \Phi(\omega,k)}{\partial k \partial k}
\end{pmatrix}
\right|_{(\omega,k) = (\omega_* , k_*)} .
\end{align}

The expression \eqref{eq:SPAm} is given in terms of the energy and angular momentum at the saddle point $(\omega_* , k_*)$. Using \eqref{eq:saddlept} and \eqref{eq:DerS}, we rewrite them such that the geodesic meaning is manifest as in \eqref{eq:2ptsemi}. As in \eqref{eq:Tnull}, we may approximate $T(\varrho \omega_*,\varrho k_*),\Theta(\varrho \omega_* , \varrho k_*) $ by the arrival time and angle of null geodesics as
\begin{align} 
\label{eq:TTnull}
\begin{aligned}
    &T(\varrho \omega_*  , \varrho k_* ) = T_\text{null} (u_*) - \frac{2  u_*}{(u_*^2 - 1)\varrho k_*} + \mathcal{O} (k_*^{-2}) , \\
    &\Theta(\varrho \omega_*  , \varrho k_*) = \Theta_\text{null} (u_*) - \frac{2  }{(u_*^2 - 1)\varrho k_*} + \mathcal{O} (k_*^{-2}),
    \end{aligned}
\end{align}
where we set $u_* = \omega_* /k_*$.
We first examine the factor $ e^{ - i \omega_* t + i k _* \theta + 2 i n S_*(0,z_-) }$.
From \eqref{eq:TTnull}, we can see that the arrival time and angle of space-like geodesic can be approximated by those of null-geodesic for large $k_*$.
Using
\begin{align}
2 S_*(0,z_-) = \omega_* T_\text{null} (u_*) - k_* \Theta_\text{null} (u_*) + \mathcal{O} \left(k_*^{-1} \right),
\end{align}
we can set
$ e^{ - i \omega_* t + i k _* \theta + 2 i n S_*(0,z_-) } \simeq 1$.
We then consider the polynomial part given by $k_*^\alpha ((\omega_*)^2 - (k_*)^2)^\nu $.
Using \eqref{eq:TTnull}, we can rewrite it as
\begin{align}
    k_*^\alpha ((\omega_*)^2 - (k_*)^2)^\nu \simeq \frac{(2 \bar \varrho n  u_* )^{\alpha + 2 \nu }}{(u_*^2 - 1)^{\alpha + \nu}} \frac{1}{( n T_\text{null} (u_*) - t)^{\alpha + 2 \nu}} .
\end{align}
Here we set $\bar \varrho \equiv \varrho^{-1} = \sqrt{\nu^2 - 1/4}$.
Finally, we evaluate $\sqrt{\text{det} \, H}$ with $H$ in \eqref{eq:H}.
Again with \eqref{eq:TTnull} we find
\begin{align}
&\left.  \frac{\partial}{\partial \omega} T(\varrho \omega  , \varrho k  ) \right|_{(\omega ,k) = (\omega_* , k_*)} \sim  \frac{1}{k_*} T_\text{null} ' (u_*) + \frac{1}{\varrho (\omega_* - k_*)^2} + \frac{1}{\varrho (\omega_* + k_*)^2} ,\\ 
&\left. \frac{\partial}{\partial k} T(\varrho \omega  , \varrho k ) \right|_{(\omega ,k) =   (\omega_* , k_*)} \sim - \frac{\omega_*}{(k_*)^2} T_\text{null} ' (u_*)- \frac{4  \omega_* k_*}{\varrho(\omega_*^2 - k_*^2)^2} ,\\ 
&\left. \frac{\partial}{\partial \omega} \Theta(\varrho\omega ,\varrho k) \right|_{(\omega ,k) = (\omega_* , k_*)} \sim \frac{\omega_*}{(k_*)^2} \Theta_\text{null} ' (u_*) +  \frac{4 \omega_* k_*}{\varrho (\omega_*^2 - k_*^2)^2} ,\\ 
&\left. \frac{\partial}{\partial k} \Theta(\varrho \omega ,\varrho k) \right|_{(\omega ,k) = (\omega_* , k_*)} \sim - \frac{(\omega_*)^2}{(k_*)^3} \Theta_\text{null} ' (u_*) - \frac{1}{\varrho(\omega_* - k_*)^2} - \frac{1}{ \varrho  (\omega_* + k_*)^2}. 
\end{align}
Here we have used 
\begin{align}
T'_\text{null} (u_*) \equiv \left. \frac{\partial}{\partial u} T_\text{null} (u) \right|_{u = u_*} , \quad \Theta '_\text{null} (u_*) \equiv  \left. \frac{\partial}{\partial u} \Theta _\text{null} (u) \right|_{u = u_*} .
\end{align}
Noticing that
\begin{align}
  \Theta_\text{null}'(u)  = u T_\text{null}'(u) ,
\end{align}
we find
\begin{align}
    \sqrt{\det H} \simeq  \sqrt{\frac{2  n^2 T_\text{null}'(u_*) }{ \varrho k_*^3}}.
\end{align}
Finally, the expression \eqref{eq:SPAm} can now be rewritten as
\begin{align} \label{eq:gRn}
    g_R^{(n)} (t, \theta ) \simeq \tilde a_n\frac{(2 \bar \varrho n u_* )^{\alpha + 2 \nu + 3/2}}{(u_*^2 - 1)^{\alpha + \nu + 3/2}} \frac{1}{( n T_\text{null} (u_*) - t)^{\alpha + 2 \nu + 3/2}} \frac{2 \pi i}{\sqrt{ 2 \bar \varrho n^2 T_\text{null}'(u_*) }}.
\end{align}
From this, we can see that the positive integer $n$ counts the number of bounce at the AdS boundary. The expression can be regarded as a more accurate version of \eqref{eq:2ptsemi} obtained by the geodesic approach.
Here we have used the explicit example of AdS-Schwarzschild black hole, but the arguments basically hold for a large class of asymptotic AdS spacetime as seen below for explicit examples. In particular, the power of singularity is 
\begin{align} \label{eq:power}
 2 \Delta - \frac{d-1}{2} = \alpha + 2 \nu + \frac32 
\end{align}
for more generic background as well.

The expression in \eqref{eq:gRn} coincides with the one in \cite{Dodelson:2023nnr} for $n=1$ up to overall factor. For instance, the power of singularity \eqref{eq:power} is reproduced. Moreover the factor depending on $T '_\text{null} (u_*) \sim e^{\gamma t}$ explains the classical Lyapunov exponent $\gamma$ at the photon sphere. In \cite{Dodelson:2023nnr}, the term proportional to $r^2$ in the potential \eqref{eq:kappa} was neglected in the Eikonal approximation. Because of this, the authors were directly working on the null limit, where the saddle point equations in \eqref{eq:saddlept} are degenerated. The saddle points are realized by pinched singularities and the condition for the pinch have been adopted. The result does not change even by including the term proportional to $r^2$ up to the overall factor. Even so, we believe that our way of computation clarifies how the singularities are developed when the space-like geodesic approaches null.

As in \eqref{eq:sumj}, the retarded Green function is given by a sum over $j$, which corresponds to the winding number. Using \eqref{eq:gRn}, we have for $0 < \theta < \pi$
\begin{align}
\label{eq:GRtth}
\begin{aligned}
    G_R ( t, \theta ) &\simeq   \sum_{n=1}^\infty \tilde a_n\frac{(2 \bar \varrho n u_* )^{\alpha + 2 \nu + 3/2}}{(u_*^2 - 1)^{\alpha + \nu + 3/2}}  \frac{2 \pi i}{\sqrt{2 \bar \varrho n^2 T_\text{null}'(u_*) }} \\ & \quad \times
    \sum_{j=1}^\infty (-1)^{jd} (X_j^{n,+} (t,\theta) + e^{i \pi \frac{d-2}{2}} X^{n,-}_j (t , \theta) ) ,
    \end{aligned}
\end{align}
where we set 
\begin{align} \label{eq:Xnpmj}
X_j^{n , \pm} (t , \theta) = 
\frac{1}{( n t_\text{BC} (2 \pi j \pm \theta) - t)^{\alpha + 2 \nu + 3/2}} 
\end{align}
with $t_\text{BC} (\theta) = T_\text{null} (u_*)$. As explained in \cite{Dodelson:2023nnr}, the structure of singularity depends on the dimension.
For $d =6,10,14,\ldots$, the behavior of singularity does not depends on $j$, so the structure is one-fold. For $d = 4,8,10$, the Green function is proportional to $\sum_j (X_k^{n,+} - X_k^{n,-})$ and the overall sign changes if we replace $+$ and $-$. Thus the structure is two-fold. For $d=3,7,11,\ldots$ and $d = 5,9,1,3\ldots$, the Green functions are proportional to $\sum_j (-1)^j (X_k^{n,+} \pm i X_k^{n,-})$, so the structures are four-fold. 
In section \ref{sec:numerical},
we will observe the four-fold and two-fold structures from the numerical results for $d=3$ and $d=4$, respectively.

\section{AdS ECOs}
\label{sec:AdSECOs}

In the previous section, we considered AdS-Schwarzschild black hole and null geodesics going around the photon sphere. In this section, we apply the analysis to AdS ECOs with photon sphere but without black hole horizon. 
We find that the new type of bulk-cone singularities appear in these cases. Furthermore, we observe bumps called echoes, which follow the bulk-cone singularities related to the photon sphere.
We first derive these signatures in the retarded Green functions for generic ECOs.
We then examine AdS gravastar and AdS wormhole as concrete examples.

\subsection{Bulk-cone singularities}
\label{sec:BCecos}

In this subsection, we point out that there are  bulk-cone singularities in the CFT correlators, which are specific to ECOs with photon sphere but without black hole horizon. 
Let us first consider the two-point function of scalar operator computed by the geodesic approach. As in \eqref{eq:2ptsemi}, there are singularities when the two boundary points $(0,0)$ and $(t,\theta)$ are related by bulk null geodesics. The equation for null geodesics can be expressed by \eqref{eq:Schcla}, where the energy and angular momentum are given by \eqref{eq:EL}.

\begin{figure}[htbp]
\centering
\includegraphics[height=4cm]{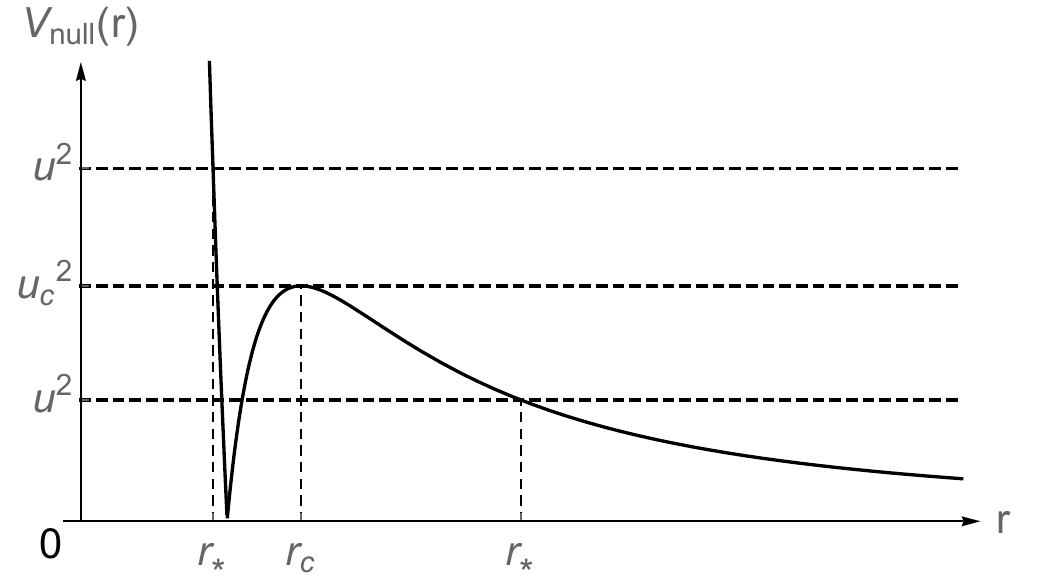}
\caption{The potential for AdS gravastar as a typical example of ECOs.
}
\label{fig:Vgravg}
\end{figure}

In the previous section, we focus on the case of AdS-Schwarzschild black hole with the metric function $f(r)$ in \eqref{eq:AdSS}.
However, the analysis holds also for generic geometry of the form \eqref{eq:metric} if $E,L,E^2 -L^2$ are large enough.
A difference arises the region near the center.
In the case of AdS-Schwarzschild black hole, the region is surrounded by the black hole horizon and cannot be accessed.
In the case of ECOs without black hole horizon, there would be a high potential at the central region due to the centrifugal force. See fig.\,\ref{fig:Vgravg} for AdS gravastar.
The structure of potential is expected be similar for other ECOs. An exception is the potential for AdS wormhole, which will be separately analyzed in subsection \ref{sec:AdSwormhole}.

We are interested in the geometry where the region outside the photon sphere is AdS-Schwarzschild spacetime.
A maximum of the potential is then given by $u_c$ in \eqref{eq:Vmax} located at the position of photon sphere, $r = r_c$, as in \eqref{eq:rc}.
For $1 < u < u_c$, the corresponding null trajectory originates from the AdS boundary, turns around near the photon sphere, and then returns to the AdS boundary. The same null trajectory exists for AdS-Schwarzschild black hole, so the CFT correlators have this type of bulk-cone singularities even for ECOs with photon sphere. A new type of null trajectories exist for $u > u_c$, which is bounced by the potential due to the centrifugal force.
This is a specific feature of geometry without black hole horizon. Once we can observe this type of bulk-cone singularities, then we can see that the dual geometry is something other than AdS-Schwarzschild black hole.

\begin{figure}[htbp]
\centering
\includegraphics[height=4cm]{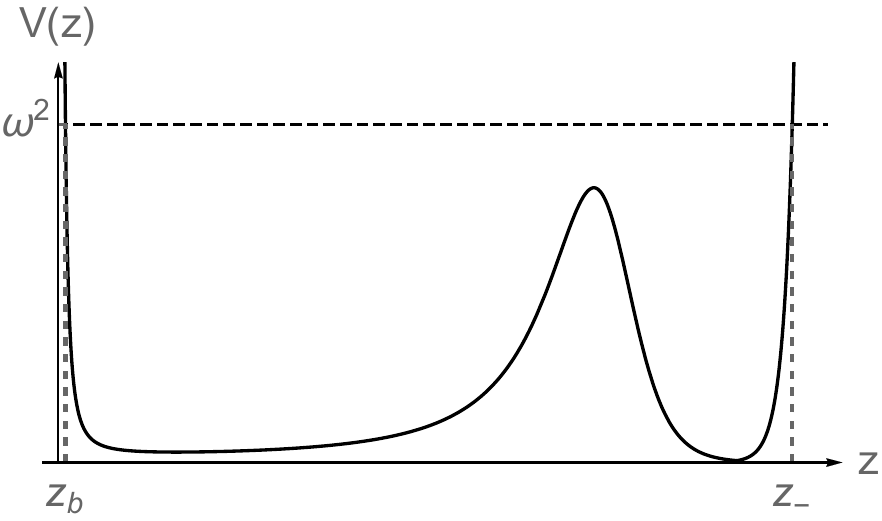}
\caption{The potential for typical ECOs. 
}
\label{fig:Vgrav2}
\end{figure}

We next move to wave analysis. The Klein-Gordon equation for bulk scalar field is reduced to the wave equation \eqref{eq:wave}, where $\kappa (z)$ and the potential $\tilde V(z)$ are given in \eqref{eq:kappa}. We have solved the wave equation in the case of AdS-Schwarzschild black hole by applying the WKB approximation. Now we would like to study ECOs. We consider the case with high potential near $z \to \infty$ due to the centrifugal force, see fig.\,\ref{fig:Vgrav2}. 
As discussed above, we expect that the CFT correlators have bulk-cone singularities specific to geometry without black hole horizon. These bulk-cone singularities are related to null trajectories bouncing at the centrifugal force. 
In the rest of this subsection,
we consider the case where the equation $\kappa (z) ^2 = 0$ has two zeros $z_b \, (\simeq 0) , z_-$. There could be the case with four zeros of $\kappa (z) ^2  = 0$, which will be analyzed in the next subsection.

We would like to compute the retarded Green function of scalar operator from the wave function of dual bulk scalar field by applying \eqref{eq:GRSon}. For AdS-Schwarzschild black hole, we have assigned the ingoing boundary condition at the horizon. For an AdS ECO, we  assign the regularity condition at the center $r \to 0$ as nothing should happen there. 
The metric function $f(r)$ is assumed to be finite as $r \to 0$, thus  $ V(z) \to  r^{-2}$ near $r \to 0$. 
For $z > z_-$, the solution satisfying the regularity condition at $r \to 0$ can be approximated by
\begin{align}
    \psi (z) \sim  \frac{1}{\sqrt{q(z)}} e^{-  \int^z_{z_-} d z' q (z')} .
\end{align}
Here $q(z)$ is defined as in \eqref{eq:Smp}.
Applying the WKB connection formula, the wave function in the region $0 \simeq z_b < z < z_-$ is obtained as
\begin{align}
    \psi (z) \sim \frac{e^{\frac{i \pi}{4} - i S(0,z_-)}}{\sqrt{\kappa(z)}} e^{i  \int^z_{0} d z' \kappa (z')} + \frac{e^{-\frac{i \pi}{4} + i S(0,z_-)}}{\sqrt{\kappa(z)}} e^{- i \int^z_{0} d z' \kappa (z')} .
\end{align}
When the wave function is of the form \eqref{eq:psito0} for $0 \simeq z_b < z < z_-$, the retarded Green function is obtained as in \eqref{eq:retardedAdSG}. Utilizing the result, the retarded Green function can be found as
\begin{align} 
\begin{aligned}
    G_R (\omega ,\ell) &=   \frac{\Gamma (- \nu)}{\Gamma(\nu)} 
    \left( \frac{\omega^2 - \ell^2}{4}\right)^\nu  \frac{  \cos (S(0,z_-) +\frac{\pi \nu }{2})  }{ \cos (S(0,z_-) - \frac{\pi \nu }{2})   } \, .
    \end{aligned}
\end{align}
As in the case of AdS-Schwarzschild black hole, it can be expanded as
\begin{align} 
    G_R (\omega ,\ell) &=   \frac{\Gamma (- \nu)}{\Gamma(\nu)} 
    \left( \frac{\omega^2 - \ell^2}{4}\right)^\nu \sum_{n=0}^\infty a_n e^{2 i n S(0,z_-)} 
\end{align}
 with
\begin{align}  
a_0 =  e^{-i\pi \nu} , \quad a_{n} = - 2 i  (-1)^n e^{ -i n \pi \nu}  \sin ( \pi \nu ) .
\end{align}
The form of the retarded Green functions is the same as that for AdS-Schwarzschild black hole. The difference is encoded in the function $S(0, z_-)$ given by \eqref{eq:S0m}, which depends on the metric function $f(r)$.
In the coordinate basis, the retarded Green function is expressed by \eqref{eq:GRtth} with \eqref{eq:Xnpmj}.
There are bulk-cone singularities when null geodesics exist between the boundary points $(0,0)$ and $(t,\theta)$ and this is consistent with the result obtained by the geodesic approach.

\subsection{Echoes}
\label{sec:Echoes}

\begin{figure}[htbp]
\centering
\includegraphics[height=4cm]{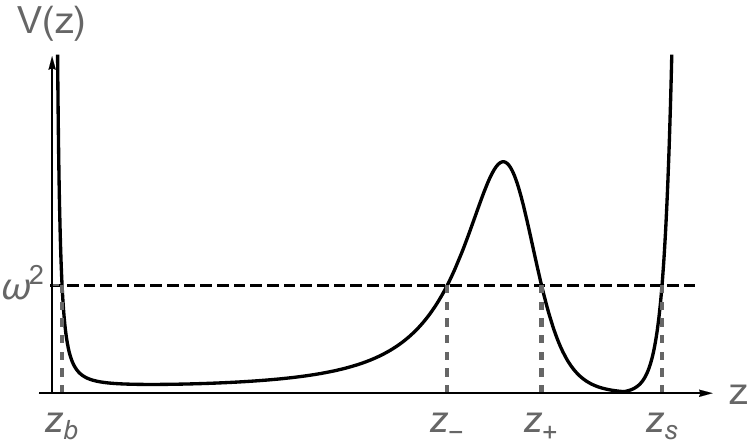}
\caption{The potential for typical ECOs.}
\label{fig:Vgrav}
\end{figure}

In this subsection, we examine the wave equation \eqref{eq:wave} where the
equation $\kappa (z) ^2 = 0$ has four zeros as in fig.\,\ref{fig:Vgrav}. 
The four zeros are denoted by $0 \simeq z_b < z_- < z_+ < z_s$. 
We thus examine the wave functions of bulk scalar field for the case.
The regarded Green functions obtained by the wave analysis can be expanded as
\begin{align} \label{eq:GRAB}
    G_R (\omega ,\ell) &= \frac{\Gamma (- \nu)}{\Gamma(\nu)} 
    \left( \frac{\omega^2 - \ell^2}{4}\right)^\nu (A + B e^{-2 S(z_- , z_+)} + \mathcal{O} (e^{-4 S(z_- , z_+)} )) .
\end{align}
The terms denoted by $A$ and $B$ are 
\begin{align} \label{eq:AB}
A = \sum_{n=0}^\infty a_n e^{2 i n S(0,z_-)} , \quad  B =  \sum_{n = 1}^\infty \sum_{p=0}^\infty b_{n,p} e^{2 i n S(0,z_-) + 2 i p S(z_+ , z_s)} .
\end{align}
The leading contributions come from the terms represented by $A$. The part is actually the same as the corresponding contributions for AdS-Schwarzschild black hole. This is because $z_-$ is now outside the photon sphere and we choose to keep the outside region to be AdS-Schwarzschild geometry. 
The terms represented by $B$ are associated with a suppression factor  $e^{- 2 S(z_-,z_+)}$ and they can be regarded as tunneling effects. 
Recall that the Sch{\"o}dinger equation comes from the Klein-Gordon equation for bulk scalar field, and the Planck constant is given by the inverse of angular momentum, $1/\ell$. In the rest of this subsection, we shall see that there are bumps arising from the part $B$. These bumps are known as the echoes, which come after the bulk-cone singularities associated with null trajectories going around the photon sphere.

We convert the momentum basis expression in \eqref{eq:GRAB} into the coordinate basis one as
\begin{align}
g_R(t,\theta)  & = \sum_{n=0}^\infty g_R^{(n)} (t , \theta) +  \sum_{n = 1}^\infty \sum_{p=0}^\infty   g_R^{(n,p)} (t , \theta) .
\end{align}
Here $g_R^{(n)} (t , \theta)$ are obtained as in \eqref{eq:gE2} and  $ g_R^{(n,p)} (t , \theta)$ are defined by
\begin{align} 
\label{eq:G2}
\begin{aligned}
& g_R^{(n,p)} (t , \theta) \\
&\simeq \tilde b_{n,p} \int_{- \infty + i \delta}^{\infty + i \delta} d \omega  \int _{- \infty + i \delta '}^{\infty + i \delta '}   d k  e^{- 2 S(z_- , z_+) } k^{\alpha}(\omega^2 - k^2)^\nu e^{- i \omega t}  e^{2 i n S(0,z_-) + 2 i p S(z_+ , z_s)} 
\end{aligned}
\end{align}
with
\begin{align}
\tilde b_{n,p} 
= \frac{\Gamma (-\nu)}{\Gamma (\nu)} 
\frac{e^{- i\frac{ \pi}{2} \alpha + i \pi \alpha \lfloor \frac{\theta}{\pi} \rfloor}}{
2^{\alpha + 1 + 2 \nu}
\pi |\sin \theta|^\alpha \Gamma (1+\alpha) 
}
b_{n,p} .
\end{align}
Here $b_{n,p}$ were introduced in \eqref{eq:AB}.

We would like to evaluate the integrals by the saddle point approximation.
The saddle point approximation can be justified for the regime where $\ell$ is  relatively large. 
Here we assume that  $|S(z_- ,z_+)| \ll |S(0,z_-)| , |S(z_+ , z_s)|$.
We look for the saddle point $(\omega ,k) = (\omega_* , k_*)$ satisfying
\begin{align} \label{eq:saddlept2}
\left.
\frac{\partial \Phi (\omega , k)}{\partial \omega}
\right|_{(\omega ,k) = (\omega_*, k_*)} = 0 , \quad 
\left.
\frac{\partial \Phi (\omega , k)}{\partial k}
\right|_{(\omega ,k) = (\omega_*, k_*)} = 0 
\end{align}
with
\begin{align}
\Phi (\omega ,k)= - \omega t + k \theta + 2 n S(0,z_-) + 2 p S(z_+ , z_s).
\end{align}
The equations \eqref{eq:saddlept2}  reduce to
\begin{align}
\begin{aligned}
&t = n T(\varrho  \omega_* , \varrho k_* ) + p \tilde T(\varrho \omega_* , \varrho k_* ) , \\ 
&\theta = n \Theta (\varrho \omega_*  , \varrho k_* )  + p \tilde \Theta (\varrho \omega_* , \varrho k_* ) .
\end{aligned}
\end{align}
Here we have introduced
\begin{align} \label{eq:DerS2}
\begin{aligned}
&2 \frac{\partial S (z_+,z_s)}{\partial \omega} = 2 \int_{r (z_s)}^{r(z_+)} \frac{\omega}{ \sqrt{\omega^2 - (k^2 - (\nu^2 - \frac14)r^2 ) \frac{f(r)}{r^2}}} \frac{dr}{f(r)}  \equiv \tilde T(\varrho \omega , \varrho k) , \\
&2 \frac{\partial  S (z_+,z_s)}{\partial k} = - 2 \int_{r (z_s)}^{r(z_+)} \frac{k}{ \sqrt{\omega^2 - (k^2 - (\nu^2 - \frac14)r^2 ) \frac{f(r)}{r^2}}} \frac{dr}{r^2}  \equiv - \tilde \Theta(\varrho \omega , \varrho k) ,
\end{aligned}
\end{align}
where $\tilde T(\varrho \omega_* , \varrho k_* )$ and $\tilde \Theta(\varrho \omega_* , \varrho k_*) $ correspond to the shift of time and angle for one period of space-like geodesic between $z_+$ and $z_s$.

For large $\omega_*, k_*$, we may set $T( \varrho \omega_* , \varrho k_* )$, $\Theta(\varrho \omega_*   , \varrho k_* ) $ as \eqref{eq:TTnull} and $\tilde T(\varrho \omega_* ,\varrho k_*)$, $\tilde \Theta(\varrho \omega_*  , \varrho k_*) $ as
\begin{align}
    \tilde T(\varrho \omega_*  , \varrho k_*) = \tilde T_\text{null} (u_*)  + \mathcal{O} (k_*^{-2}) , \quad     \tilde \Theta(\varrho \omega_* , \varrho k_*) = \tilde \Theta_\text{null} (u_*) + \mathcal{O} (k_*^{-2}).
\end{align}
Here we defined
\begin{align} 
\label{eq:TThetat}
\begin{aligned}
&\tilde T_\text{null}(u) = 2 u \int_{r(z_s)}^{r(r_+)} \frac{dr}{f(r)} \frac{1}{\sqrt{u^2 -V_\text{null} (r) }} , \\
&\tilde \Theta_\text{null}(u) = 2  \int_{r(z_s)}^{r(z_+)} \frac{dr}{r^2} \frac{1}{\sqrt{u^2 -V_\text{null} (r) }} ,
\end{aligned}
\end{align}
which correspond, respectively, to the shift of time and angle for one period of null geodesic between $z_+$ and $z_s$.
Following the analysis in subsection \ref{sec:coordinate}, we arrive at
\begin{align} \label{eq:gRnt}
\begin{aligned}
    & g_R^{(n,p)} (t, \theta )  \simeq \tilde b_{n,p} e^{- 2 S(z_-,z_+)} \frac{(2 \bar \varrho n u_* )^{\alpha + 2 \nu + 3/2}}{(u_*^2 - 1)^{\alpha + \nu + 3/2}} \\
   & \quad  \times \frac{1}{( n T_\text{null} (u_*) + p \tilde T_\text{null} (u_*) - t)^{\alpha + 2 \nu + 3/2}} \frac{2 \pi i }{\sqrt{ 2 \bar \varrho n (n T_\text{null}'(u_*) + p \tilde T _\text{null} ' (u_*) )}}.
   \end{aligned}
\end{align}
In the above expression, we have singularities when $n T_\text{null} (u_*) + p \tilde T_{\null} (u_*) - t= 0$. 
Here $S(z_- ,z_+)$ in \eqref{eq:G2} has been evaluated at the saddle point $(\omega , k) = (\omega_*, k_*)$.
We may treat $S(z_- ,z_+)$  in a similar way as done for $S(0,z_-) , S(z_+,z_s)$.
This shifts the position of singularity to a complex value as 
\begin{align} \label{eq:gRnt2}
\begin{aligned}
    & g_R^{(n,p)} (t, \theta )  \simeq \tilde b_{n,p}  \frac{(2 \bar \varrho  n  u_* )^{\alpha + 2 \nu + 3/2}}{(u_*^2 - 1)^{\alpha + \nu + 3/2}} \\
   & \quad  \times \frac{1}{( n T_\text{null} (u_*) + p \tilde T_\text{null} (u_*) - t + i \epsilon)^{\alpha + 2 \nu + 3/2}} \frac{2 \pi i }{\sqrt{ 2 \bar \varrho  n (n T_\text{null}'(u_*) + p \tilde T _\text{null} ' (u_*) )}}.
   \end{aligned}
\end{align}
Here $\epsilon$ takes a small real value  and the singularity is replaced by a bump whose center is at $t = n T_\text{null} (u_*) + p \tilde T_{\null} (u_*)$. 
The bulk-cone singularities associated with null trajectories going around the photon sphere are with $p=0$. The expression in \eqref{eq:gRnt2} suggests that there are bumps after the bulk-cone singularities with the period of $\tilde  T (u_*)$. These bumps can be identified as echoes, which were discussed in \cite{Cardoso:2016rao,Cardoso:2016oxy} for the case with asymptotically flat spacetime.

\subsection{AdS gravastar}
\label{sec:gravastar}

As explained above, there are two signatures indicating that the geometry is an ECO without black hole horizon. One is the presence of bulk-cone singularities associated with null trajectories that travel into the region inside the photon sphere. The other is the appearance of echoes, which follow the bulk-cone singularities associated with null trajectories that circulate around the photon sphere. In this subsection, we examine this by analyzing a concrete example of ECOs, i.e., AdS gravastar.
In the next subsection, we consider AdS wormhole as another concrete example of ECOs.

\subsubsection{Construction of the geometry}
\label{sec:junction}

We first construct a gravastar solution as an asymptotic AdS spacetime by extending the four-dimensional geometry constructed in \cite{Chen:2025cee}. See \cite{Mazur:2001fv,Visser:2003ge,Pani:2009ss,Cardoso:2014sna} for the asymptotic flat case.
We place a shell at $r= r_0$ and glue the two geometries with the metrics
\begin{align} \label{eq:ansatz}
    ds^2_\pm=-f_\pm(r_\pm)dt_\pm^2+g_\pm(r_\pm)dr_\pm^2+r_\pm^2d\Omega^2_{d-1}.
\end{align}
The subscripts $+$ and $-$ indicate that the metrics and coordinates are those for the exterior and interior regions, respectively.
We require that the induced metrics 
$h_\pm$,
\begin{align} \label{eq:induced}
    h_\pm=-f_\pm(r_0) dt_\pm^2+r_0^2d\Omega^2_{d-1},
\end{align}
to be continuous, i.e., $h_+=h_-\equiv h$.
The condition gives the relation between the time coordinates,
\begin{align} \label{eq:cond_ind}
    \frac{dt_+^2}{dt_-^2}=\frac{f_-(r_0)}{f_+(r_0)}.
\end{align}
We further require the junction condition
\begin{align} \label{eq:cond_energy}
    [[\chi]]-[[\text{tr}\, \chi]]h=-8\pi \mathcal{S} ,
\end{align}
where $\chi$ is the extrinsic curvature on the time-like hypersurface and the bracket $[[A]]=A_+-A_-$ denotes the jump of a quantity $A$ across the shell.
We set the form of surface stress-energy tensor as
\begin{align}
\mathcal{S}=\sigma d\tau^2+\bar P r_0^2d\Omega^2_{d-1},
\end{align}
where $\sigma$ and $\bar P$ are the surface energy density and the surface tension, respectively. Moreover, $\tau$ is the proper time on the shell satisfying $d\tau\propto dt_\pm$.

We are interested in the case where the exterior region is given by AdS-Schwarzschild geometry and the interior region is described by dS spacetime.
In order to realize these geometries, we set the metric functions as
\begin{align} \label{eq:fpfm}
\begin{aligned}
    f_+(r)&=g_+^{-1}(r)=1-\frac{\mu}{r^{d-2}}+ \frac{r^2}{R_\mathrm{AdS}^2},\\
    f_-(r)&=g_-^{-1}(r)=1-\frac{r^2}{R_\mathrm{dS}^2},
\end{aligned}
\end{align}
where $R_\mathrm{dS}$ and $R_\mathrm{AdS}$ are the dS and AdS length, respectively, and we take $R_\mathrm{AdS}=1$ throughout this paper.
The mass parameter $\mu$ is related to the ADM mass $M$ as 
\begin{align}
\mu=w_d M , \quad w_d= \frac{16\pi G M}{(d-1)\Omega_{d-1}} .
\end{align}
Here $G$ is the gravitational constant and $\Omega_{d-1}$ is the area of a unit $(d-1)$-sphere.
Additionally, we need to specify the equation of state 
of the shell, $\bar P =\bar P (\sigma)$, to construct the AdS gravastar.
In order to make the analysis simpler, we set the surface energy density as $\sigma = 0$ in the following.

We can put the metric of AdS gravastar into the form  \eqref{eq:metric}, where the metric function $f(r)$ is
\begin{align}
\label{eq:gravastar-f}
    f(r)=\left\{\begin{array}{ll}
    1- \frac{\mu}{r}-\frac{2\Lambda}{d(d-1)} r^2 & (r\ge r_0), \\
    1-\left(\frac{8\pi \rho}{d} +\frac{2\Lambda}{d(d-1)}\right) r^2 & (r<r_0),
    \end{array}\right.
\end{align}
where $\rho=(d/(8\pi))(R_\mathrm{dS}^{-2}-R_\mathrm{AdS}^{-2})$ is the vacuum energy density in the dS region on the AdS background and $\Lambda=-d(d-1)/2R_\mathrm{AdS}^2$.
We assume that the function is continuous at the position of the shell, i.e., $r=r_0$, which leads to
\begin{align}
\label{eq:r0-mu-rho}
    r_0= \left( \frac{d\mu}{8\pi \rho} \right)^\frac1d.
\end{align}
We further set the shell to be located inside the photon sphere, i.e., $r_0 < r_c$, where $r_c$ is given by \eqref{eq:rc}. We would like to have a geometry without black hole horizon, which requires that $r_0 > r_h$. Here $r_h$ is the radius of horizon of the original AdS-Schwarzschild black hole satisfying \eqref{eq:rh}.

\subsubsection{Retarded Green functions}
\label{sec:RGFgravastar}

We have computed the retarded Green function in the CFT dual to a generic ECO. 
Here we explicitly examine it in the case of AdS gravastar constructed above. 
We thus solve the wave equation \eqref{eq:wave} in the WKB approximation with the potential $\tilde V(z)$ in \eqref{eq:kappa}.
Note that the metric function $f(r)$ is now given by \eqref{eq:gravastar-f}.

We first consider the case where the equation $\kappa (z)^2 = 0$ has two zeros at $z_b, z_-$ $(0 \simeq z_b < z_-)$, see fig.\,\ref{fig:Vgrav2}. The retarded Green function can be obtained as in \eqref{eq:GRtth} with \eqref{eq:Xnpmj}.
The functions $T_\text{null} (u_*)$ and $\Theta_\text{null} (u_*)$ are computed with $V_\text{null}(r) = f(r) r^{-2}$ and \eqref{eq:gravastar-f} as in \eqref{eq:Schcla}.
Here $u_*$ is a solution of $2 \pi j \pm \theta = n \Theta (u_*)$. 
The bulk-cone singularities are located at  $t = n T_\text{null} (u_*)$ and the geometry can be probed from the information. As will be numerically confirmed below, the retarded Green functions have bulk-cone singularities at the arrival time of null geodesics traveling the region inside the photon sphere. From this, we can distinguish the dual geometry from AdS-Schwarzschild black hole.

We then move to the case with four zeros in $\kappa (z) ^2  = 0$, see fig.\,\ref{fig:Vgrav}.
We denote these zeros by $z= z_b, z_\pm , z_s$ $(0 \simeq z_b < z_- < z_+ < z_s)$.
The wave function for $z > z_s$ is
\begin{align}
    \psi (z) \sim \frac{1}{\sqrt{q(z)}} e^{ -  \int^z_{z_s} d z' q (z')} 
\end{align}
as explained above. We connect it to the region $z_+ < z < z_s$ as 
\begin{align}
    \psi (z) \sim \frac{e^{\frac{i \pi}{4} - i S(z_+,z_s)}}{\sqrt{\kappa(z)}} e^{i  \int^z_{z_+} d z' \kappa (z')} + \frac{e^{-\frac{i \pi}{4} + i S(z_+,z_s)}}{\sqrt{\kappa(z)}} e^{- i  \int^z_{z_+} d z' \kappa (z')} .
\end{align}
We further continue the wave function to the region $0 \simeq z_b < z_-$  as
\begin{align}
    \psi (z) \sim \frac{ C_+}{\sqrt{\kappa(z)}} e^{i  \int^z_{0} d z' \kappa (z')} + \frac{ C_-}{\sqrt{ \kappa(z) }} e^{- i \int^z_{0} d z'  \kappa (z') }
\end{align}
with
\begin{align}
   &  C_+ = e^{\frac{i \pi}{4} - i S(z_+,z_s)} C_+^{(1)} + e^{-\frac{i \pi}{4} + i S(z_+,z_s)} C_-^{(2)} , \\
   & C_- =  e^{\frac{i \pi}{4} - i S(z_+,z_s)} C_- ^{(1)} +  e^{-\frac{i \pi}{4} + i S(z_+,z_s)} C_+ ^{(2)}  .
\end{align}
Here we set
\begin{align} \label{eq:C12pm}
\begin{aligned}
&C_+ ^{(1)}= \left (e^{S(z_- , z_+)} + \frac14e^{-S(z_- , z_+)} \right ) e^{-i S(0,z_-)}, \\ 
&C_-^{(1)}  = \left (e^{S(z_- , z_+)} - \frac14e^{-S(z_- , z_+)} \right )e^{ - \frac12 i \pi}  e^{i S(0,z_-)} , \\
&C_+ ^{(2)}= \left (e^{S(z_- , z_+)} + \frac14e^{-S(z_- , z_+)} \right ) e^{i S(0,z_-)}, \\ 
&C_-^{(2)}  = \left (e^{S(z_- , z_+)} - \frac14e^{-S(z_- , z_+)} \right )e^{  \frac12 i \pi}  e^{-i S(0,z_-)} .
\end{aligned}
\end{align}

The retarded Green function is given by \eqref{eq:retardedAdSG} for the wave function with the asymptotic behavior \eqref{eq:psito0}.
We thus find
\begin{align} 
\begin{aligned}
    G_R (\omega ,\ell) &=  \frac{\Gamma (- \nu)}{\Gamma(\nu)} 
    \left( \frac{\omega^2 - \ell^2}{4}\right)^\nu \\
    & \quad \times \frac{ D_+ \cos( S(0,z_-)+ \frac{ \pi \nu}{2}  ) - \frac{i}{4} e^{- 2 S(z_- ,z_+)} D_- \sin( S(0,z_-)+ \frac{ \pi \nu}{2}  )  }{ D_+ \cos (  S(0,z_-) - \frac{ \pi \nu}{2}  )  - \frac{i}{4} e^{- 2 S(z_- ,z_+)} D_-
    \sin( S(0,z_-) - \frac{ \pi \nu}{2}  ) } 
    \end{aligned}
\end{align}
with
\begin{align}
    D_\pm = 1 \pm e^{ 2 i S(z_+ ,z_s)} .
\end{align}
It is convenient to expand the retarded Green function in the form of \eqref{eq:GRAB} with \eqref{eq:AB}.
The coefficients are computed as
\begin{align}  \label{eq:GRtegravastar1}
a_0 =  e^{-i\pi \nu} , \quad  a_{n} = - 2 i  (-1)^n e^{ -i n \pi \nu}  \sin ( \pi \nu ) 
\end{align}
for $n=1,2,3,\ldots$
and
\begin{align}\label{eq:GRtegravastar2}
b_{n,0} =  i (-1)^{n} n e^{- i n \pi \nu} \sin ( \pi \nu) , \quad b_{n,p} =  2 i (-1)^{n + p} n  e^{- i n \pi \nu}  \sin ( \pi \nu) 
\end{align}
for $p=1,2,3,\ldots$.
As explained above, the contribution from the terms in $A$ is the same as that for AdS-Schwarzschild black hole. Furthermore, the contribution from the terms in $B$ explains the echoes which follow bulk-cone singularities arising from the terms in $A$.

\subsection{AdS thin-shell wormholes}
\label{sec:AdSwormhole}

In the previous subsection, we examined AdS gravastar as a representative of ECOs with photon sphere but without black hole horizon. In the case of AdS gravastar, the potential for the Sch{\"o}dinger equation has a high wall near the center of geometry due to the centrifugal force. This is a typical situation for ECOs but there could be examples not sharing such a property. An example is given by a wormhole solution, where the region inside the photon sphere is replaced by another region outside the horizon of AdS-Schwarzschild geometry. In this case, null trajectories are not reflected by the high potential near the center but go through the throat and arrive at the opposite AdS boundary. In this subsection, we study AdS wormhole solution as an example of atypical ECOs.

\subsubsection{Construction of the geometry}

We construct a wormhole solution as an asymptotic AdS spacetime by gluing two AdS-Schwarzschild geometries by a thin shell. See, e.g., \cite{Visser:1989kg,Morris:1988tu} for asymptotic flat wormholes.
We use the form of metrics for the two geometries by
\begin{align} 
    ds_\pm^2 = - f_\pm(r_\pm) dt^2 + f (r_\pm)^{-1} dr_\pm^2 + r_\pm^2 d \Omega_{d-1}^2,
\end{align}
where the metric functions are (see eq.\,\eqref{eq:AdSS}) 
\begin{equation}
    f_\pm (r_\pm) \, (\equiv f(r_\pm))=  r_\pm^2 + 1 - \frac{\mu}{r_\pm^{d-2}}.
\end{equation}
In order to construct the wormhole geometry, we first remove the region $r_+ < r_0$. 
Here we assume that $r_h < r_0 < r_c$, where $r_h$ and $r_c$ are the positions of black hole horizon and photon sphere, respectively.
We further remove the region $r_- < r_0$ for the other black hole and glue the two geometries at the hypersurface $\Sigma$ described by $r_+ = r_- = r_0$.
The junction condition leads that the surface stress-energy tensor $\mathcal{S}$ at the shell is determined by the jump in the extrinsic curvature as in \eqref{eq:cond_energy}.
For a static shell at $r_\pm = r_0$, the surface energy density $\sigma$ and pressure $\bar P$ are given by
\begin{align}
    \sigma &= -\frac{(d-1)}{4\pi r_0} \sqrt{f(r_0)}, \\
    \bar P &= \frac{1}{8\pi} \left[ \frac{f'(r_0)}{\sqrt{f(r_0)}} + \frac{(d-2)}{r_0} \sqrt{f(r_0)} \right].
\end{align}
In particular, $\sigma$ is negative and the weak energy condition is violated \cite{Morris:1988tu}, see also \cite{Kokubu:2014vwa,Kokubu:2015spa,Kokubu:2020lxs}.

In order to compute the retarded Green function of scalar operator, we need to evaluate the wave equation  \eqref{eq:wave}. The potential $\tilde V(z)$ is \eqref{eq:kappa} but now the metric function is that for AdS thin-shell wormhole.
We define the coordinate $z$ by $d z  = - dr_+/f (r_+)$ for $0 < z < z_0 $, where the integration constant is set as $z(r_0) = z_0$. We further set $z$ by $d z = dr_-/f (r_-)$ for $ z_0 < z < 2 z_0 $.
There are two conformal boundaries at $z \to 0$ and $z \to 2 z_0$ but only the correlation functions of those operators living on the boundary at $z \to 0$ are considered in this paper. For the cases with correlation functions involving both boundaries, see, e.g., \cite{Gao:2016bin,Maldacena:2017axo,Maldacena:2018lmt}.
At the junction $z =  z_0$ ($r_\pm = r_0$), the bulk scalar field must satisfy the continuity conditions
\begin{align}
\label{eq:continuity}
    \psi^+(r_0) = \psi^-(r_0), \quad
    \partial_z \psi^+(r_0) = \partial_z \psi^-(r_0) .
\end{align}
We also need to consider the two AdS boundaries at  $z \to 0$ and $z \to 2 z_0$. 
At $z \to 2 z_0$, we assign the Dirichlet boundary condition as
\begin{align} \label{eq:bc-inf}
    \psi (z) \sim  (2 z_0 - z)^{\frac{1}{2} + \nu} .
\end{align}
At the other boundary at $z \to 0$, we read off
the asymptotic behaviors \eqref{eq:waveSon} in order to compute the retarded Green functions as in \eqref{eq:GRSon}.

\subsubsection{Retarded Green functions}
\label{sec:RGTwormhole}

\begin{figure}
\centering
\includegraphics[height=4cm]{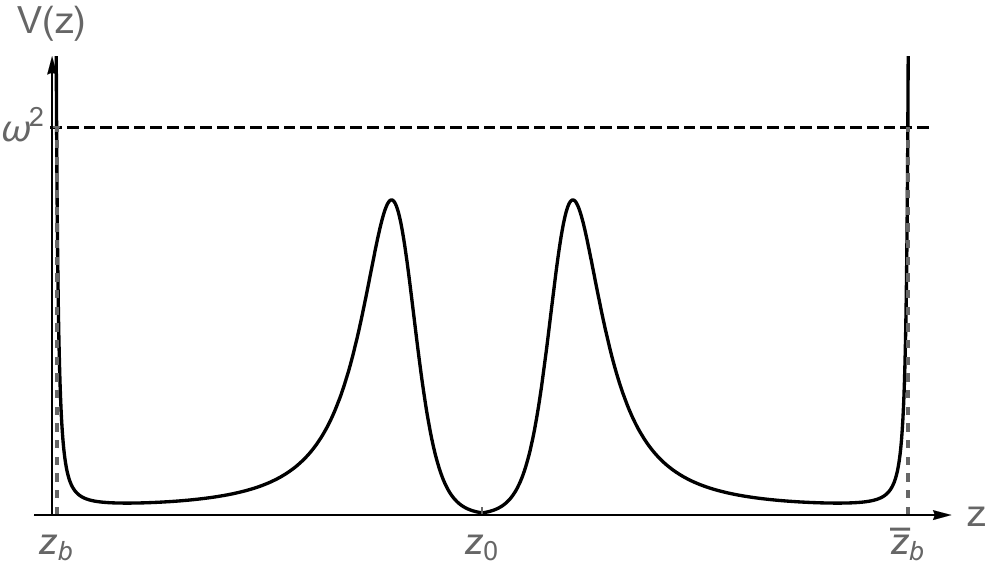}
\includegraphics[height=4cm]{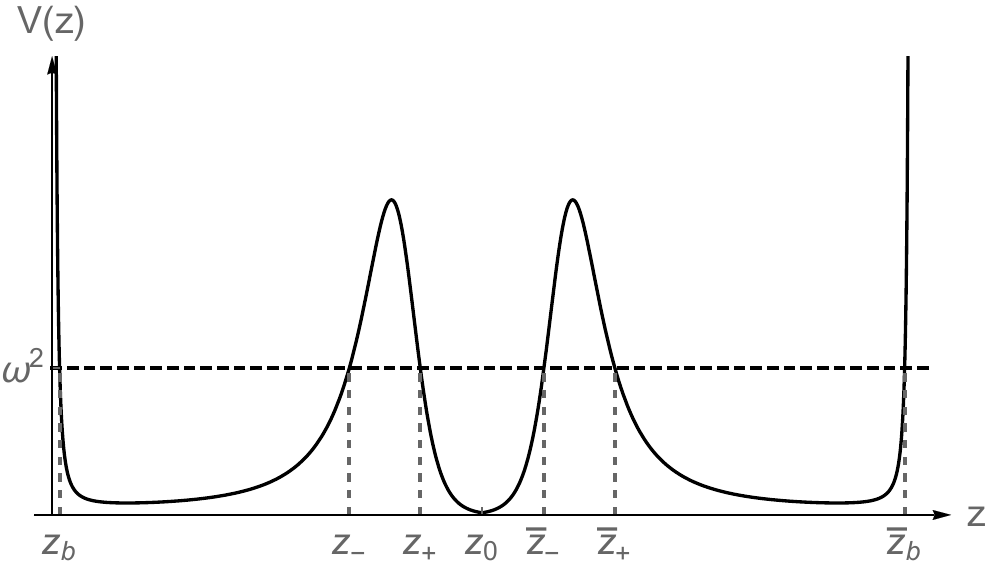}
\caption{The potential for AdS wormhole with two and four zeros of $\kappa^2(z)=0$.}
\label{fig:Vworm}
\end{figure}

We are now prepared to solve the wave equation \eqref{eq:wave} in the WKB approximation. 
The equation $\kappa (z)^2 = 0$ has several zeros and the number of zeros depends on the energy $\omega$ and the angular momentum $\ell$. We first consider the case where the equation  $\kappa (z)^2 = 0$ has  two zeros $z_b, \bar z_b $ with $\bar z_b  = 2 z_0 - z_b$, see the left panel of fig.\,\ref{fig:Vworm}. 
We will see that the retarded Green function exhibits the bulk-cone singularities associated with null trajectories that pass through the wormhole throat, are reflected at the opposite AdS boundary, and then return to the original AdS boundary. We then move to the case where the equation  $\kappa (z)^2 = 0$ has six zeros $z_b , z_\pm , \bar z_\pm  , \bar z_b $ where $z_b < z_- < z_+ < z_0 < \bar z_- < \bar z_+  < \bar z_b$. Here we set $\bar z_-   = 2 z_0 - z_+ $ and $\bar z_+ = 2 z_0 - z_-  $. See the right panel of fig.\,\ref{fig:Vworm}.
We shall observe the bulk-cone singularities corresponding to null trajectories going around the photon sphere and echoes following the singularities.

As mentioned above, we first consider the case with two zeros $z_b, \bar z_b $, where $z_b \simeq 0$ and $\bar z_b  \simeq 2 z_0$ for large $\ell$.
We start from the region near the opposite AdS boundary with $\hat z \, ( \equiv 2 z_0 - z) < 1/\ell $. The wave function satisfying the boundary condition \eqref{eq:bc-inf} is given by
\begin{align} \label{eq:psiJ}
    \psi (z) = \sqrt{2 \pi \ell \hat z} J_\nu (\ell \sqrt{u^2 -1} \hat z) .
\end{align}
Note that the Bessel function behaves as 
\begin{align}
    J_\nu (x) \sim \sqrt{\frac{2}{\pi x}} \cos \left( x - \frac{\nu \pi}{2} - \frac{\pi}{4}\right)
\end{align}
for large $x$.
With the help of the asymptotic behavior, 
we obtain the wave function for $0 \simeq z_b < z < \bar z_b  \simeq 2 z_0$  as
\begin{align} \label{eq:wavewhormhole}
\begin{aligned}
    \psi (z) &\sim \frac{ e^{\frac{i \pi}{4}+\frac{i \nu \pi}{2}}}{\sqrt{ \kappa(z) }} e^{- i \int^{2 z_0}_{z} d z'  \kappa (z') } + \frac{  e^{-\frac{i \pi}{4}-\frac{i \nu \pi}{2}} }{\sqrt{\kappa(z)}} e^{i  \int^{2z_0}_{z} d z' \kappa (z')}  \\
    &\sim \frac{ e^{\frac{i \pi}{4}+\frac{i \nu \pi}{2} - i S(0,2z_0)}}{\sqrt{ \kappa(z) }} e^{i \int^{z}_{0} d z'  \kappa (z') } + \frac{  e^{-\frac{i \pi}{4}-\frac{i \nu \pi}{2} +  i S(0,2z_0)} }{\sqrt{\kappa(z)}} e^{ - i  \int^{z}_{0} d z' \kappa (z')} .
    \end{aligned}
\end{align}
Using the expression \eqref{eq:retardedAdSG} for the wave function with \eqref{eq:psito0}, we find
\begin{align} 
\begin{aligned}
    G_R (\omega ,\ell) &=  \frac{\Gamma (- \nu)}{\Gamma(\nu)} 
    \left( \frac{\omega^2 - \ell^2}{4}\right)^\nu  \frac{  \cos (S(0,2z_0))  }{ \cos (S(0,2z_0) - \pi \nu )   } ,
    \end{aligned}
\end{align}
which can be expanded as in \eqref{eq:GRte} with
\begin{align}  \label{eq:an2}
a_0 =  e^{-i\pi \nu} , \quad a_{n} = - 2 i  (-1)^n e^{ - 2 i n \pi \nu}  \sin ( \pi \nu ) .
\end{align}
In the coordinate basis, the retarded Green function has bulk-cone singularities as in \eqref{eq:GRtth} with \eqref{eq:Xnpmj}. In this case, the singularities for $n =1$ correspond to the null trajectories bouncing at the opposite AdS boundary and coming back to the original AdS boundary. As we will see numerically below, the arrival time of the null trajectory is about twice of that of null trajectory going around the photon sphere and coming back to the original AdS boundary. Thus we can easily distinguish the two signals and this is a main difference from the case of AdS gravastar. For $n > 1$, the null trajectories bounce $n$ times at the opposite AdS boundary and $(n-1)$ times at the original AdS boundary.

Next we consider the case with six zeros $z_b , z_\pm , \bar z_\pm  , \bar z_b $ with $0 \simeq z_b < z_- < z_+ < z_0 <\bar z_- < \bar z_+  < \bar z_b \simeq 2 z_0$. The wave function for the region 
$\bar z_+ < z < \bar z_b  \simeq 2 z_0$ is given by
\begin{align}
    \psi (z) &\sim \frac{ e^{\frac{i \pi}{4}+\frac{i \nu \pi}{2} - i S(\bar z_+,2z_0)}}{\sqrt{ \kappa(z) }} e^{i \int^{z}_{\bar z_+} d z'  \kappa (z') } + \frac{  e^{-\frac{i \pi}{4}-\frac{i \nu \pi}{2} +  i S(\bar z_+,2z_0)} }{\sqrt{\kappa(z)}} e^{ - i  \int^{z}_{\bar z_+} d z' \kappa (z')} 
\end{align}
as in \eqref{eq:wavewhormhole}.
We continue the wave function to the region $z_+ < z < \bar z_-$ as
\begin{align}
    \psi (z) \sim \frac{ C_+}{\sqrt{\kappa(z)}} e^{i  \int^z_{z_+} d z' \kappa (z')} + \frac{ C_-}{\sqrt{ \kappa(z) }} e^{- i \int^z_{z_+} d z'  \kappa (z') }
\end{align}
with
\begin{align}
\begin{aligned}
    C_+ & = e^{\frac{i \pi}{4} +\frac{i \nu \pi}{2} - i S(\bar z_+,2z_0) } \left (e^{S(\bar z_- , \bar z_+ )} + \frac14e^{-S(\bar z_-  , \bar z_+ )} \right )e^{i S(z_+, \bar z_- )} \\
   & \quad +  e^{-\frac{i \pi}{4}-\frac{i \nu \pi}{2} + i S(\bar z_+,2z_0) }  \left (e^{S(\bar z_- , \bar z_+)} -\frac14e^{-S( \bar z_- , \bar z_+ )} \right ) e^{\frac12 i \pi} e^{i S(z_+, \bar z_-)} , \\
    C_- & = e^{\frac{i \pi}{4}+ \frac{i \nu \pi}{2} - i S(\bar z_+,2z_0)  }  \left (e^{S(\bar z_- , \bar z_+)} - \frac14e^{-S(\bar z_- , \bar z_+ )} \right ) e^{-\frac12 i \pi} e^{-i S(z_+, \bar z_-)}  \\
      & \quad + e^{-\frac{i \pi}{4} -\frac{i \nu \pi}{2} + i S(\bar z_+,2z_0) } \left (e^{S(\bar z_- , \bar z_+ )} + \frac14e^{-S(\bar z_-  ,\bar  z_+ )} \right )  e^{-i S(z_+,\bar z_-)} .
      \end{aligned}
\end{align}
We then move to the region $0 \simeq z_b < z < z_-$, where the wave function is given by
\begin{align}
    \psi (z) \sim \frac{ \tilde C_+}{\sqrt{\kappa(z)}} e^{i  \int^z_{0} d z' \kappa (z')} + \frac{\tilde C_-}{\sqrt{ \kappa(z) }} e^{- i \int^z_{0} d z'  \kappa (z') }
\end{align}
with
\begin{align}
   & \tilde C_+ = C_+ C_+^{(1)} + C_- C_-^{(2)} , \\
   & \tilde C_- =  C_+ C_- ^{(1)} +  C_- C_+ ^{(2)}  .
\end{align}
Here we set $C^{(1)}_\pm$ and $C^{(2)}_\pm$ as in \eqref{eq:C12pm}.
Using the expression \eqref{eq:retardedAdSG} for the wave function with \eqref{eq:psito0}, we obtain
\begin{align} 
    G_R (\omega ,\ell) &=  \frac{\Gamma (- \nu)}{\Gamma(\nu)} 
    \left( \frac{\omega^2 - \ell^2}{4}\right)^\nu \frac{\tilde C_+ e^{-\frac{i \pi \nu}{2} + \frac{i \pi}{4}} - \tilde C_- e^{\frac{i \pi \nu}{2} - \frac{i \pi}{4}}}{\tilde C_+ e^{\frac{i \pi \nu}{2} + \frac{i \pi}{4}} - \tilde C_- e^{-\frac{i \pi \nu}{2} - \frac{i \pi}{4}}}  .
\end{align}
With the help of $S(z_- , z_+) = S(\bar z_- , \bar z_+ )$,
we can rewrite it as \eqref{eq:GRAB} with \eqref{eq:AB}, where
\begin{align} \label{eq:GRtewormhole1}
a_0 =  e^{-i\pi \nu} , \quad  a_{n} = - 2 i  (-1)^n e^{ -i n \pi \nu}  \sin ( \pi \nu ) 
\end{align}
for $n=1,2,3,\ldots$
and
\begin{align} \label{eq:GRtewormhole2}
b_{n,0} =  - i (-1)^{n}  n e^{-  i n \pi \nu} \sin (\pi \nu) , \quad b_{n,p} = -2 i (-1)^{n + p }   n e^{-  i n \pi \nu} \sin (\pi \nu) 
\end{align}
for $p=1,2,3,\ldots$.
The part with $A$ is the same as that of AdS-Schwarzschild black hole. Therefore, there are bulk-cone singularities shared by the cases of AdS-Schwarzschild black hole and AdS wormhole. The part with $B$ leads to echoes following the bulk-cone singularities arising from the part with $A$.

\section{Numerical analysis}
\label{sec:numerical}

In this section, we numerically examine the bulk-cone singularities for black holes and ECOs.
We first solve the ordinary differential equation~\eqref{eq:zwaveeq} for $\psi_{\omega\ell}(z)$ for each
$(\omega,\ell)$.
The boundary conditions depend on the underlying geometry.
For a black hole, we impose the ingoing boundary condition~\eqref{eq:bc-ingoing} at the horizon.
For the gravastar, we impose regularity at the center $r=0$.
It is convenient to translate this regularity condition to the boundary condition at the gravastar surface by using the exact solution of the scalar field in the dS region:
\begin{align}
\psi(z_0)=\psi_\mathrm{dS}(z_0), \quad
\left.\frac{d\psi}{dz}\right|_{z=z_0}=\left.\frac{d\psi_\mathrm{dS}}{dz}\right|_{z=z_0},
\end{align}
where $z=z_0$ at $r=r_0$ and $\psi_\mathrm{dS}(z)$ is the solution in the dS region $z>z_0$ found in eq.~\eqref{eq:psi}.
See Appendix~\ref{sec:bc-gravastar} for the derivation.
For the thin-shell wormhole, we impose the Dirichlet boundary condition~\eqref{eq:bc-inf} at the AdS boundary in the opposite side together with the continuity condition~\eqref{eq:continuity} at the throat.

Near the AdS boundary $z\sim0$, the solution is given as the superposition,
\begin{align}
\psi(z)=\mathcal{A}(\omega,\ell)\psi_\mathcal{A}(z)+\mathcal{B}(\omega,\ell)\psi_\mathcal{B}(z) ,
\end{align}
of the two linearly independent solutions, $\psi_\mathcal{A}(z)\sim z^{\frac12-\nu}$ and $\psi_\mathcal{B}(z)\sim z^{\frac12+\nu}$ as in \eqref{eq:waveSon}.
We read off the coefficients $\mathcal{A}$ and $\mathcal{B}$ from the numerical solution and obtain $G_R(\omega,\ell)$ from eq.~\eqref{eq:GRSon}.
Note that we need to expand $\psi_\mathcal{A}(z)$ and $\psi_\mathcal{B}(z)$ in $z$ so that we can accurately read off the coefficient $\mathcal{B}$ of the damping function $\psi_\mathcal{B}(z)$.
For example, for $d=3$ and $\nu=3/2$, they are expanded as%
\footnote{The asymptotic behavior $\psi_\mathcal{A}(z)\sim z^{\frac12-\nu}$ does not completely specify the function $\psi_\mathcal{A}(z)$ as any linear combination of $\psi_\mathcal{A}$ and $\psi_\mathcal{B}$ satisfies the condition. Here we assume that $\psi_\mathcal{A}$ does not includes $\psi_B$ in its component.}
\begin{align}
    \psi_\mathcal{A}(z)&=z^{-1}+C_{\mathcal{A}1}z+C_{\mathcal{A}3}z^3+C_{\mathcal{A}4}z^4+\mathcal{O}(z^5) , \\
    \psi_\mathcal{B}(z)&=z^{2}+C_{\mathcal{B}4}z^4+\mathcal{O}(z^5),
\end{align}
where the coefficients $C_{\mathcal{A}n}$ and $C_{\mathcal{B}n}$ depend on $\omega$ and $\ell$.
By the inverse Fourier transformation and the inverse spherical harmonics transformation, we numerically get $G_R(t,\theta)$ in the coordinate basis.
To reduce the bulk-cone singularities to finite bumps, we include the Gaussian smearing factor in the inverse transformation as
\begin{align}
    G_R(t,\theta)=\int_{-\infty + i \delta}^{+\infty + i \delta}d\omega e^{-i\omega t}\sum_{\ell=0}^{\infty}Y_{\ell 0}(\theta)G_R(\omega,\ell)e^{-\frac{\mathrm{Re}(\omega)^2}{\omega_c^2}-\frac{\ell^2}{\ell_c^2}}
\end{align}
with the cut-off parameters $\omega_c$ and $\ell_c$~\cite{Dodelson:2023nnr}.
As there might be poles on or slightly below the real axis, we shift the contour of the $\omega$-integration slightly above the real axis by $\delta > 0$.
Using the fact $G_R(-\omega,\ell)=G_R(\omega,\ell)^*$, we have
\begin{align}
    G_R(t,\theta)=\int_{0+i\delta}^{\infty+i\delta}d\omega \sum_{\ell=0}^{\infty}Y_{\ell0}(\theta)\left[G_R(\omega,\ell)e^{-i\omega t}+G_R(\omega,\ell)^*e^{i\omega^* t}\right]e^{-\frac{\mathrm{Re}(\omega)^2}{\omega_c^2}-\frac{\ell^2}{\ell_c^2}}.
\end{align}
We perform the integration over $\omega$ and the summation over $\ell$ up to the sufficiently large values  $\omega_\mathrm{max}\,(\gg\omega_c)$ and $\ell_\mathrm{max}\,(\gg \ell_c)$, respectively.

\subsection{AdS-Schwarzschild black hole}

\begin{figure}[H]
 \centering
 \includegraphics[width=16cm]{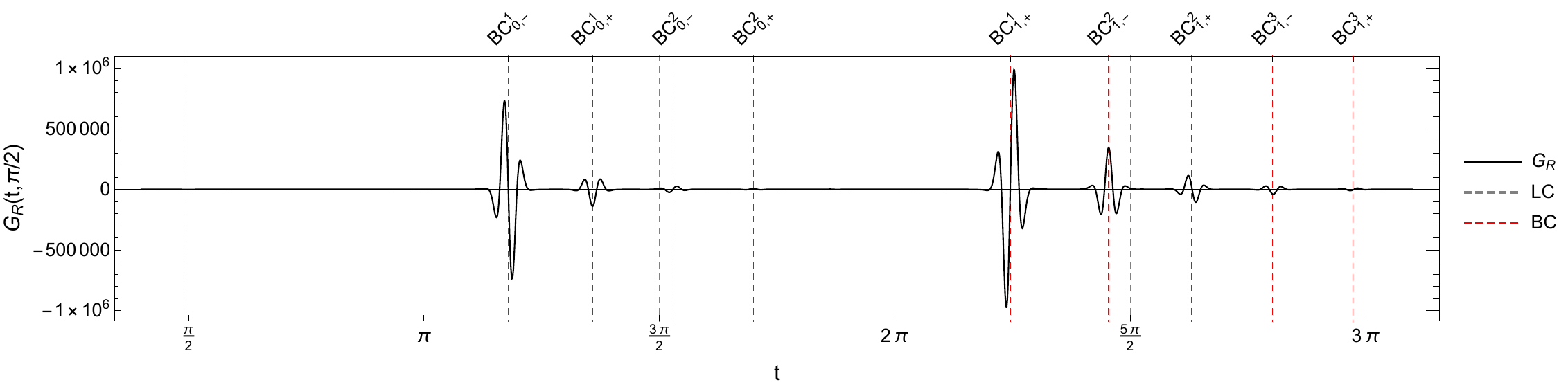}
 \caption{The Green function $G_R(t,\pi/2)$ for AdS Schwarzschild black hole with $d=3$ and $\mu=1/15$. From the left, the bulk-cone singularities are labeled as $\mathrm{BC}^1_{0,-}$, $\mathrm{BC}^1_{0,+}$, $\mathrm{BC}^2_{0,-}$, $\mathrm{BC}^2_{0,+}$, $\mathrm{BC}^1_{1,+}$, $\mathrm{BC}^2_{1,-}$, $\mathrm{BC}^2_{1,+}$, $\mathrm{BC}^3_{1,-}$ and $\mathrm{BC}^3_{1,+}$.}
 \label{fig:GRt_BH_3d}
 \end{figure}

\begin{figure}[H]
 \centering
 \includegraphics[height=3.5cm]{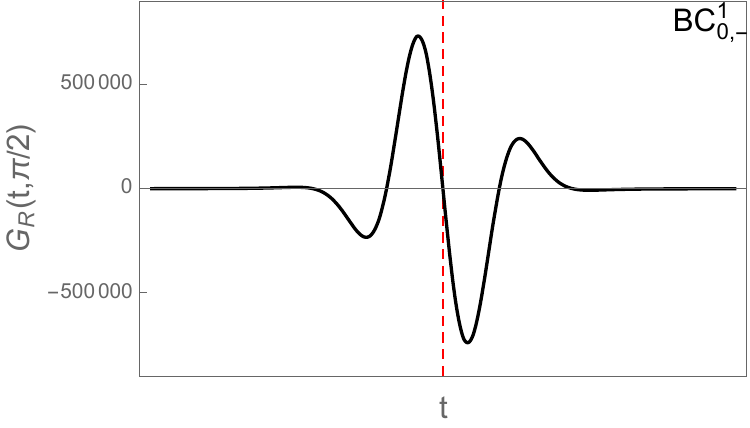}
 \includegraphics[height=3.5cm]{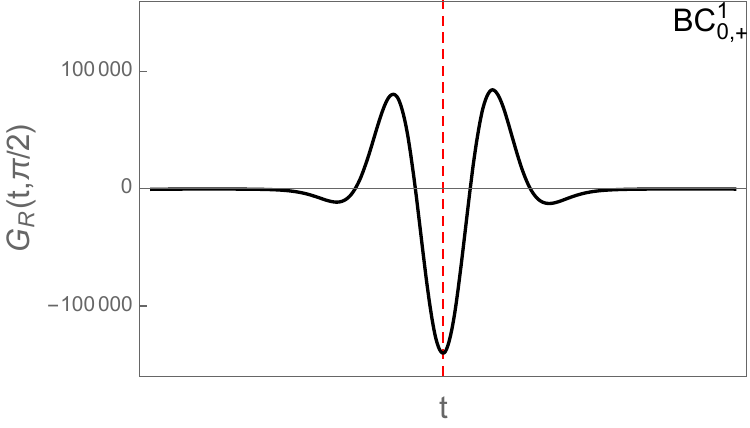}\\
 \includegraphics[height=3.5cm]{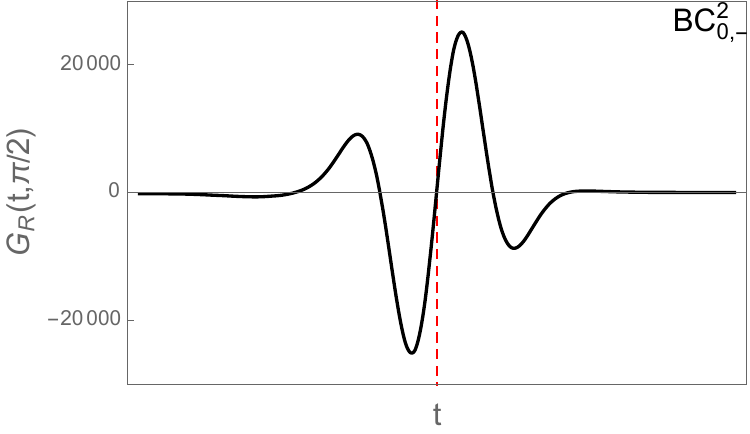}
 \includegraphics[height=3.5cm]{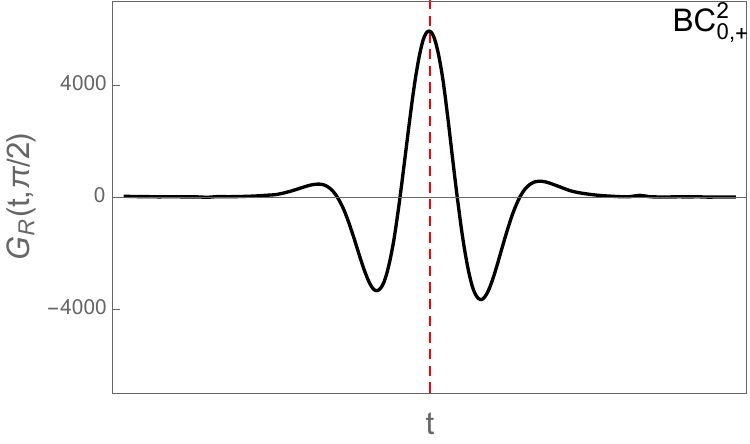}
 \caption{The four-fold structure of $G_R(t,\pi/2)$ in $d=3$. The first four bulk-cone bumps $\mathrm{BC}_{0,-}^1$, $\mathrm{BC}_{0,+}^1$, $\mathrm{BC}_{0,+}^2$, $\mathrm{BC}_{0,-}^2$ from the first sequence ($n=1$) of fig.\,\ref{fig:GRt_BH_3d} appear with the four different shapes.}
 \label{fig:GRt_BH_3d_4fold}
 \end{figure}

\begin{figure}[H]
 \centering
 \includegraphics[height=3.5cm]{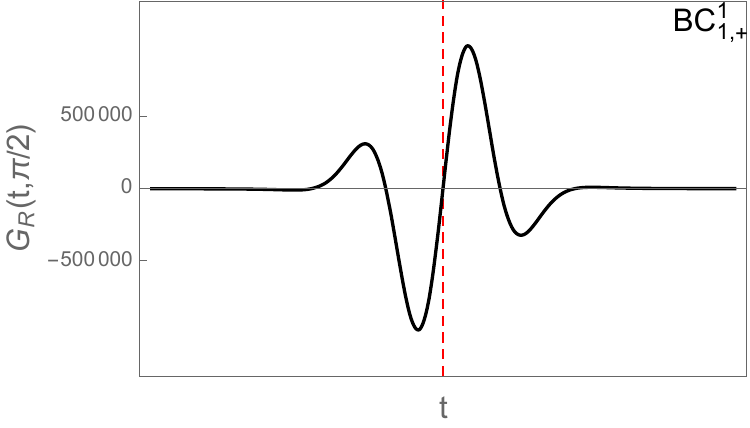}
 \includegraphics[height=3.5cm]{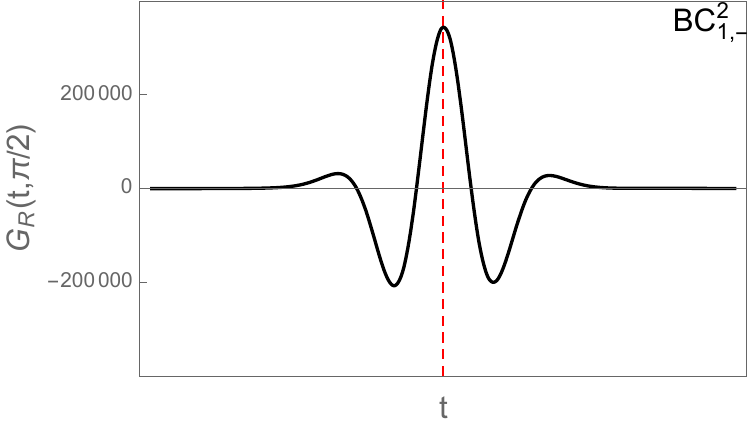}\\
 \includegraphics[height=3.5cm]{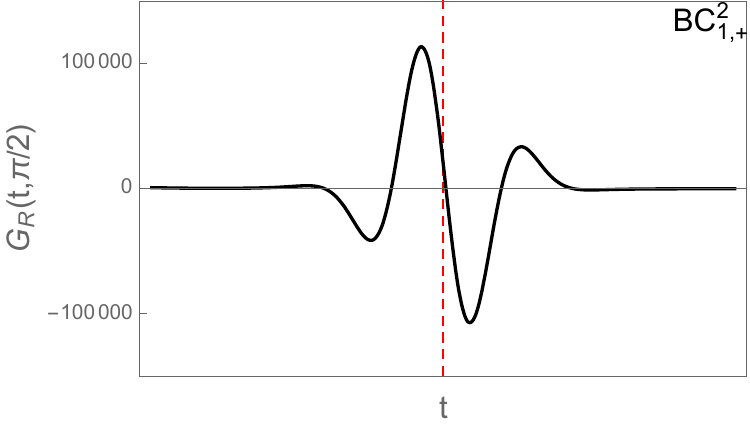}
 \includegraphics[height=3.5cm]{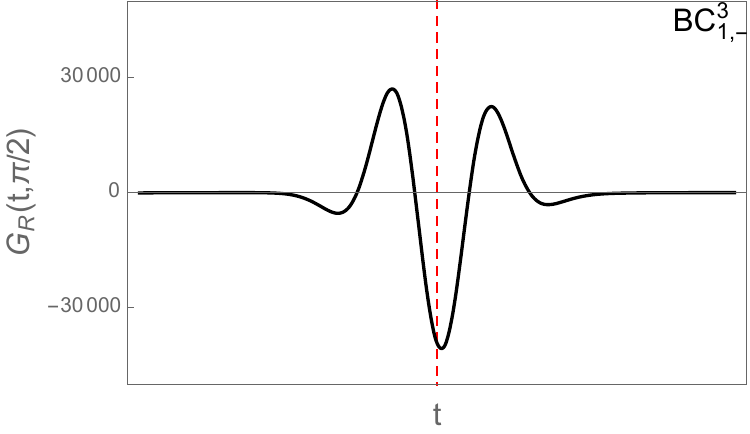}
 \caption{The first four bulk-cone bumps $\mathrm{BC}_{1,+}^1$, $\mathrm{BC}_{1,-}^2$, $\mathrm{BC}_{1,+}^2$, $\mathrm{BC}_{1,-}^3$ from the second sequence ($n=2$) of fig.\,\ref{fig:GRt_BH_3d}.}
 \label{fig:GRt_BH_3d_4fold_n2}
 \end{figure}

The Green function for the AdS-Schwarzschild black hole in the spatial dimension $d=3$ is reviewed 
in the left panel of fig.~\ref{fig:BCstructure_BH_4d} and fig.~\ref{fig:GRt_BH_3d}.
The mass parameter is taken as $\mu=1/15$.
We further fix $\theta=\pi/2$.
The parameters for the integration are taken as $\omega_c=\ell_c=40$, $\omega_\mathrm{max}=\ell_\mathrm{max}=150$, and $\delta=0.2$.
The red dashed lines indicate the arrival time expected from the geodesic analysis for the first few singularities $\mathrm{BC}^j_{n-1,\pm}$ with $n=1,2$.
There are strong bumps at the arrival time and their amplitude decreases in time for each $n$ as expected from the semi-analytic formula~\eqref{eq:GRtth}.
We can see the four-fold structure of the bump shapes appearing in each sequence in $d=3$.
As depicted in fig.~\ref{fig:GRt_BH_3d_4fold}, the four shapes are classified into even and odd types, together with their inverted counterparts.
The formula~\eqref{eq:GRtth} also predicts the relative shape for the different values of $n$. 
Due to the phase factor $a_n$ in \eqref{eq:an}, the bulk-cone singularities of $\mathrm{BC}_{0,\pm}^{j}$ and $\mathrm{BC}_{1,\pm}^{j}$ acquire a relative phase
$a_2/a_1=-e^{-i\pi \nu}= - i$. 
This relative phase indicates that one of them has an even-type shape, while the other is of odd type.
For example, the second bump of the first sequence is $\mathrm{BC}_{0,+}^{1}$ and the first bump of the second sequence is $\mathrm{BC}_{1,+}^{1}$.
Indeed, $\mathrm{BC}_{0,+}^{1}$ is even-shaped and $\mathrm{BC}_{1,+}^{1}$ is odd-shaped as in figs.~\ref{fig:GRt_BH_3d}--\ref{fig:GRt_BH_3d_4fold_n2}. 
We can also verify the consistency between the semi-analytic expression \eqref{eq:GRtth} and the numerical results for the bumps labeled by $\mathrm{BC}_{n-1,-}^1$ and $\mathrm{BC}_{n-1,-}^2$ with $n=1,2$.

\begin{figure}[H]
 \centering
 \includegraphics[width=16cm]{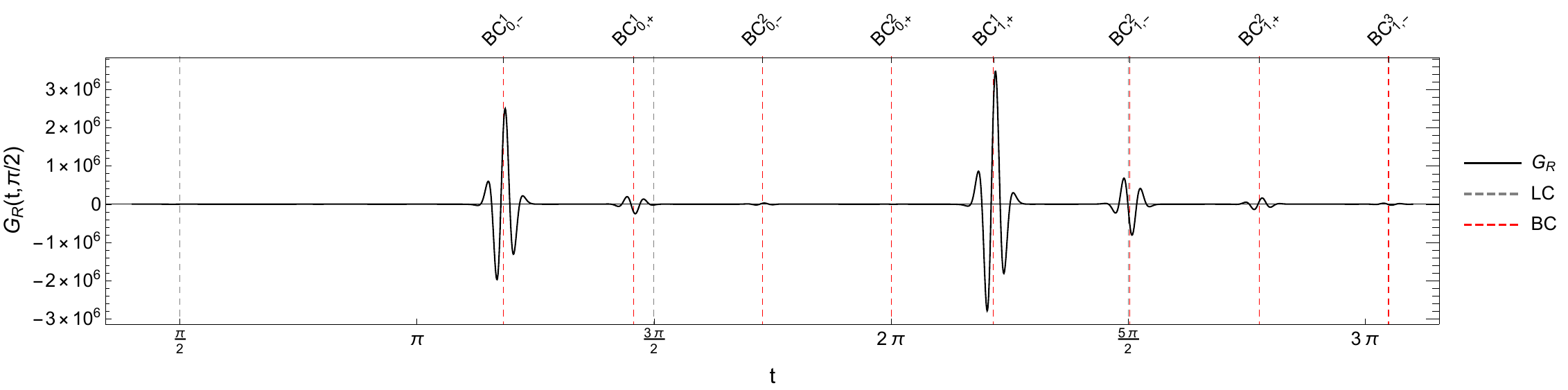}
 \caption{The Green function $G_R(t,\pi/2)$ for AdS-Schwarzschild black hole with $d=4$ and $\mu=1/50$. From the left, the bulk-cone singularities are labeled as $\mathrm{BC}^1_{0,-}$, $\mathrm{BC}^1_{0,+}$, $\mathrm{BC}^2_{0,-}$, $\mathrm{BC}^2_{0,+}$, $\mathrm{BC}^1_{1,+}$, $\mathrm{BC}^2_{1,-}$, $\mathrm{BC}^2_{1,+}$, and $\mathrm{BC}^3_{1,-}$. 
 }
 \label{fig:GRt_BH_4d}
 \end{figure}

\begin{figure}[H]
 \centering
 \includegraphics[height=3.5cm]{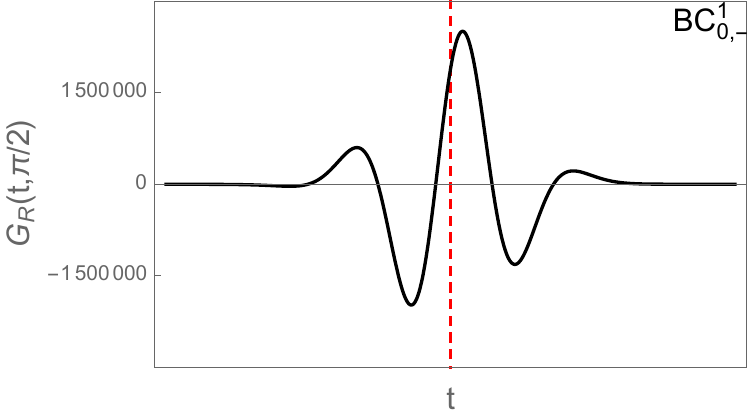}
 \includegraphics[height=3.5cm]{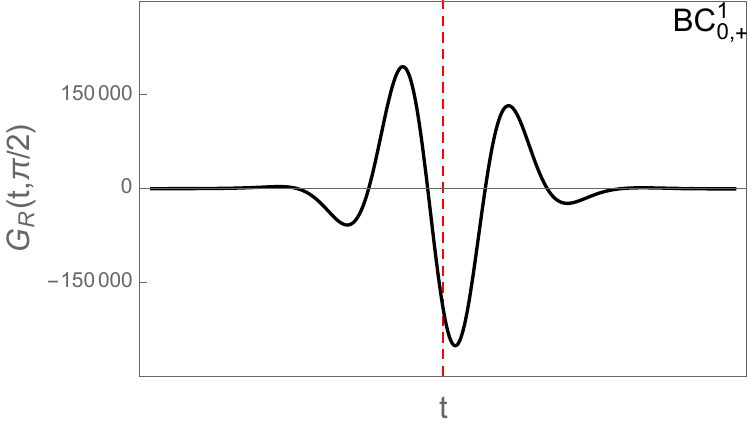}
 \caption{The two-fold structure of $G_R(t,\pi/2)$ in $d=4$. The first two bulk-cone bumps $\mathrm{BC}_{0,-}^1$ and $\mathrm{BC}_{0,+}^1$ from the first sequence ($n=1$) of fig.\,\ref{fig:GRt_BH_4d} appear with the two different shapes.}
 \label{fig:GRt_BH_4d_2fold}
 \end{figure}

\begin{figure}[H]
 \centering
 \includegraphics[height=3.5cm]{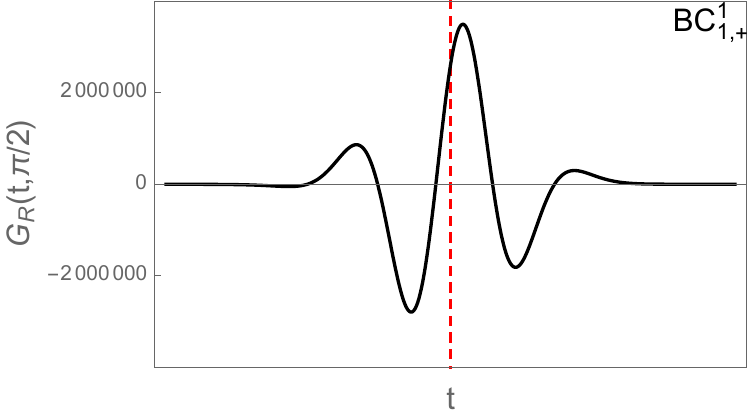}
 \includegraphics[height=3.5cm]{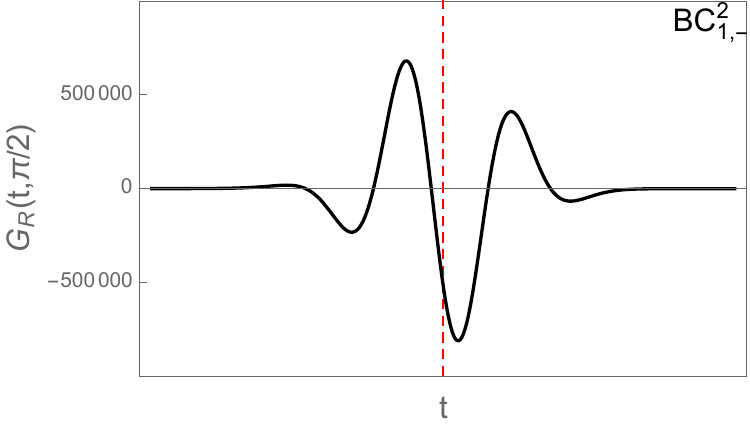}
 \caption{The first two bulk-cone bumps $\mathrm{BC}_{1,+}^1$ and $\mathrm{BC}_{1,-}^2$ from the second sequence ($n=2$) of fig.\,\ref{fig:GRt_BH_4d}.}
 \label{fig:GRt_BH_4d_2fold_n2}
 \end{figure}

The result for the black hole in $d=4$ is shown in the right panel of fig.~\ref{fig:BCstructure_BH_4d} and fig.~\ref{fig:GRt_BH_4d}.
The mass parameter is taken as $\mu=1/50$.
We fix $\theta=\pi/2$.
The parameters for the integration are taken as $\omega_c=\ell_c=35$, $\omega_\mathrm{max}=\ell_\mathrm{max}=150$, and $\delta=0.2$.
We observe qualitatively the same behavior as in the $d=3$ case.
However, in $d=4$, the two-fold structure appears in the bumps as in fig.\,\ref{fig:GRt_BH_4d_2fold}.
They are both (nearly) even-shaped but one is inversion of the other one.
In fig.\,\ref{fig:GRt_BH_4d}, the second bump of the first sequence is $\mathrm{BC}_{0,+}^{1}$ and the first one of the second sequence is $\mathrm{BC}_{1,+}^{1}$.
Their shape is upside down and this is consistent with their relative phase $a_2/a_1=-e^{-i\pi \nu}=-1$ predicted from the semi-analytic formula \eqref{eq:GRtth}.%
\footnote{The phase factor $a_n$ vanishes if we naively insert $\nu = d/2 = 2$. However, the retarded Green function is proportional to $\Gamma (- \nu)$ as in \eqref{eq:GRte} and the combination $\Gamma (- \nu) \sin (\pi \nu)$ remains finite for $\nu = 2$.}
See also figs.\,\ref{fig:GRt_BH_4d_2fold} and\,\ref{fig:GRt_BH_4d_2fold_n2}.

 \subsection{AdS gravastar}
 \label{sec:numerical-grav}

\begin{figure}[H]
 \centering
 \includegraphics[height=4.5cm]{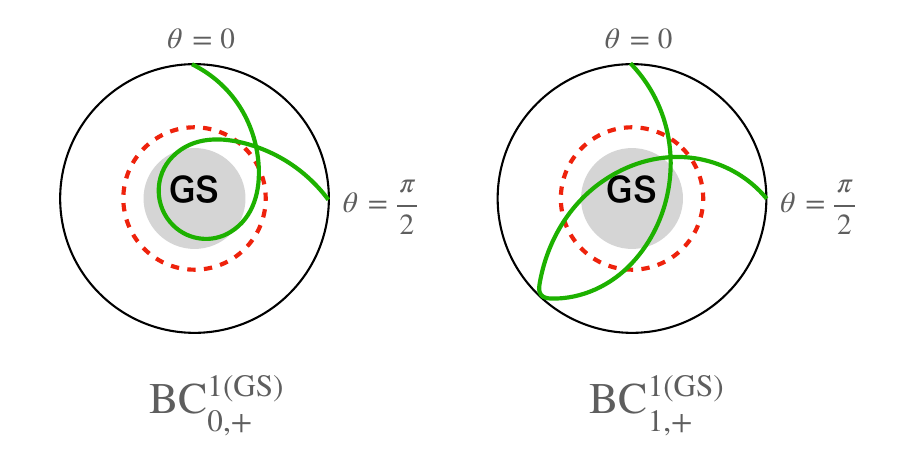}
 \caption{Illustration of null geodesics passing through the gravastar interior. The trajectories 
 $\mathrm{BC}_{0+}^{1\mathrm{(GS)}}$ and $\mathrm{BC}_{1+}^{1\mathrm{(GS)}}$ go through the dS region with zero and one bounce at the AdS boundary, respectively.}
 \label{fig:obt_gs}
 \end{figure}

\begin{figure}[H]
 \centering
 \includegraphics[height=7cm]{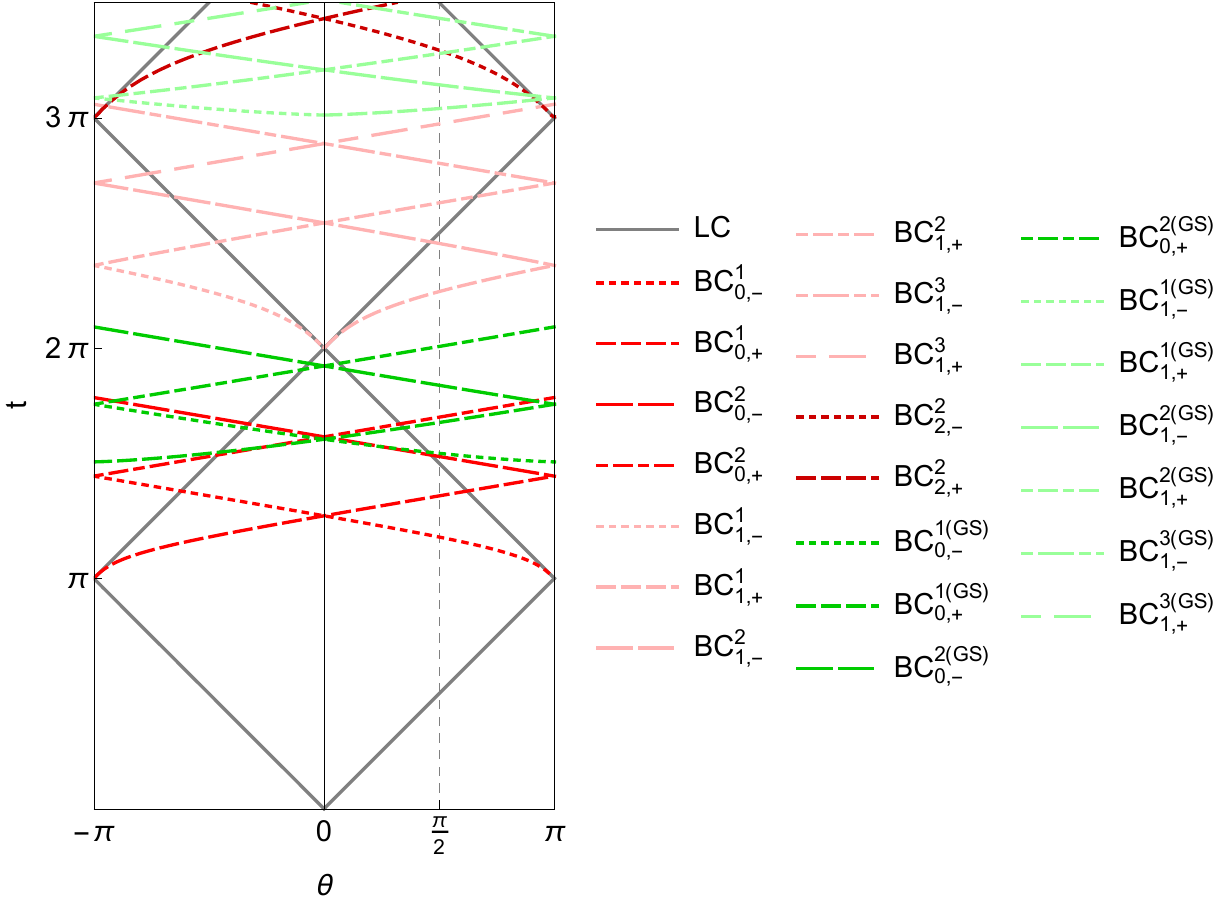}
 \caption{The structure of bulk-cone singularities  for AdS gravastar with $d=3$, $\mu=1/15$, and $r_0=1.001r_h$.}
 \label{fig:BCstructure_GV_3d}
 \end{figure}

 Here we examine the AdS gravastar with $d=3$, $\mu=1/15$ and $r_0=1.001r_h$, where $r_h$ is the horizon radius (if exists) expected from the outside AdS-Schwarzschild geometry.
 We expect there are null geodesics passing through the gravastar interior as in fig.~\ref{fig:obt_gs}.
 The spacetime diagram for the bulk-cone singularities from geodesic analysis is shown in fig.\,\ref{fig:BCstructure_GV_3d}.
 The red lines show $\mathrm{BC}_{n,\pm}^j$ common to the black hole case with the same $\mu$.
 The green lines show the singularities $\mathrm{BC}_{n,\pm}^{j\mathrm{(GS)}}$ corresponding to null geodesics passing through the gravastar interior.
 These geodesics are delayed according to the geometry.
 In particular, the gravastar surface $r_0$ is close to the horizon radius $r_h$ and the lapse function $f(r)$ is very small there.
 Geodesics passing through the gravastar interior spend time during crossing the surface.
 
\begin{figure}[H]
 \centering
 \includegraphics[width=16cm]{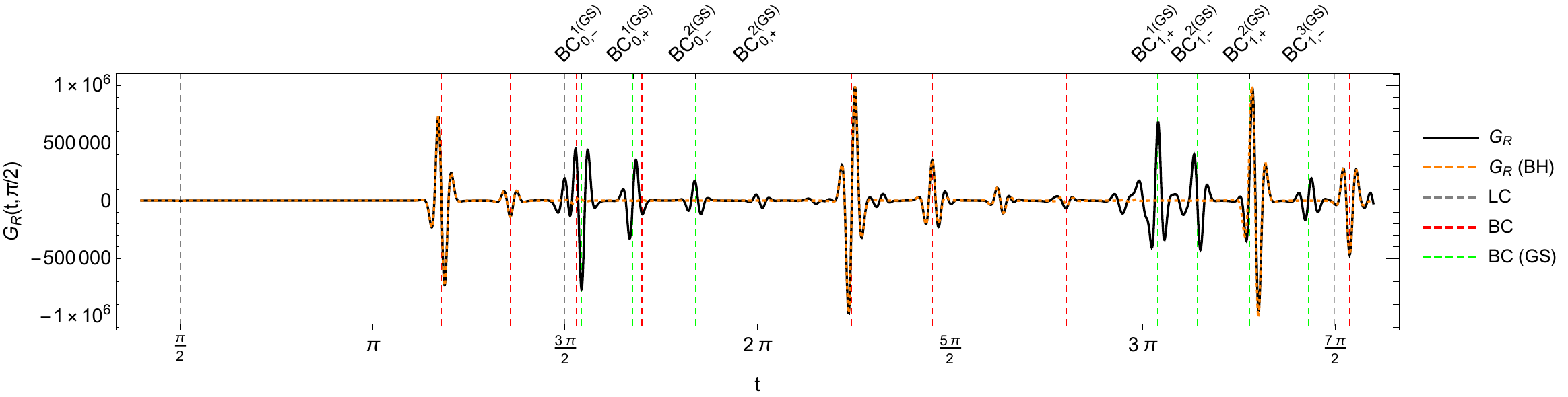}
 \caption{The Green function $G_R(t,\pi/2)$ for AdS gravastar with $d=3$, $\mu=1/15$, and $r_0=1.001r_h$. 
 The green dashed lines show $\mathrm{BC}_{0,-}^{1\mathrm{(GS)}}$, $\mathrm{BC}_{0,+}^{1\mathrm{(GS)}}$, $\mathrm{BC}_{0,-}^{2\mathrm{(GS)}}$, $\mathrm{BC}_{0,+}^{2\mathrm{(GS)}}$, $\mathrm{BC}_{1,+}^{1\mathrm{(GS)}}$, $\mathrm{BC}_{1,-}^{2\mathrm{(GS)}}$, $\mathrm{BC}_{1,+}^{2\mathrm{(GS)}}$, $\mathrm{BC}_{1,-}^{3\mathrm{(GS)}}$ from left to right.
 The red dashed lines show $\mathrm{BC}^1_{0,-}$, $\mathrm{BC}^1_{0,+}$, $\mathrm{BC}^2_{0,-}$, $\mathrm{BC}^2_{0,+}$, $\mathrm{BC}^1_{1,+}$, $\mathrm{BC}^2_{1,-}$, $\mathrm{BC}^2_{1,+}$, $\mathrm{BC}^3_{1,-}$, $\mathrm{BC}^2_{2,-}$, and $\mathrm{BC}^2_{2,+}$.
 }
 \label{fig:GRt_GV_3d}
 \end{figure}

 The numerical result for the retarded Green function $G_R(t,\theta)$ at $\theta=\pi/2$ is shown in fig.\,\ref{fig:GRt_GV_3d}.
 The black solid curve is for the gravastar and the orange dashed curve is for the black hole with the same $\mu$, i.e.~the one shown in fig.\,\ref{fig:GRt_BH_3d}.
The times of the bulk-cone singularities expected from the geodesic analysis are shown by the vertical dashed lines.
 The red ones are those for $\mathrm{BC}_{n,\pm}^{j}$ and green ones are for $\mathrm{BC}_{n,\pm}^{j\mathrm{(GS)}}$.
 For each arrival time, we can see a strong bump.
 Thus, the geodesic analysis agrees well with the full numerical computation of the bulk-cone singularities.
 The bumps at the red dashed lines are common to both cases of the gravastar and black hole, implying that they are bulk-cone singularities characterized by the geometry outside the photon sphere.
 On the other hand, the bumps at green lines around $3\pi/2\lesssim t\lesssim 2\pi$ and $3\pi\lesssim t\lesssim 7\pi/2$ are specific to the gravastar case and, in particular, their amplitude is as strong as the bumps common to the black hole.
 This result is consistent with our semi-analytic formula~\eqref{eq:GRtth}, which states that any bulk-cone singularity has the power $\alpha+2\nu+3/2=2\Delta-(d-1)/2$.

\begin{figure}[H]
 \centering
 \includegraphics[width=7.9cm]{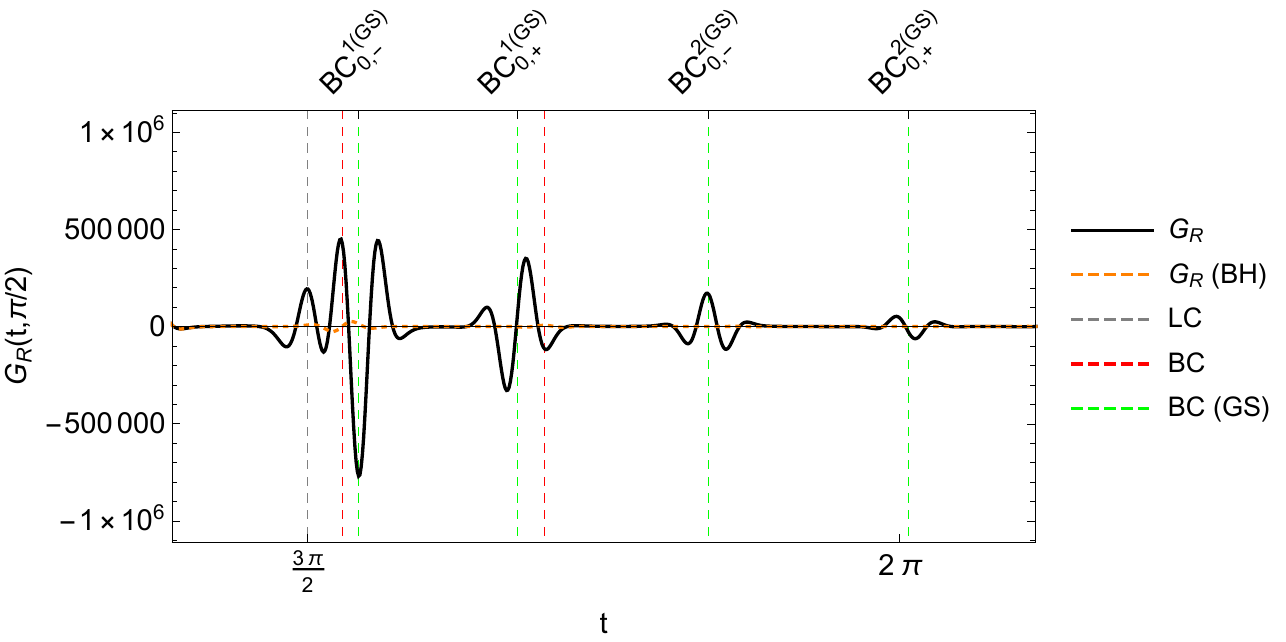}
 \includegraphics[width=7.9cm]{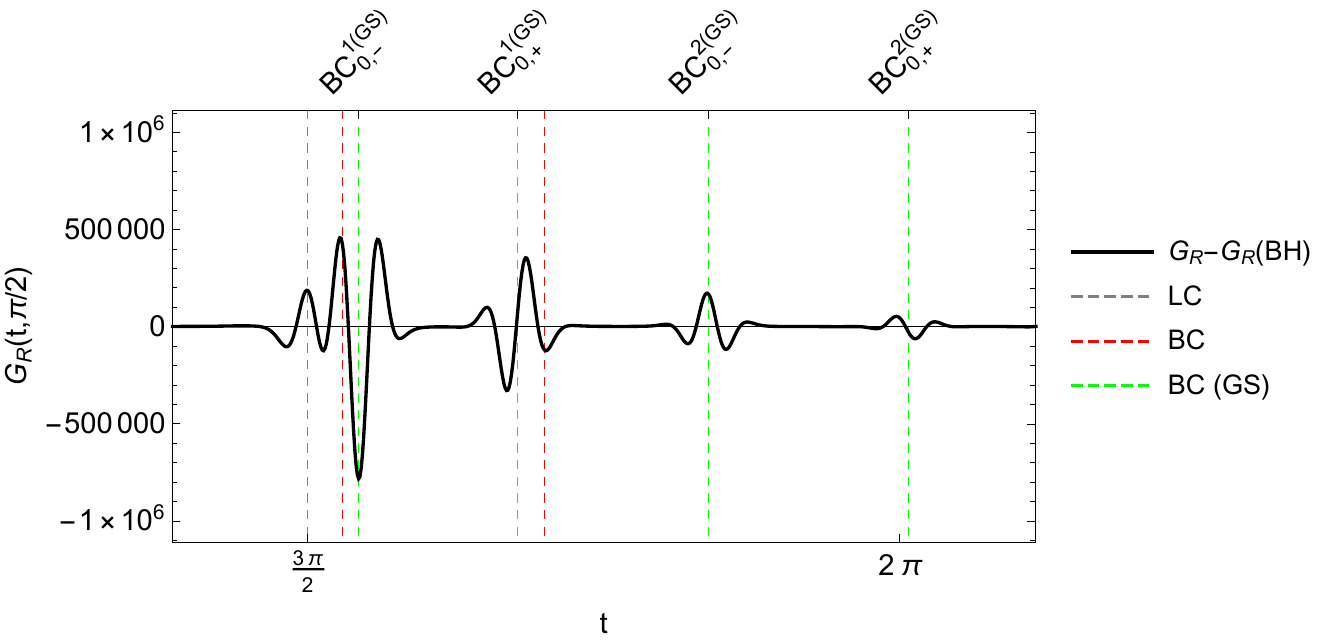}
 \caption{The Green function $G_R(t,\pi/2)$ for the gravastar in the region $1.5 \pi \lesssim t \lesssim 2\pi$. The green dashed lines show $\mathrm{BC}_{0,-}^{1\mathrm{(GS)}}$, $\mathrm{BC}_{0,+}^{1\mathrm{(GS)}}$, $\mathrm{BC}_{0,-}^{2\mathrm{(GS)}}$, and $\mathrm{BC}_{0,+}^{2\mathrm{(GS)}}$ from left to right. 
 Note that the bumps of $G_R(\mathrm{BH})$ in this range are so small that the difference is subtle between the left and right panels.
 }
 \label{fig:GRt_GV_3d_1stseq}
 \end{figure}

\begin{figure}[H]
 \centering
 \includegraphics[width=7.9cm]{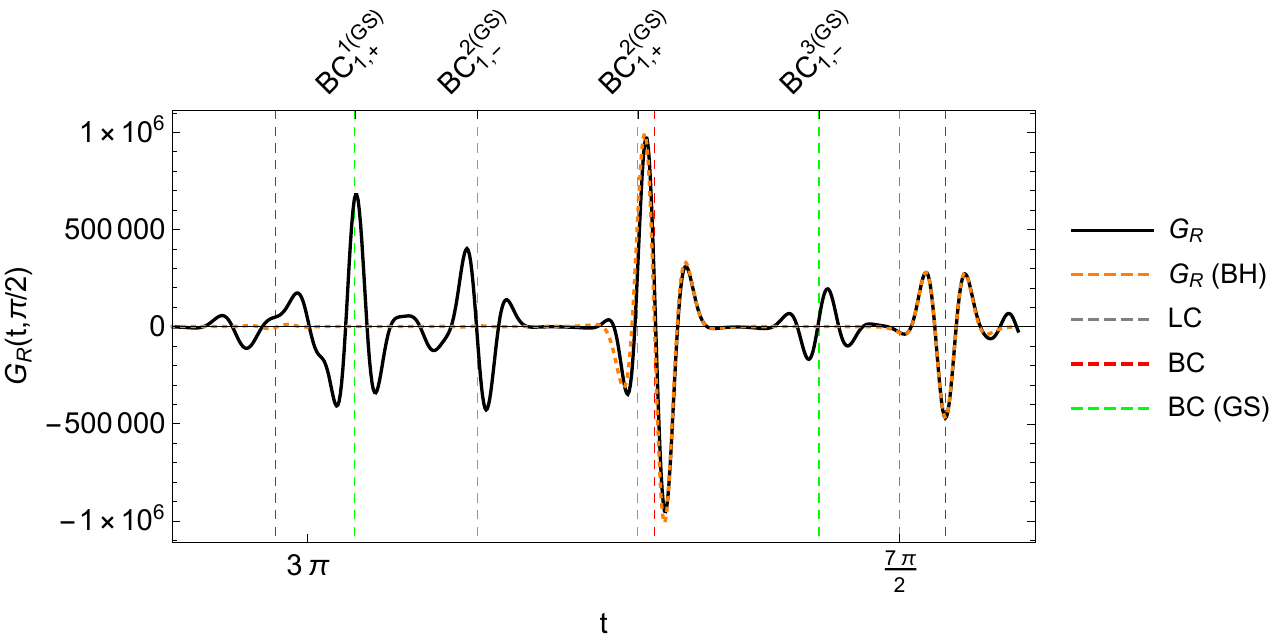}
 \includegraphics[width=7.9cm]{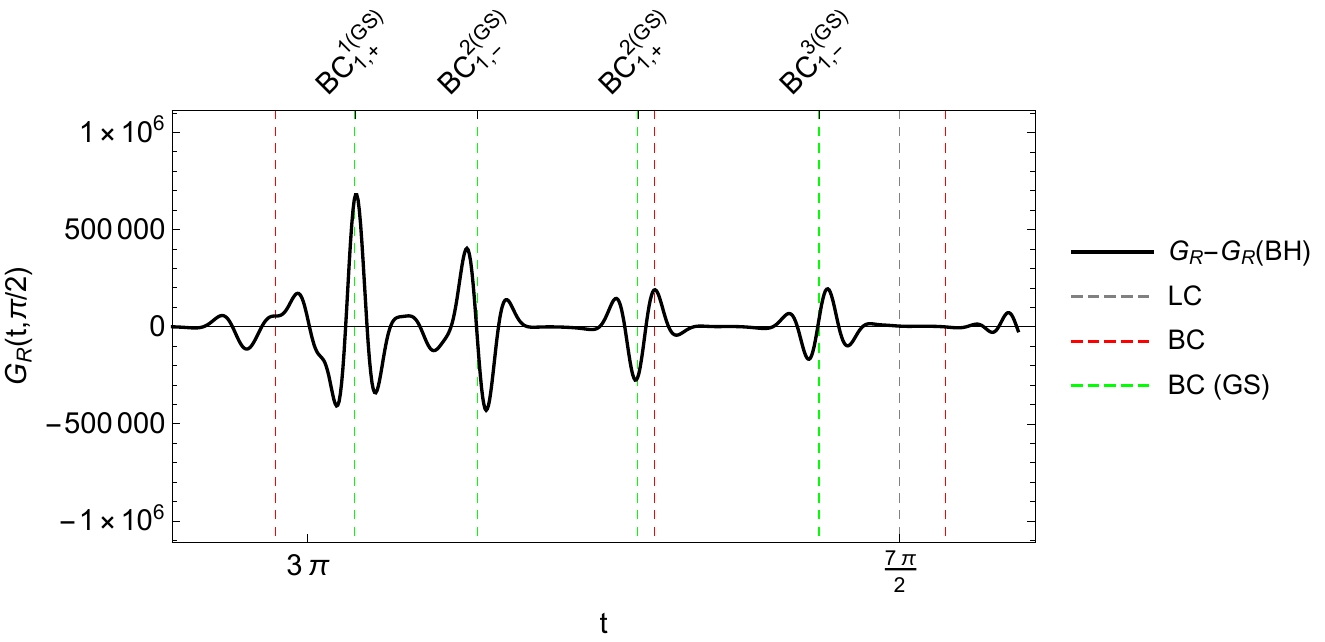}
 \caption{The Green function $G_R(t,\pi/2)$ for AdS gravastar in the region $3 \pi \lesssim t \lesssim 3.5\pi$. The green dashed lines show $\mathrm{BC}_{1,+}^{1\mathrm{(GS)}}$, $\mathrm{BC}_{1,-}^{2\mathrm{(GS)}}$, $\mathrm{BC}_{1,+}^{2\mathrm{(GS)}}$, and $\mathrm{BC}_{1,-}^{3\mathrm{(GS)}}$ from left to right.}
 \label{fig:GRt_GV_3d_2ndseq}
 \end{figure}

\begin{figure}[H]
 \centering
 \includegraphics[height=3.5cm]{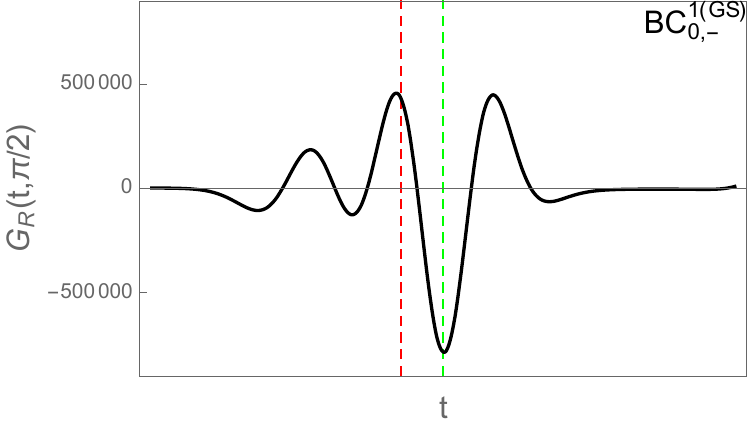}
 \includegraphics[height=3.5cm]{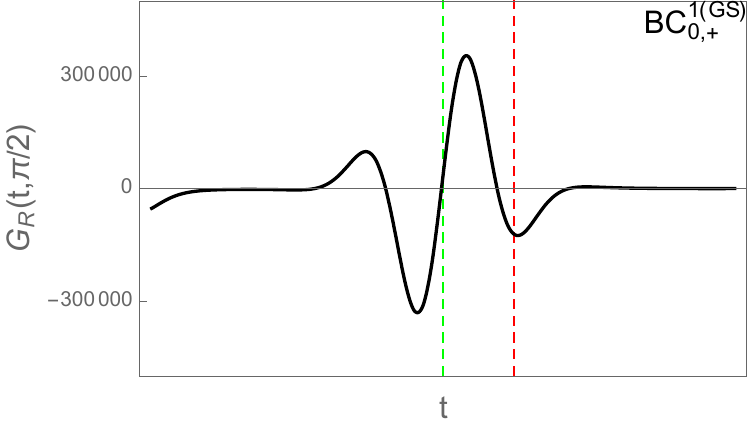}\\
 \includegraphics[height=3.5cm]{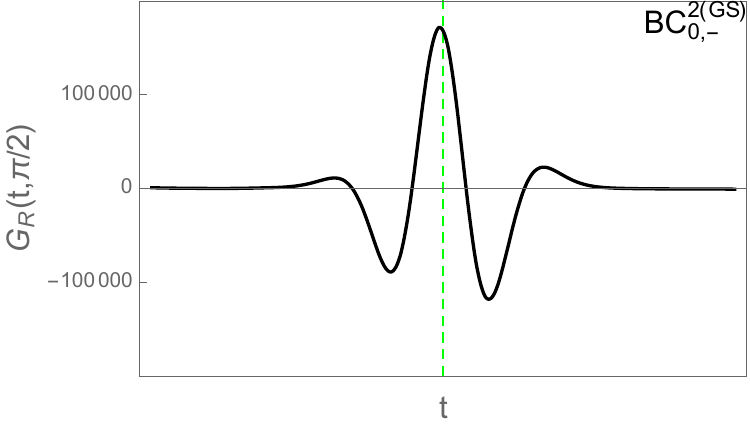}
 \includegraphics[height=3.5cm]{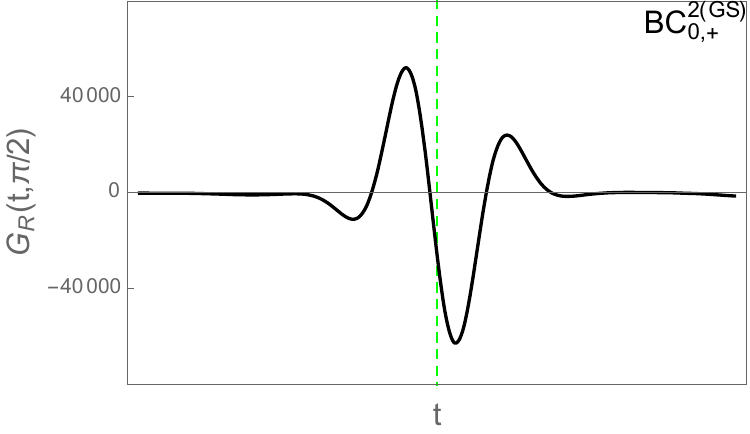}
 \caption{The four-fold structure for the gravastar in $d=3$. The four subsequent bulk-cone bumps $\mathrm{BC}_{0,-}^{1\mathrm{(GS)}}$, $\mathrm{BC}_{0,+}^{1\mathrm{(GS)}}$, $\mathrm{BC}_{0,-}^{2\mathrm{(GS)}}$, $\mathrm{BC}_{0,+}^{2\mathrm{(GS)}}$ of $G_R\mathrm{(GS)}-G_R\mathrm{(BH)}$ from fig.\,\ref{fig:GRt_GV_3d_1stseq} appear with the four different shapes.}
 \label{fig:GRt_GS_3d_4fold}
 \end{figure}

The bumps of $\mathrm{BC}_{n,\pm}^{j\mathrm{(GS)}}$ are also presented in figs.~\ref{fig:GRt_GV_3d_1stseq} and~\ref{fig:GRt_GV_3d_2ndseq}.
The left panels show magnified views of fig.~\ref{fig:GRt_GV_3d}, while the right panels show the difference between the gravastar Green function and the black hole one, i.e., $G_R\mathrm{(GS)}-G_R\mathrm{(BH)}$.
The bumps of $\mathrm{BC}_{n,\pm}^{j\mathrm{(GS)}}$ clearly show the monotonically decreasing behavior.
 This result is also expected from our semi-analytic formula, whose factor gives the relative amplitude of bumps.
 We will revisit the relative amplitude and its relation to the Lyapunov exponent in subsection~\ref{sec:lyapunov}. 
It is also remarkable that, once the bumps common to the black hole are subtracted, the remaining series of bumps exhibits the four-fold structure as shown in fig.~\ref{fig:GRt_GS_3d_4fold} for $n=1$. This is also predicted from the formula~\eqref{eq:GRtth} even for the gravastar case.
The bump shapes for different $n$ are also consistent with the formula.
For example, 
 $\mathrm{BC}_{0,+}^{1\mathrm{(GS)}}$ in fig.\,\ref{fig:GRt_GV_3d_1stseq} is odd-shaped and $\mathrm{BC}_{1,+}^{1\mathrm{(GS)}}$ in fig.\,\ref{fig:GRt_GV_3d_2ndseq} is even-shaped.
 The analytic formula predicts a relative phase $a_2/a_1=-e^{-i\pi\nu}=-i$, which leads to change of the parity from even to odd or vice versa.

 It is also interesting to compare the shape of the two types of bulk-cone bumps, $\mathrm{BC}_{n-1,\pm}^{j}$ and $\mathrm{BC}_{n-1,\pm}^{j\mathrm{(BS)}}$.
 From fig.~\ref{fig:GRt_GV_3d}, the first ones from each type of bumps, $\mathrm{BC}_{0,-}^{1}$ and $\mathrm{BC}_{0,-}^{1\mathrm{(GS)}}$, are odd and even-shaped, respectively.
 Even though they have the same label of $j,n,\pm$, their shape parity is different.
 The semi-analytic formula~\eqref{eq:GRtth} explains its reason.
 The parameters $\nu,\alpha$ and labels $j,n,\pm$ determine the phase, real or pure imaginary, of the factor of eq.~\eqref{eq:GRtth} except for the contribution from $1/\sqrt{T_\mathrm{null}'(u_*)}$.
 For $\mathrm{BC}_{n-1,\pm}^{j}$, $u_*<u_c$ and $T_\mathrm{null}(u_*)\to\infty$ as $u_*\to u_c-0$ because they correspond to null geodesics winding around the photon sphere.
 It implies $T_\mathrm{null}'(u_*)>0$ and the factor $1/\sqrt{T_\mathrm{null}'(u_*)}$ does not change the phase of the factor.
 On the other hand, for $\mathrm{BC}_{n-1,\pm}^{j\mathrm{(GS)}}$, $u_*>u_c$ and $T_\mathrm{null}(u_*)\to\infty$ as $u_*\to u_c+0$.
 It implies $T_\mathrm{null}'(u_*)<0$ and $1/\sqrt{T_\mathrm{null}'(u_*)}=-i/\sqrt{|T_\mathrm{null}'(u_*)|}$ multiplies the phase $-i$ to the overall factor.
 This extra phase changes the parity of bumps from even to odd or vice versa.
 Accordingly, $\mathrm{BC}_{n-1,\pm}^{j}$ and $\mathrm{BC}_{n-1,\pm}^{j\mathrm{(GS)}}$ for the same $j,n,\pm$ have different parity.

\begin{figure}[H]
 \centering
 \includegraphics[height=7cm]{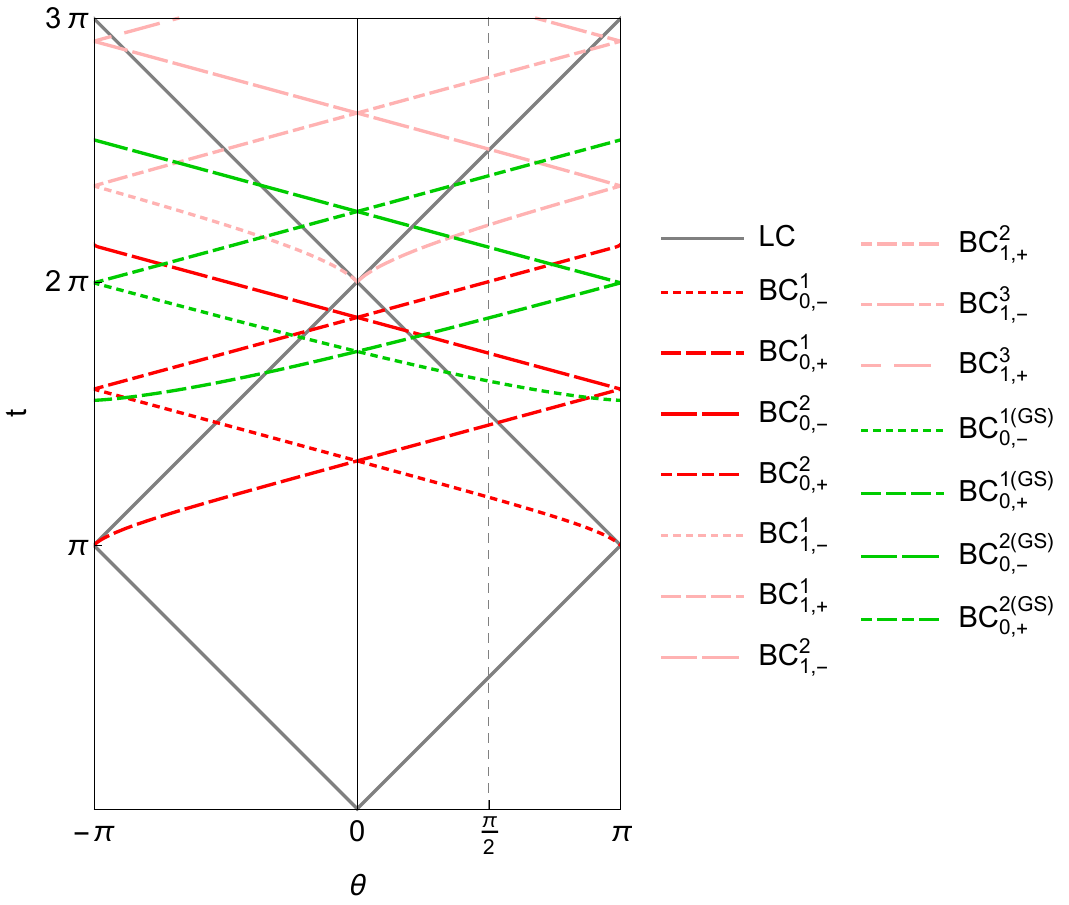}
 \caption{The structure of bulk-cone singularities  for AdS gravastar with $d=4$, $\mu=1/50$, and $r_0=1.001r_h$.}
 \label{fig:BCstructure_GV_4d_rh}
 \end{figure}

\begin{figure}[H]
 \centering
 \includegraphics[width=16cm]{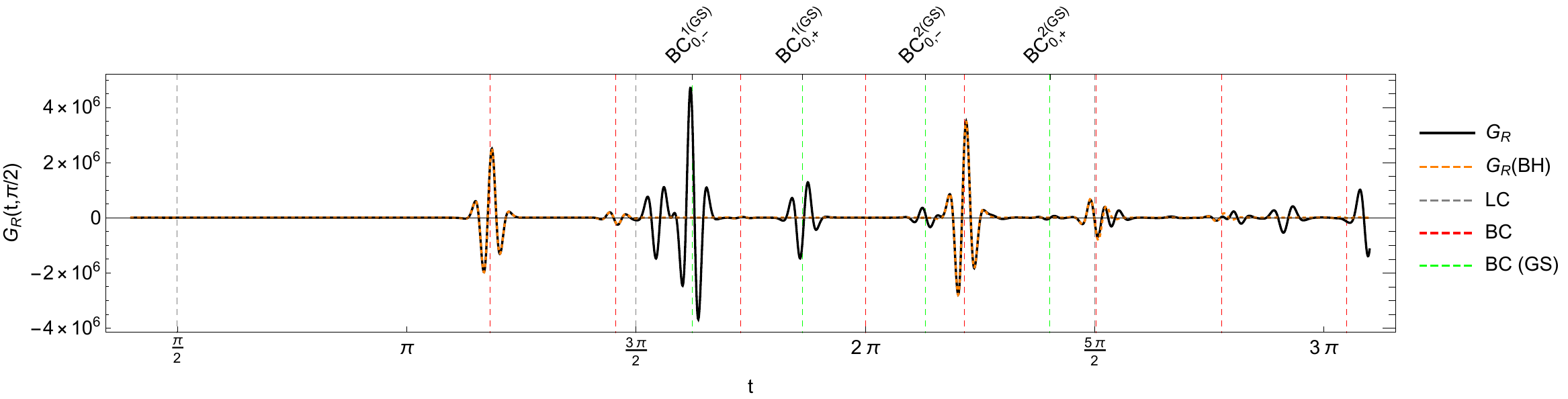}
 \caption{The Green function $G_R(t,\pi/2)$ for AdS gravastar with $d=4$, $\mu=1/50$, and $r_0=1.001r_h$.
 The green dashed lines show $\mathrm{BC}_{0,-}^{1\mathrm{(GS)}}$, $\mathrm{BC}_{0,+}^{1\mathrm{(GS)}}$, $\mathrm{BC}_{0,-}^{2\mathrm{(GS)}}$, and $\mathrm{BC}_{0,+}^{2\mathrm{(GS)}}$ from left to right.
 The red dashed lines show $\mathrm{BC}^1_{0,-}$, $\mathrm{BC}^1_{0,+}$, $\mathrm{BC}^2_{0,-}$, $\mathrm{BC}^2_{0,+}$, $\mathrm{BC}^1_{1,+}$, $\mathrm{BC}^2_{1,-}$, $\mathrm{BC}^2_{1,+}$, and $\mathrm{BC}^3_{1,-}$.
 }
 \label{fig:GRt_GV_4d_rh}
 \end{figure}

\begin{figure}[H]
 \centering
 \includegraphics[width=7.9cm]{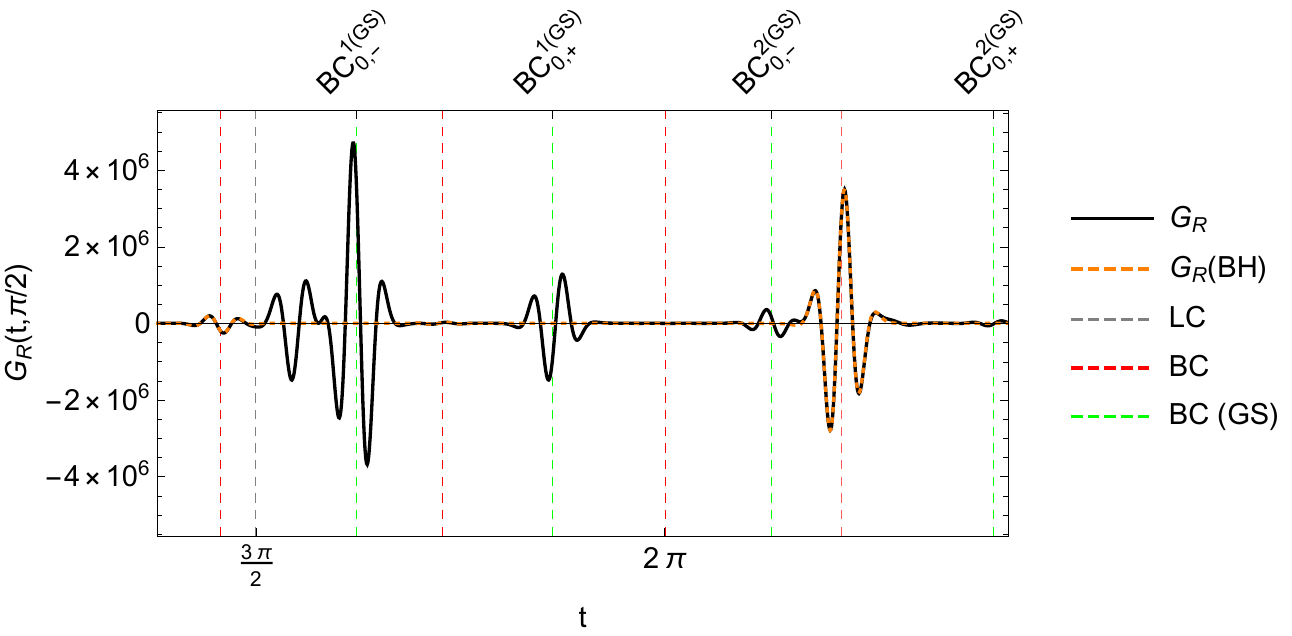}
 \includegraphics[width=7.9cm]{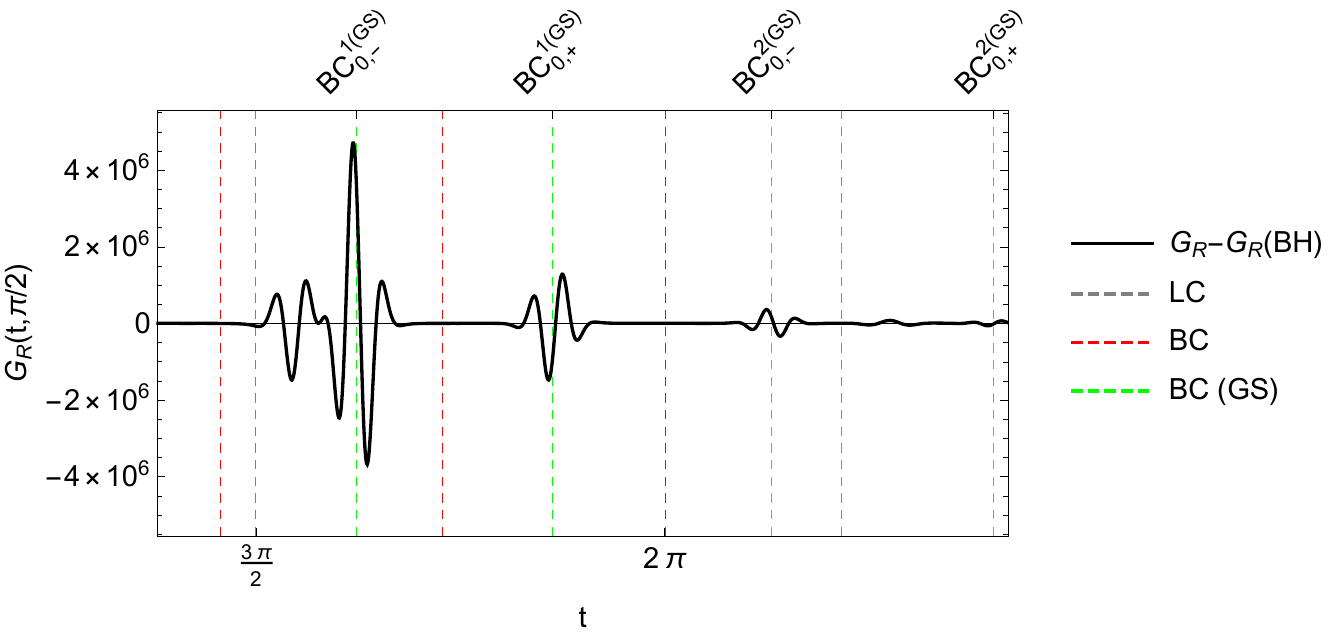}
 \caption{The Green function $G_R(t,\pi/2)$ for AdS gravastar in the region $3\pi/2 \lesssim t \lesssim 5\pi/2$.
 The green dashed lines show $\mathrm{BC}_{0,-}^{1\mathrm{(GS)}}$, $\mathrm{BC}_{0,+}^{1\mathrm{(GS)}}$, $\mathrm{BC}_{0,-}^{2\mathrm{(GS)}}$, and $\mathrm{BC}_{0,+}^{2\mathrm{(GS)}}$ from left to right.
 }
 \label{fig:GRt_GV_4d_rh_1stseq}
 \end{figure}

 Next we examine the AdS gravastar with $d=4$, $\mu=1/50$, and $r_0=1.001r_h$.
 We observe results qualitatively similar to the previous $d=3$ case.
 The structure of the bulk-cone singularities is shown in fig.~\ref{fig:BCstructure_GV_4d_rh}.
 The specific singularities $\mathrm{BC}_{n-1,\pm}^{j\mathrm{(GS)}}$ appear after $\mathrm{BC}_{n-1,\pm}^{j}$.
 The numerical result for the retarded Green function $G_R(t,\theta)$ is shown in figs.~\ref{fig:GRt_GV_4d_rh} and~\ref{fig:GRt_GV_4d_rh_1stseq}, which is consistent with the geodesic analysis.
 From the right panel of fig.\,\ref{fig:GRt_GV_4d_rh_1stseq}, we can read off the series of $\mathrm{BC}_{n-1,\pm}^{j\mathrm{(GS)}}$ for $n=1$ with the decreasing amplitude.
 As in the black hole case with $d=4$, the two-fold structure is also observed.
 
 Interestingly, all the bumps of $\mathrm{BC}_{n-1,\pm}^{j\mathrm{(GS)}}$ are odd-shaped, while all those of $\mathrm{BC}_{n-1,\pm}^{j}$ are even-shaped. 
 In $d=4$, the parity of the shape clearly distinguishes the two types of bulk-cone singularities. 
 The fact 
 can be also read off from the semi-analytic formula~\eqref{eq:GRtth}.
 For $d=4$, the factor becomes real or pure-imaginary depending on $\alpha$, $\nu$, and the sign of $T_\mathrm{null}'(u_*)$.
 Since $T_\mathrm{null}'(u_*)>0$ for singularities $\mathrm{BC}_{n-1,\pm}^j$, the factor $1/\sqrt{T_\mathrm{null}'(u_*)}$ does not contribute to the phase and all of the singularities have common parity, even or odd.
 On the other hand, since $T_\mathrm{null}'(u_*)<0$ for singularities $\mathrm{BC}_{n-1,\pm}^{j\mathrm{(GS)}}$, the factor $1/\sqrt{T_\mathrm{null}'(u_*)}=-i/\sqrt{|T_\mathrm{null}'(u_*)|}$ gives additional phase $-i$ and all of these singularities have common parity that is different from $\mathrm{BC}_{n-1,\pm}^{j}$.
 This argument on the parity relation of the two types of bulk-cone bumps would apply to general ECOs with a regular center in $d=4$, and the emergence of bulk-cone bumps of both parities can be their universal feature. 
 The exception is the wormhole case that we will see next.
 
 Note that there is a small bump just after $t=3\pi/2$ and before the first bulk-cone singularity $\mathrm{BC}_{0,-}^{1\mathrm{(GS)}}$.
 This signal would be a wave packet partially reflected at the gravastar surface, at which the potential $V(r)$ has a cusp due to the junction of geometries.
 We have discussed this feature with fictitious geodesics in~\cite{Chen:2025cee}.
 It would disappear for smooth ECOs other than our gravastar model.

 \subsection{AdS wormhole}
 \label{sec:numerical-worm}

\begin{figure}[H]
 \centering
 \includegraphics[height=6cm]{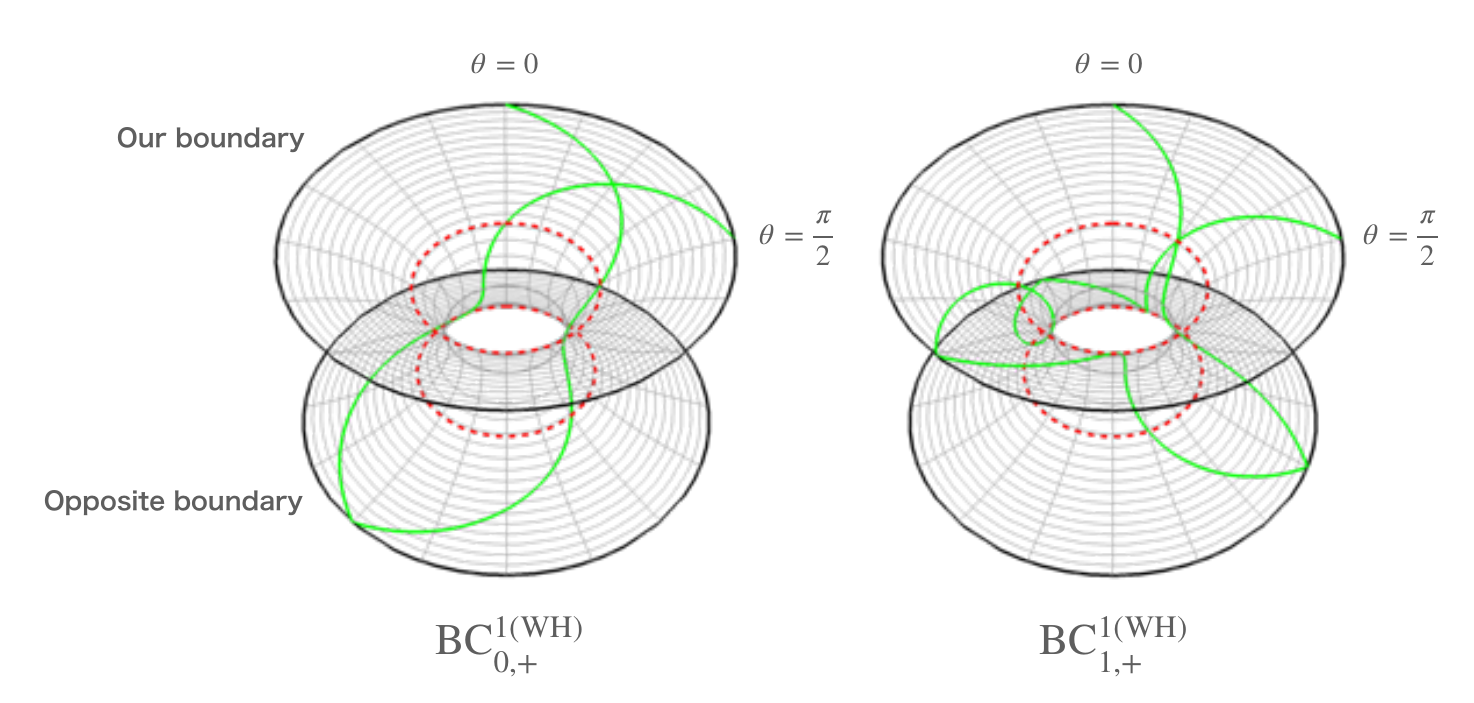}
 \caption{Illustration of null geodesics passing through the wormhole throat. They are also bounced at the AdS boundary in the opposite side.}
 \label{fig:obt_wh}
 \end{figure}

\begin{figure}[H]
 \centering
 \includegraphics[height=7cm]{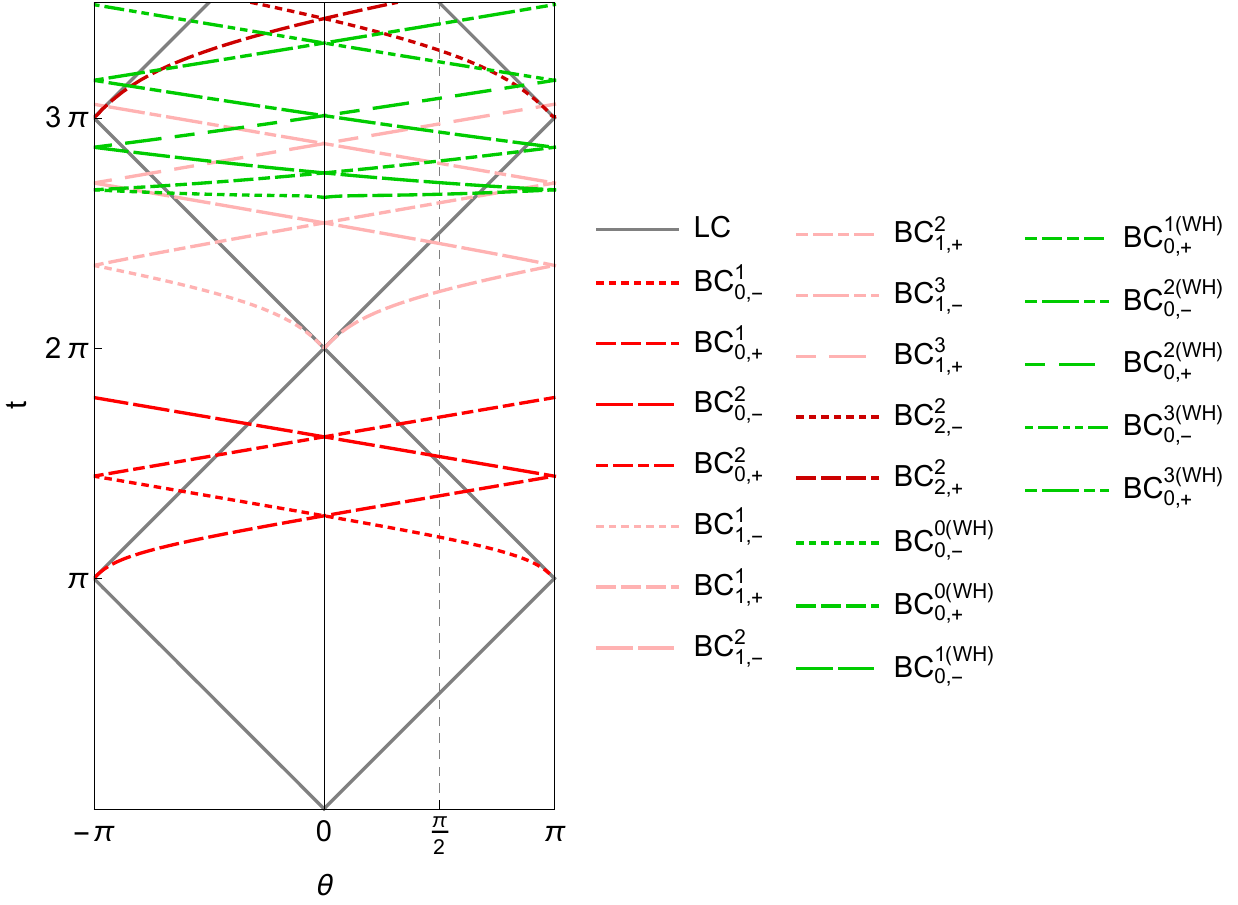}
 \caption{The structure of bulk-cone singularities  for AdS wormhole with $d=3$, $\mu=1/15$, and $r_0=1.001r_h$. 
 }
 \label{fig:BCstructure_WH_3d}
 \end{figure}

We examine the AdS wormhole with $d=3$, $\mu=1/15$, and $r_0=1.001r_h$.
There are null geodesics passing through the wormhole throat, which are bounced at the AdS boundary in the opposite side, as in fig.~\ref{fig:obt_wh}.
The structure of bulk-cone singularities is shown in fig.\,\ref{fig:BCstructure_WH_3d}.
Compared with the case of AdS gravastar in the same dimension (see fig.\,\ref{fig:BCstructure_GV_3d}), the bulk-cone singularities labeled by $\mathrm{BC}_{n-1,\pm}^{j\mathrm{(WH)}}$ appear at later times, because the corresponding bulk null geodesics propagate to the opposite side of the geometry, $z>z_0$, and are reflected by the other conformal boundary before returning to our side.
Note that the earliest $\mathrm{BC}_{n-1,\pm}^{j\mathrm{(WH)}}$ appears at $\theta=0$, in contrast to the gravastar case. This is because a radial (zero-angular-momentum) null geodesic with $\theta=0$ reaches the opposite conformal boundary and is reflected back while keeping $\theta=0$, whereas in the gravastar case the corresponding geodesic passes through the center 
$r=0$ and then continues toward 
$\theta=\pi$.

\begin{figure}[H]
 \centering
 \includegraphics[width=16cm]{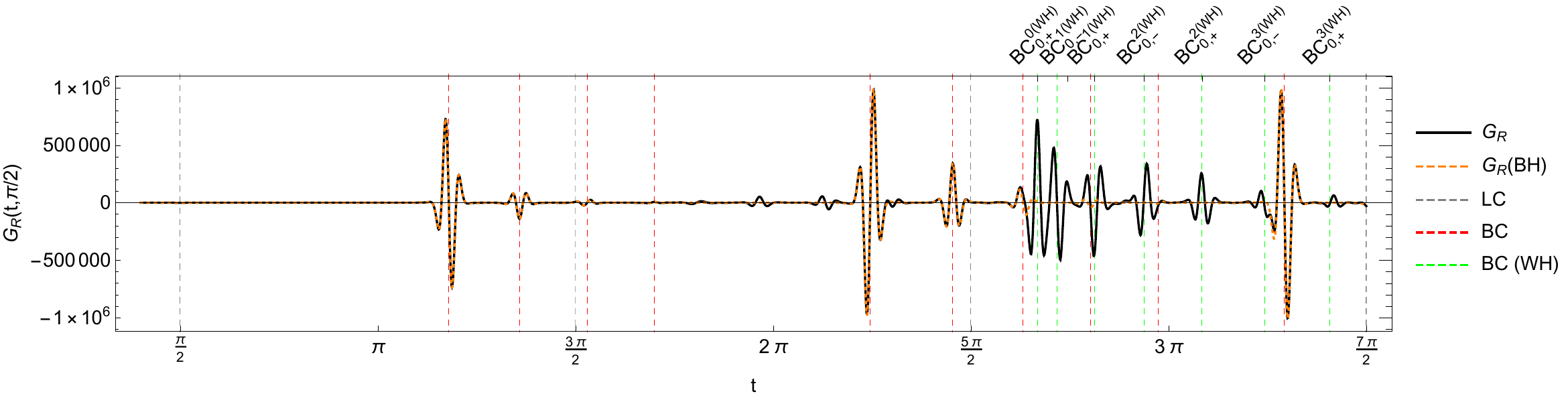}
 \caption{The Green function $G_R(t,\pi/2)$ for AdS wormhole with $d=3$, $\mu=1/15$, and $r_0=1.001r_h$. 
 The green dashed lines show $\mathrm{BC}_{0,+}^{0\mathrm{(WH)}}$, $\mathrm{BC}_{0,-}^{1\mathrm{(WH)}}$, $\mathrm{BC}_{0,+}^{1\mathrm{(WH)}}$, $\mathrm{BC}_{0,-}^{2\mathrm{(WH)}}$, $\mathrm{BC}_{0,+}^{2\mathrm{(WH)}}$, $\mathrm{BC}_{0,-}^{3\mathrm{(WH)}}$, and $\mathrm{BC}_{0,+}^{3\mathrm{(WH)}}$ from left to right.
 The red dashed lines show $\mathrm{BC}^1_{0,-}$, $\mathrm{BC}^1_{0,+}$, $\mathrm{BC}^2_{0,-}$, $\mathrm{BC}^2_{0,+}$, $\mathrm{BC}^1_{1,+}$, $\mathrm{BC}^2_{1,-}$, $\mathrm{BC}^2_{1,+}$, $\mathrm{BC}^3_{1,-}$, and $\mathrm{BC}^2_{2,-}$.
 }
 \label{fig:GRt_WH_3d}
 \end{figure}

\begin{figure}[H]
 \centering
 \includegraphics[width=7.9cm]{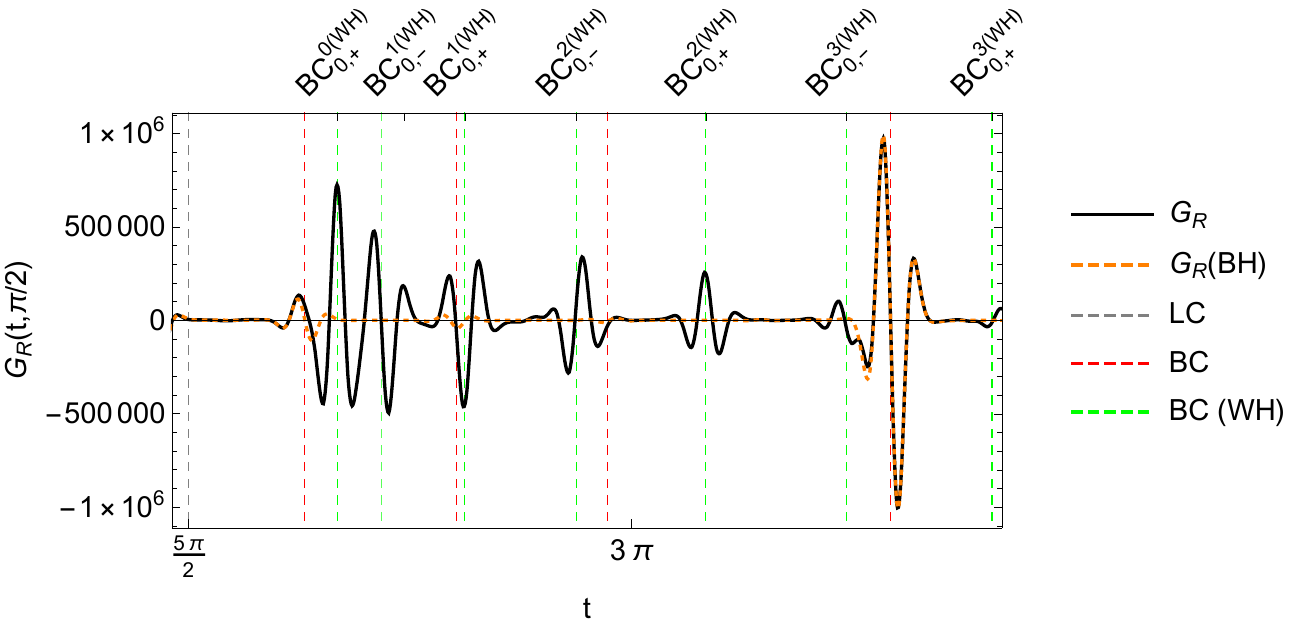}
 \includegraphics[width=7.9cm]{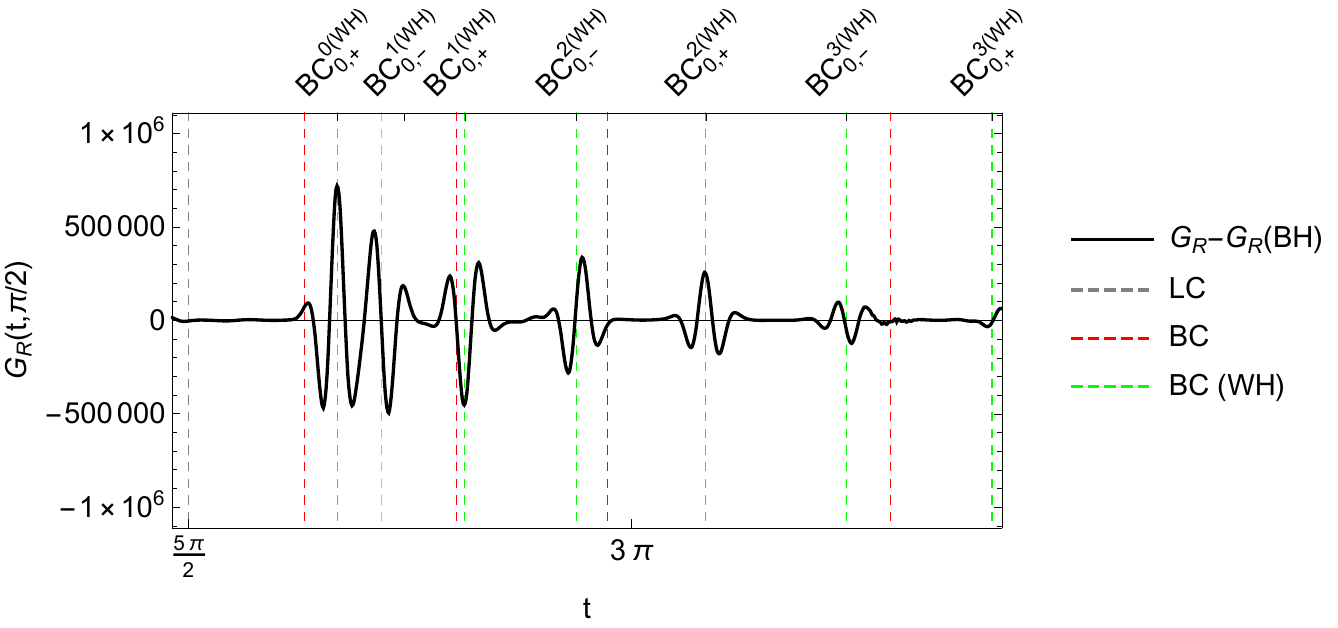}
 \caption{The Green function $G_R(t,\pi/2)$ for AdS wormhole in the region $ 5\pi/2 \lesssim t \lesssim 2\pi$. The green dashed lines show $\mathrm{BC}_{0,+}^{0\mathrm{(WH)}}$, $\mathrm{BC}_{0,-}^{1\mathrm{(WH)}}$, $\mathrm{BC}_{0,+}^{1\mathrm{(WH)}}$, 
 $\mathrm{BC}_{0,-}^{2\mathrm{(WH)}}$, $\mathrm{BC}_{0,+}^{2\mathrm{(WH)}}$, $\mathrm{BC}_{0,-}^{3\mathrm{(WH)}}$, and $\mathrm{BC}_{0,+}^{3\mathrm{(WH)}}$ from left to right.
 }
 \label{fig:GRt_WH_3d_1stseq}
 \end{figure}

Figs.~\ref{fig:GRt_WH_3d} and~\ref{fig:GRt_WH_3d_1stseq} show the retarded Green function $G_R(t,\theta)$ for $\theta=\pi/2$ and the part for the characteristic bumps in $5\pi/2\lesssim t\lesssim 7\pi/2$.
 The bumps are as strong as those common to the black hole case.
 The subtraction of the common bumps $G_R(\mathrm{WH})-G_R(\mathrm{BH})$ is shown in the right panel of fig.~\ref{fig:GRt_WH_3d_1stseq}.
 First, we can observe that the decay rate of the amplitude of the bumps is smaller than those of the black hole and gravastar cases.
 The first seven bumps of $\mathrm{BC}_{n-1,\pm}^{j\mathrm{(WH)}}$ for $n=1$ are clearly visible in the same scale, while only three or four of the common bumps $\mathrm{BC}_{n-1,\pm}^{j}$ are apparent for each $n$.
 In section~\ref{sec:lyapunov}, we will argue this point and find that the WKB analysis suggests that the decay rate is $\gamma/4$ rather than $\gamma$.
 Second, we can see the four-fold structure of $\mathrm{BC}_{n-1,\pm}^{j\mathrm{(WH)}}$ as well.
 Moreover, the shape parity in the sequence is also consistent with the semi-analytic formula.
 For example, the numerical results in figs.~\ref{fig:GRt_WH_3d} and \ref{fig:GRt_WH_3d_1stseq} show that both of the bumps $\mathrm{BC}_{0-}^{1}$ and $\mathrm{BC}_{0-}^{1\mathrm{(WH)}}$ are odd-shaped.
 The semi-analytic formula~\eqref{eq:GRtth} predicts that, for the same labels $j,n,\pm$, $\mathrm{BC}_{n-1,\pm}^{j}$ and $\mathrm{BC}_{n-1,\pm}^{j\mathrm{(WH)}}$ have the same phase except for the contributions from $a_n$ and $1/\sqrt{T_\mathrm{null}'(u_*)}$.
 For the bulk-cone singularities common to a black hole, $a_n$ is obtained from eq.~\eqref{eq:an} and $T_\mathrm{null}'(u_*)>0$.
 For those of the wormhole, $a_n$ is obtained from eq.~\eqref{eq:an2} and $T_\mathrm{null}'(u_*)<0$.
 Accordingly, the relative phase between $\mathrm{BC}_{n-1,\pm}^{j}$ and $\mathrm{BC}_{n-1,\pm}^{j\mathrm{(WH)}}$ is obtained as $(a_n^{\mathrm{BH}}/a_n^{\mathrm{WH}})\times i=ie^{in\pi\nu}$.
 For $n=1$ and $\nu=d/2=3/2$, it becomes $1$.
 Thus, the analytic formula predicts the same phase for $\mathrm{BC}_{0,\pm}^{j}$ and $\mathrm{BC}_{0,\pm}^{j\mathrm{(WH)}}$ implying the same shape and parity for the corresponding bumps as in the numerical result.
 
 Note that there are small signals around $t\sim 2\pi$.
 Again, we infer that these features originate from waves that are partially reflected at the junction surface 
$r=r_0$  due to the cusp-like potential.
 We expect that such signals would disappear if the spacetime is constructed without a thin-shell and smooth.

\begin{figure}[H]
 \centering
 \includegraphics[height=7cm]{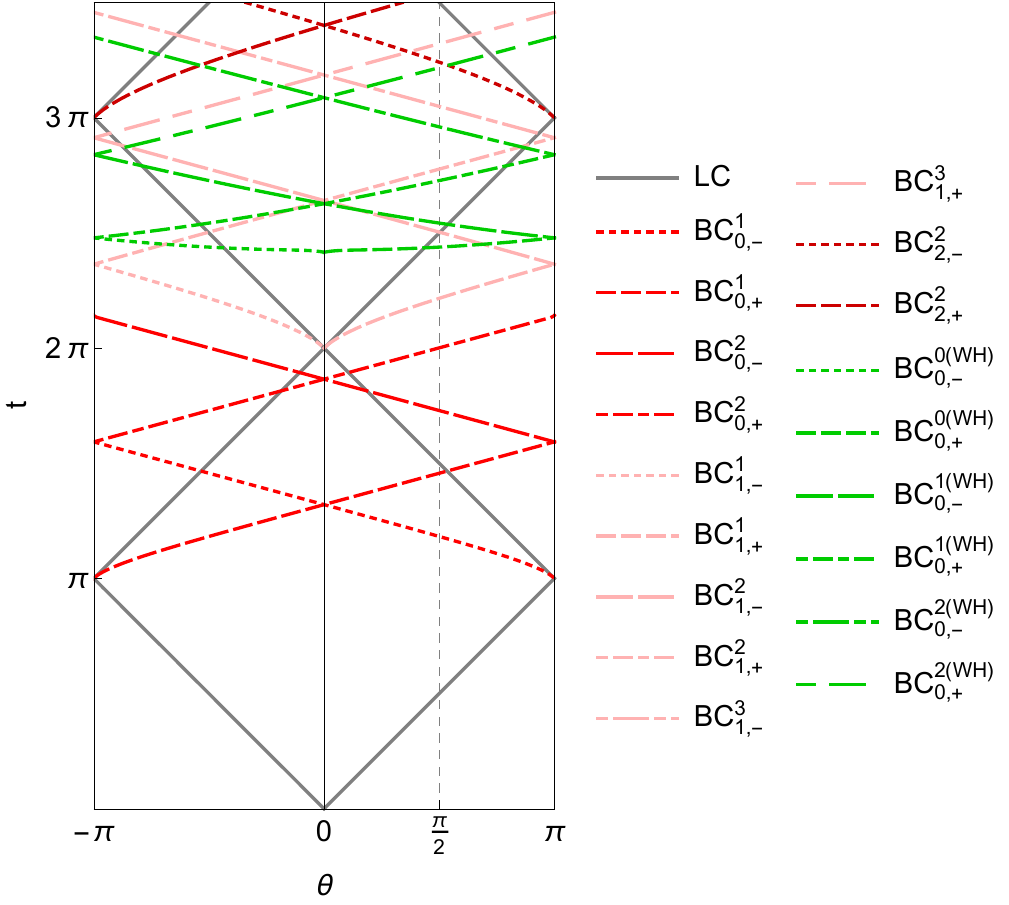}
 \caption{The structure of bulk-cone singularities  for AdS wormhole with $d=4$, $\mu=1/50$, and $r_0=1.001r_h$.}
 \label{fig:BCstructure_WH_4d}
 \end{figure}

\begin{figure}[H]
 \centering
 \includegraphics[width=16cm]{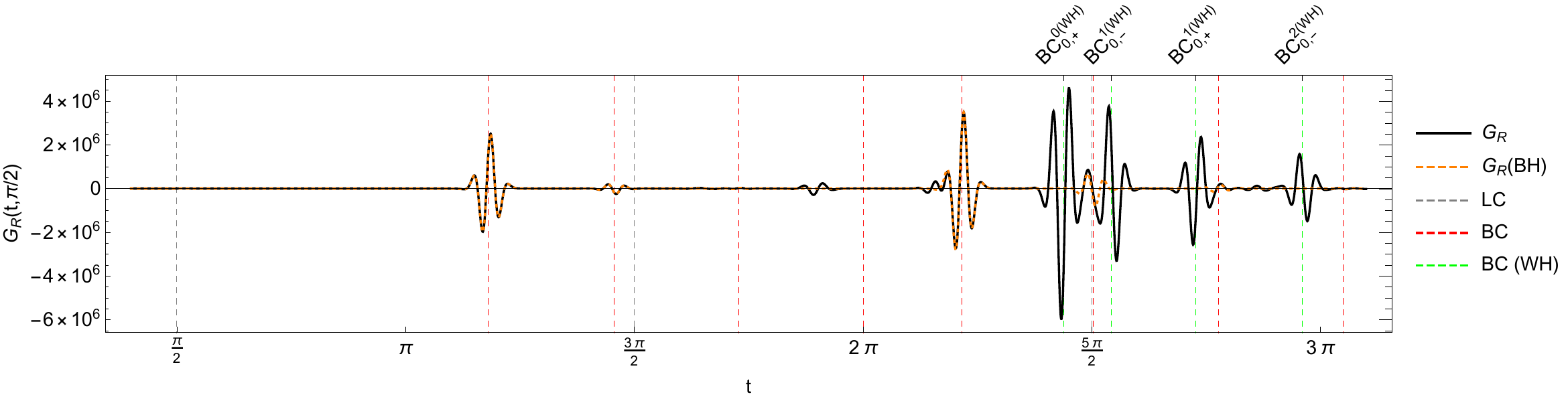}
 \caption{The Green function $G_R(t,\pi/2)$ for AdS wormhole with $d=4$, $\mu=1/50$, and $r_0=1.001r_h$. 
 The green dashed lines show $\mathrm{BC}_{0,+}^{0\mathrm{(WH)}}$, $\mathrm{BC}_{0,-}^{1\mathrm{(WH)}}$, $\mathrm{BC}_{0,+}^{1\mathrm{(WH)}}$, and $\mathrm{BC}_{0,-}^{2\mathrm{(WH)}}$ from left to right.
 The red dashed lines show $\mathrm{BC}^1_{0,-}$, $\mathrm{BC}^1_{0,+}$, $\mathrm{BC}^2_{0,-}$, $\mathrm{BC}^2_{0,+}$, $\mathrm{BC}^1_{1,+}$, $\mathrm{BC}^2_{1,-}$, $\mathrm{BC}^2_{1,+}$, $\mathrm{BC}^3_{1,-}$.
 }
 \label{fig:GRt_WH_4d}
 \end{figure}

\begin{figure}[H]
 \centering
 \includegraphics[width=7.9cm]{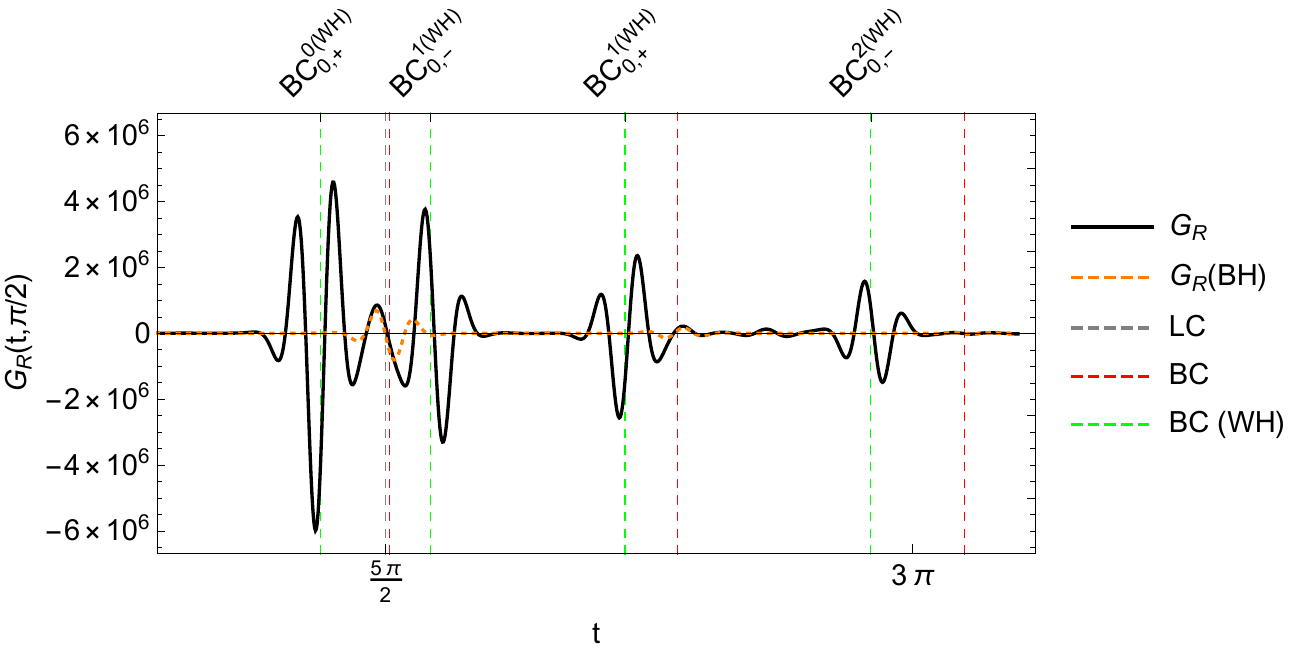}
 \includegraphics[width=7.9cm]{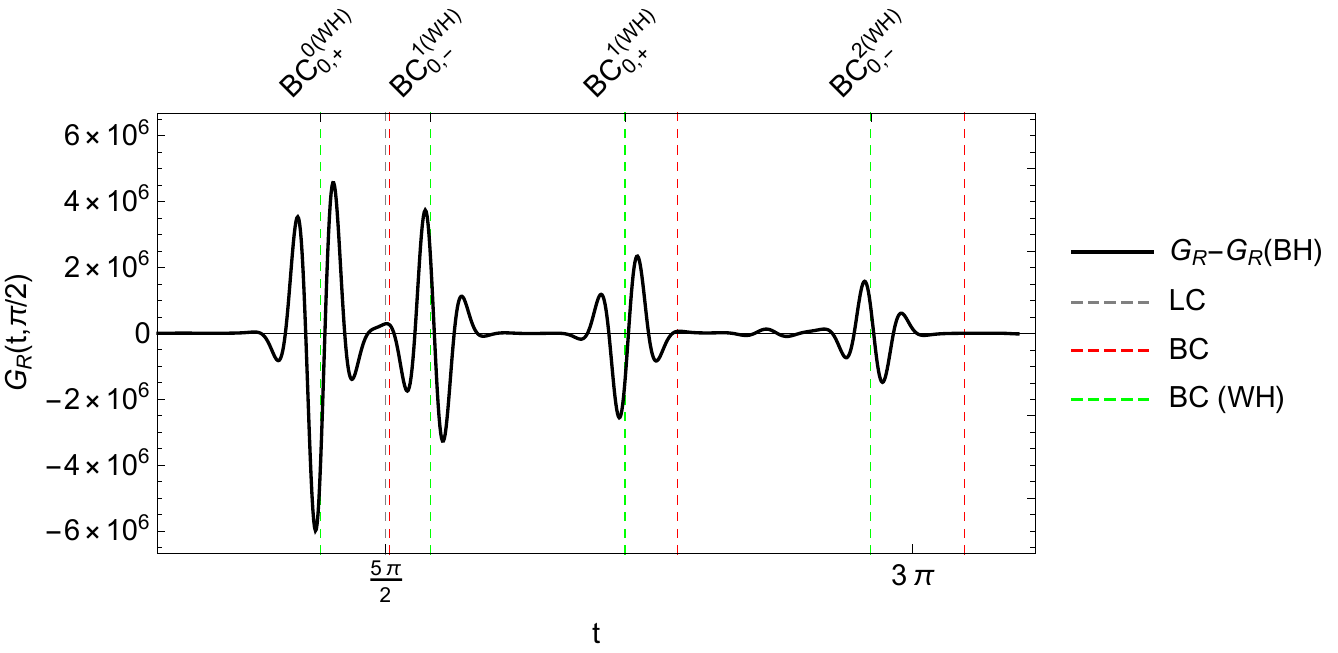}
 \caption{The Green function $G_R(t,\pi/2)$ for AdS wormhole in the region $ 5\pi/2 \lesssim t \lesssim 2\pi$.  The green dashed lines show $\mathrm{BC}_{0,+}^{0\mathrm{(WH)}}$, $\mathrm{BC}_{0,-}^{1\mathrm{(WH)}}$, $\mathrm{BC}_{0,+}^{1\mathrm{(WH)}}$, and
 $\mathrm{BC}_{0,-}^{2\mathrm{(WH)}}$ from left to right.
 }
 \label{fig:GRt_WH_4d_1stseq}
 \end{figure}

 Finally, we examine the AdS wormhole with $d=4$, $\mu=1/50$, and $r_0=1.001r_h$.
 The structure of bulk-cone singularities is shown in fig.~\ref{fig:BCstructure_WH_4d}.
 The retarded Green function $G_R(t,\theta)$ for $\theta=\pi/2$ is shown in figs.~\ref{fig:GRt_WH_4d} and~\ref{fig:GRt_WH_4d_1stseq}.
We observe features qualitatively similar to the $d=3$ case. The bumps of  $\mathrm{BC}_{n-1,\pm}^{j\mathrm{(WH)}}$ exhibit a slower decay in amplitude and show the two-fold structure in their shape.
Although the first bump $\mathrm{BC}_{0,+}^{0\mathrm{(WH)}}$ is subtle, all the other bumps $\mathrm{BC}_{0,+}^{j\mathrm{(WH)}}$ in the right panel of fig.~\ref{fig:GRt_WH_4d_1stseq} are clearly odd-shaped, whereas all the bumps of $\mathrm{BC}_{0,+}^{j}$ are even-shaped.
As argued in the previous $d=3$ wormhole case, the semi-analytic formula suggests their relative phase as $ie^{in\pi\nu}$.
For $n=1$ and $\nu=d/2=2$, it becomes $-i$ implying that $\mathrm{BC}_{0,+}^{j}$ and $\mathrm{BC}_{0,+}^{j\mathrm{(WH)}}$ have different parity and the numerical result is consistent with the prediction.
 
\subsection{Echoes of AdS ECOs}
\label{sec:echoes}

In \cite{Cardoso:2016oxy}, gravitational echoes are observed for ECOs.
The echoes appear in the radial wave equation for a fixed angular momentum $\ell$, where the authors put a Gaussian profile for the initial value.
That is, they inject a spherical wave with $\ell=1$ toward the central object and it is radially localized with a finite thickness.
Here we generalize their configuration such as to be suitable for the AdS/CFT context.
We expect that the inverse Fourier transformation of $G_R(\omega,\ell)$ without summation over $\ell$,
\begin{align}
G_R(t,\ell)=\int^{+\infty+i\delta}_{-\infty +i\delta} d\omega e^{-i\omega t}G_R(\omega,\ell)e^{-\frac{\mathrm{Re}(\omega)^2}{\omega_c^2}}
\end{align}
realizes a spherical wave with finite thickness.%
\footnote{The smeared Green function is related to the one-point function
\(\langle \mathcal{O}(t,\ell) \rangle_{J_{\mathcal O}}\)
in the presence of a non-trivial source \(J_{\mathcal O}\).
The source \(J_{\mathcal O}\), which couples to the operator \(\mathcal{O}\),
is with fixed \(\ell\) and has a Gaussian profile centered around \(t=0\).
In the dual gravity description, the scalar field takes boundary value proportional to \(J_{\mathcal O}\) near the AdS boundary, and hence our setup is a direct generalization of that in \cite{Cardoso:2016oxy}. 
In this sense, the smeared Green function can be regarded as a physical observable, see, e.g., \cite{Hashimoto:2018okj,Hashimoto:2019jmw}.
We would like to thank the referee for suggesting this point.
}
The cutoff $\omega_c$ should give the time interval $\delta t\sim 1/\omega_c$ in which the source at the boundary is switched on.
The time interval would be translated to the typical length of the radial localization of the wave as $\delta r\sim \delta t$.

In order to clearly observe the echoes, we need to choose an appropriate cutoff $\omega_c\lesssim \sqrt{V_\ell(r_c)}$ such that the waves with $\omega<\omega_c$  can tunnel through the photon-sphere potential.
According to \cite{Cardoso:2016oxy}, the typical time scale of the echo interval is given by
\begin{align}
    \Delta t_\mathrm{echo}\sim 2\int_{r_0}^{r_c}\frac{dr}{f(r)}.
\end{align}
This is twice the propagation time of a null geodesic from the ECO surface to the photon sphere radius.
Additionally, we would also see that bumps repeatedly appear due to reflection by the AdS boundary.
Their typical time interval is given by twice the propagation time of a null trajectory from the AdS boundary to the photon sphere,%
\footnote{
The time scales $\Delta t_\mathrm{echo}$ and $\Delta t_\mathrm{bdry}$ roughly estimate $\tilde{T}_\mathrm{null}(u)$ of eq.~\eqref{eq:TThetat} and $T_\mathrm{null}(u)$, respectively.
Although it is difficult to directly compare the semi-analytic formula for $G_R(t,\theta)$ and the numerical result for $G_R(t,\ell)$, the typical time scales agree with each other.
}
\begin{align}
    \Delta t_\mathrm{bdry}\sim 2\int_{r_c}^{\infty}\frac{dr}{f(r)}.
\end{align}

\subsubsection{AdS gravastar}

\begin{figure}[H]
\centering
\includegraphics[width=7.9cm]{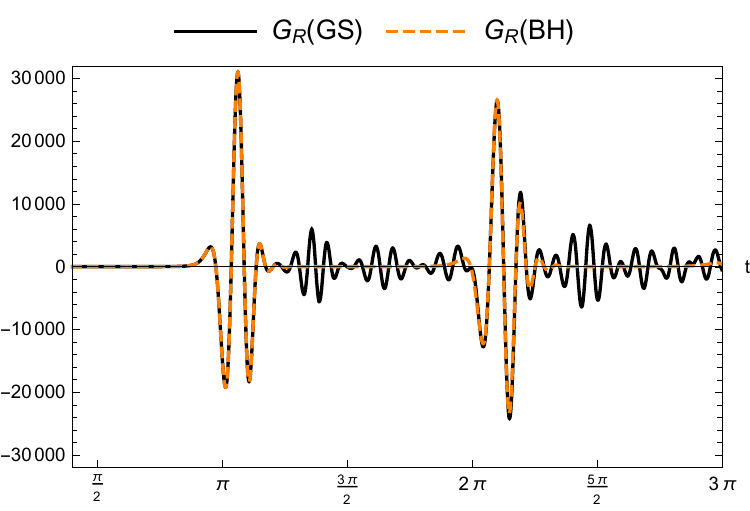}
\includegraphics[width=7.9cm]{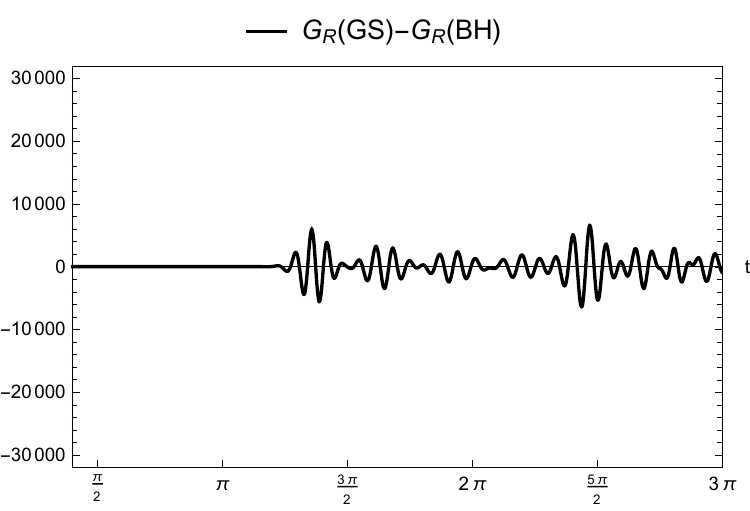}
\caption{Echoes of $G_R(t,\ell)$ for AdS gravastar with $d=3$, $\mu=1/50$, and $r_0=1.000001r_h$. We take $\ell=1$ and $\omega_c=15$.}
\label{fig:GRtl_GS_d3_mu1_50}
\end{figure}

For the AdS gravastar with $d=3$, $\mu=1/50$, and $r_0=1.000001r_h$, we obtain the results for $\ell=1$ as in fig.~\ref{fig:GRtl_GS_d3_mu1_50}.
The time scales are given as $\Delta t_\mathrm{echo}\simeq0.544\ll\Delta t_\mathrm{bdry}\simeq3.243$.
The left panel shows the Green functions of the gravastar and the black hole with the same $\mu$.
The strong bumps completely coincide and they are the waves reflected by the photon sphere.
The second one is the wave once reflected by the AdS boundary.
The right panel shows their difference and the remaining weak waves are the echoes.
The first four waves have decreasing amplitude as the typical echo signals~\cite{Cardoso:2016oxy}.
At the fifth wave around $t\sim 5\pi/2$, the amplitude becomes larger.
This should correspond to the beginning of the second echo sequence, which has undergone a single reflection at the AdS boundary.
In fact, the interval of the beginning of the first and second echo sequences, $\Delta t_\text{int} \sim \pi$, coincides with that of the strong bumps in the left panel.
The result is consistent with the hierarchy, $\Delta t_\mathrm{echo}\ll \Delta t_\mathrm{bdry}$.

\begin{figure}[H]
\centering
\includegraphics[width=7.9cm]{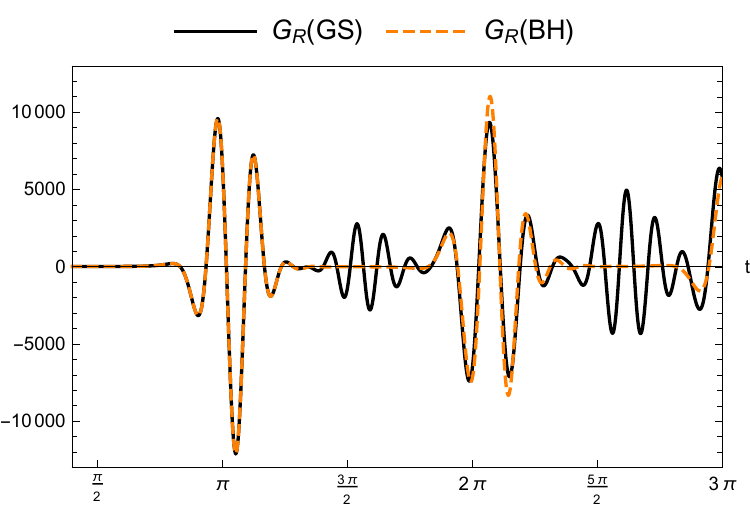}
\includegraphics[width=7.9cm]{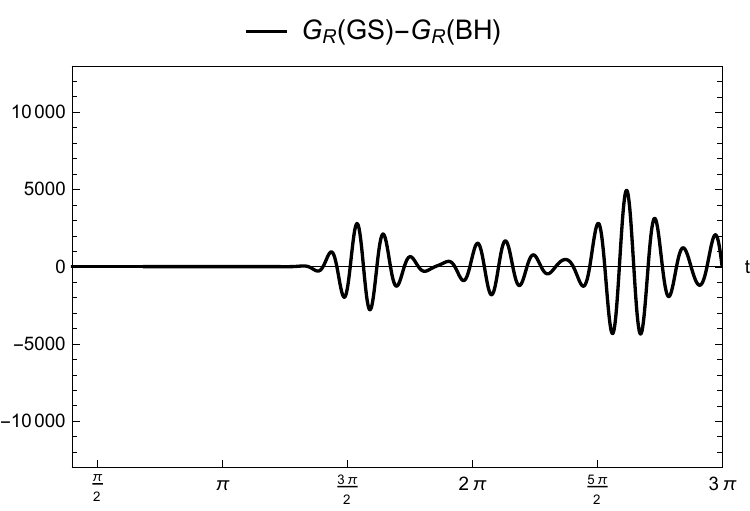}
\caption{Echoes in $G_R(t,\ell)$ for AdS gravastar with $d=4$, $\mu=1/300$, and $r_0=1.000001r_h$. We take $\ell=1$ and $\omega_c=9$.}
\label{fig:GRtl_GS_d4_mu1_300}
\end{figure}

For $d=4$, $\mu=1/300$, $r_0=1.000001r_h$, $\ell=1$, and $\omega_c=9$, the results are shown in fig.~\ref{fig:GRtl_GS_d4_mu1_300}.
The time scales are $\Delta t_\mathrm{echo}\simeq0.778\ll\Delta t_\mathrm{bdry}\simeq3.065$.
We can see three echoes, where the first two are of the first echo sequence and the third is of the second sequence.

\subsubsection{AdS wormhole}
\begin{figure}[H]
\centering
\includegraphics[width=7.9cm]{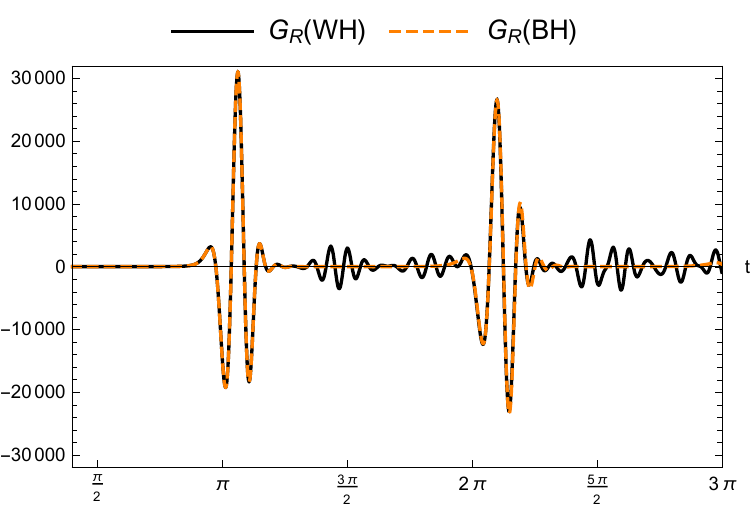}
\includegraphics[width=7.9cm]{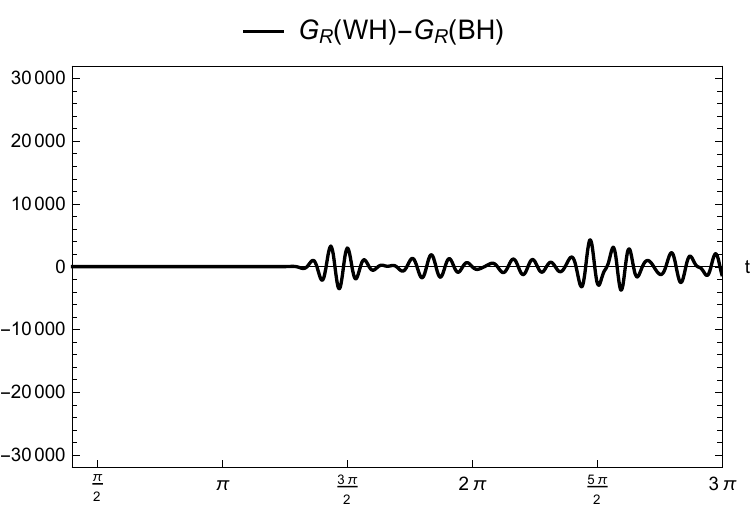}
\caption{Echoes in $G_R(t,\ell)$ for AdS wormhole with $d=3$, $\mu=1/50$ and $r_0=1.000001r_h$. We take $\ell=1$ and $\omega_c=15$.}
\label{fig:GRtl_WH_d3_mu1_50}
\end{figure}

\begin{figure}[H]
\centering
\includegraphics[width=7.9cm]{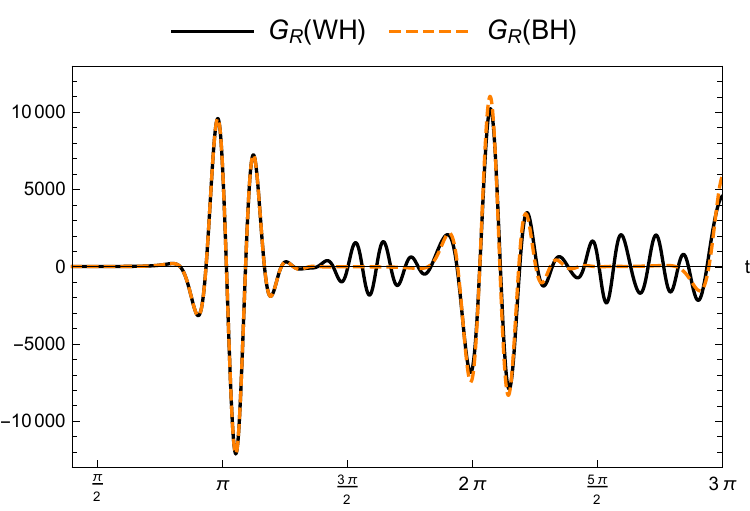}
\includegraphics[width=7.9cm]{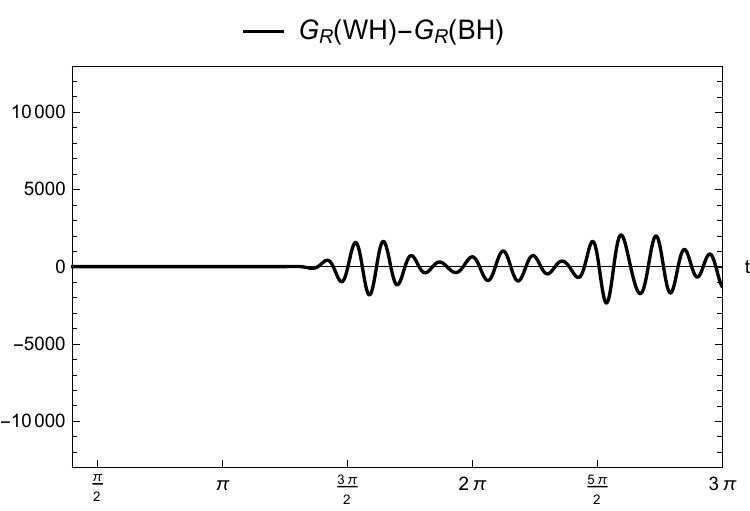}
\caption{Echoes in $G_R(t,\ell)$ for AdS wormhole with $d=4$, $\mu=1/300$ and $r_0=1.000001r_h$. We take $\ell=1$ and $\omega_c=9$.}
\label{fig:GRtl_WH_d4_mu1_300}
\end{figure}

For the AdS wormhole with $d=3$, $\mu=1/50$, and $r_0=1.000001r_h$, we obtain the results for $\ell=1$ as in fig.~\ref{fig:GRtl_WH_d3_mu1_50}.
The time scales are given as $\Delta t_\mathrm{echo}\simeq0.544\ll\Delta t_\mathrm{bdry}\simeq3.243$.
The cutoff frequency is taken as $\omega_c=15$.
In this case, three signals from the first echo sequence are observed and the second echo sequence follows them from $t\sim 5\pi/2$.
For $d=4$, $\mu=1/300$, $r_0=1.000001r_h$, $\ell=1$, and $\omega_c=9$, the results are shown in fig.~\ref{fig:GRtl_WH_d4_mu1_300}.
The time scales are given as $\Delta t_\mathrm{echo}\simeq0.778\ll\Delta t_\mathrm{bdry}\simeq3.065$.
In this case, we can also see echoes, where the first two echoes are of the first sequence and the third is of the second sequence.


Note that, in the AdS wormhole, waves can also be reflected by the AdS boundary on the opposite side of the geometry.
The propagation time of such waves would be $2(\Delta t_\mathrm{bdry}+\Delta t_\mathrm{echo})\sim 2\pi$.
So, their signal might exist in figs.~\ref{fig:GRtl_WH_d3_mu1_50} and~\ref{fig:GRtl_WH_d4_mu1_300}.
However, they should have experienced tunneling at least four times.
We expect that their amplitude is highly suppressed and invisible in the present numerical results.

\subsection{Decay rate and Lyapunov exponent}
\label{sec:lyapunov}
According to eq.~\eqref{eq:GRtth}, the decay rate of the bulk-cone singularities is controlled by $T_\mathrm{null}'(u_*)$.
For the bulk-cone singularities reflected by the photon sphere, it is related to the classical Lyapunov exponent 
$\gamma$ as $T_\mathrm{null}'(u_*)\sim \exp (\gamma t)$.
This divergence corresponds to the limit $u_*\to u_c-0$ in eq.~\eqref{eq:TTheta} with the potential of fig.~\ref{fig:AdSSnull}, where the orbit winds around the photon sphere infinite times from outside.
In~\cite{Dodelson:2023nnr}, it was shown that this prediction agrees with the decreasing amplitude of the bumps in the numerical results.

For the bulk-cone singularities specific to our ECOs, the decay rate is also related to the Lyapunov exponent.
For the AdS gravastar, null geodesics passing through its interior have a supercritical parameter $u>u_c$ and approach 
$u\to u_c+0$  at late arrival times, as suggested by the potential in fig.~\ref{fig:Vgravg}.
Evaluating 
in the limit, we find that the (half) exponent behaves as 
$|T_\mathrm{null}'(u_*)|\sim \exp( \gamma t /2 )$.
This is because that the null geodesics pass through the photon sphere twice, when going into and coming back from the photon sphere inside.
The time scale for staying around the photon sphere is given by $T_\mathrm{null}\sim 1/\gamma$, and staying there twice means that the arrival time becomes twice $T_\mathrm{null}\sim 2/\gamma$, leading to the half exponent.
Note that the divergence $T(u)\to\infty$ for $u\to u_c+0$ implies $T_\mathrm{null}'(u_*)<0$.
For the wormhole case, there are two photon spheres with the same potential height as in fig.~\ref{fig:Vworm}.
A supercritical null geodesic with $n=1$ (no bounce at our AdS boundary and one bounce at the opposite boundary) passes through the local maxima of the potential four times, leading to the quarter exponent 
$|T_\mathrm{null}'(u_*)|\sim \exp (\gamma t /4)$.
The bulk-cone singularities for the ECOs have weaker decay rate.

\begin{figure}[H]
 \centering
 \includegraphics[height=6cm]{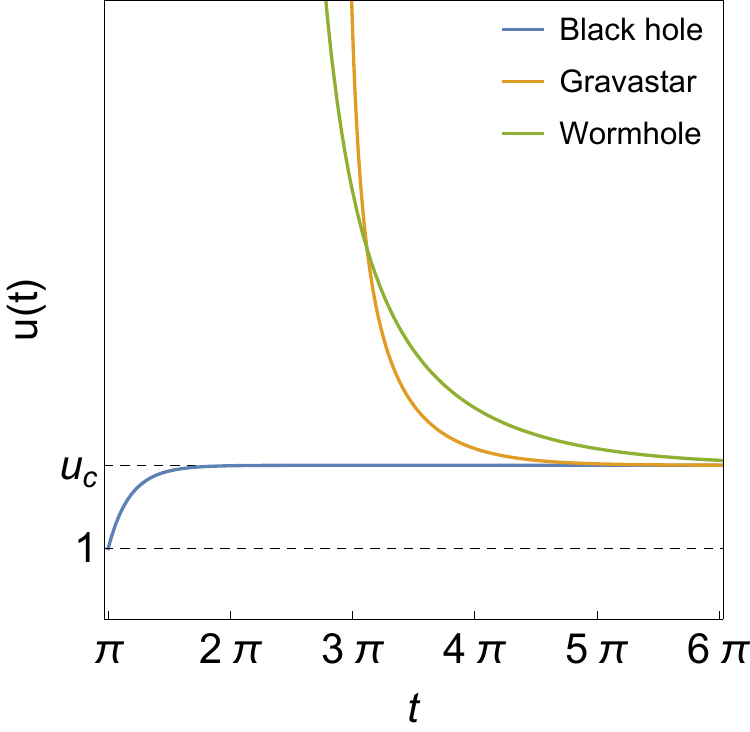} \hspace{2cm}
 \includegraphics[height =6cm]{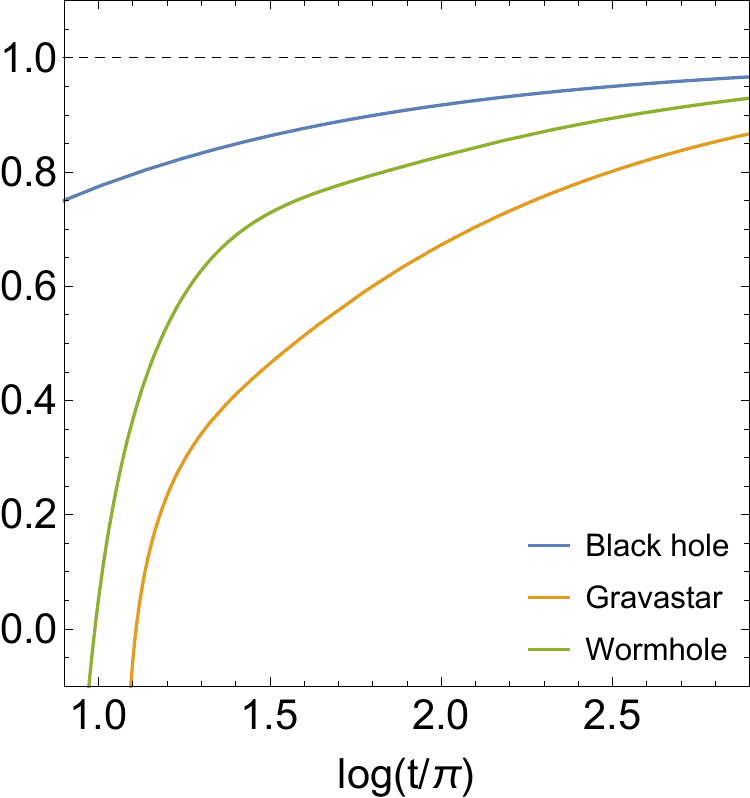}
 \caption{Plot of $u(t)$ and $T_\mathrm{null}'(u(t))$ for the black hole, gravastar and wormhole.
 The right panel shows $\frac{1}{\gamma t}\log |T_\mathrm{null}'(u(t))|$, $\frac{2}{\gamma t}\log |T_\mathrm{null}'(u(t))|$, and $\frac{4}{\gamma t}\log |T_\mathrm{null}'(u(t))|$, respectively.
 They have the same ADM mass $\mu=1$ in $d=4$ and both ECO radii are $r_0=1.001 r_h$.}
 \label{fig:lyapunov}
 \end{figure}

Fig.~\ref{fig:lyapunov} shows the behavior of $u(t)$ and $|T_\mathrm{null}'(u(t))|$ for the black hole and ECOs with the same mass $\mu=1$, where $u(t)$ is the inverse function of $T_\mathrm{null}(u)$.
In the late time, we have $u(t)\to u_c \pm 0$.
The convergence of $|T_\mathrm{null}'(u(t))|$ to 
$\exp(\gamma t)$, $\exp ( \gamma t / 2 )$, and $\exp ( \gamma t /4)$ 
are confirmed for the black hole, gravastar, and wormhole, respectively.
The slower decay of the bulk-cone singularities for the gravastar and wormhole agrees with the numerical results in sections~\ref{sec:numerical-grav} and~\ref{sec:numerical-worm}.
We have observed that the amplitude of the bumps of $\mathrm{BC}_{n-1,\pm}^{j\mathrm{(GS)}}$ and $\mathrm{BC}_{n-1,\pm}^{j\mathrm{(WH)}}$ for ECOs decreases more slowly than those of $\mathrm{BC}_{n-1,\pm}^{j}$ for the black holes.

\section{Conclusion and discussions}
\label{sec:conclusion}

We have evaluated the retarded Green function \eqref{eq:defretarded} of scalar operator in a CFT from the bulk scalar wave function in the dual asymptotical AdS geometries. 
The CFT correlation function has bulk-cone singularities and their locations provide the information about bulk null geodesics. From this information, we can read off the structure of bulk geometry. 
We have shown this by geodesic approximation as in \eqref{eq:2ptsemi} and by wave function analysis in 
\eqref{eq:GRtth} with \eqref{eq:Xnpmj}.
The wave function analysis is based on the WKB approach to the wave equation for bulk scalar field as well as numerical computations. The results extend the previous ones obtained in \cite{Dodelson:2020lal,Dodelson:2023nnr} for AdS-Schwarzschild black hole such that the analysis is applicable to generic asymptotic AdS geometry. In particular, we have shown that the power of bulk-cone singularity in \eqref{eq:power} holds generically. Moreover, the structure of bulk-cone singularities depends on the dimensionality as in the case of AdS-Schwarzschild black hole.

We can construct geometry with photon sphere but without black hole horizon. Such geometries are given by ECOs and AdS gravastar and AdS wormhole have been analyzed as the concrete examples. We examined the structure of these geometries by applying the semi-analytical method we have developed. We have shown that the retarded Green functions have the bulk-cone singularities associated with null geodesics traveling into the region inside the photon sphere as well as those going around the photon sphere. In the case of ECOs, there are wave functions localized inside the photon sphere and we have shown that the retarded Green functions have echoes following the bulk-cone singularities related to null geodesics going around the photon sphere. Using the bulk-cone singularities and echoes, we can clearly distinguish AdS-Schwarzschild black hole and its alternatives, such as, AdS gravastar and AdS wormhole. We regard these two geometries as the two representatives of ECOs. We have focused on geometries without rotation or charge, and it is important to investigate what happens when more generic configurations are considered.

In this paper, we have examined the bulk wave equations to obtain CFT correlation functions. It is beneficial if the similar results can be obtained directly from the CFT viewpoints. The bulk-cone singularities are expected to be strong coupling effects of CFT. The bulk analysis suggests that the power of bulk-cone singularities given in \eqref{eq:power} appears independent of the detailed structure of dual geometry. It is desired to understand the power in terms of strongly coupled phenomena. 
On the other hand, the echoes are signals for the existence of localized modes near the minimum of the potential inside the photon sphere. 
The condensation of the localized modes is expected to lead to the decay of an ECO into a black hole, see, e.g., \cite{Cardoso:2014sna}. This decay is caused by nonlinear instabilities and it is difficult to analyze the phenomena directly from gravity theory. The decay into black hole is known to be described as thermalization in dual CFT. Therefore, the dual CFT viewpoints are expected to be useful to understand the instabilities of ECOs. The dual CFT also suggests that the singularities in the bulk would be resolved by gravitational and stringy effects and it is interesting to study them in the current setup, see, e.g., \cite{Dodelson:2020lal,Dodelson:2023nnr}.

We examined AdS gravastar and AdS wormhole as simple thin-shell models for which the retarded Green functions of the dual CFT can be computed in a controlled manner. It is important to construct such kind of geometries in more realistic setups.
See, e.g., \cite{Ogawa:2024joy} for the case of asymptotic flat gravastar. AdS wormhole solutions in superstring theory have been constructed as in \cite{Maldacena:2004rf} and attract much attentions recently in the context of AdS/CFT correspondence, see, e.g., \cite{Gao:2016bin,Maldacena:2017axo,Maldacena:2018lmt}. 
Recently, there are several works on examining the black hole singularity from the dual CFT viewpoints \cite{Afkhami-Jeddi:2025wra,Ceplak:2025dds,Dodelson:2025jff,Chakravarty:2025ncy,Jia:2025jbi} (see, e.g., \cite{Kraus:2002iv,Fidkowski:2003nf,Festuccia:2005pi} for previous works). We hope that the techniques developed in this paper are useful for these problems as well.

\subsection*{Acknowledgements}

We are grateful to Koji Hashimoto, Yasuyuki Hatsuda, Takaaki Ishii, Keiju Murata, Naritaka Oshita, and Seiji Terashima for useful discussions. The work of H.\,Y.\,C. is supported in part by Ministry of Science and Technology (MOST) through the grant 114-2112-M-002-022-. 
The work of Y.\,H. is supported by JSPS KAKENHI Grant Numbers JP21H05187 and JP23K25867. The work of Y.\,K. is supported by JSPS KAKENHI Grant Numbers JP21K20367 and JP23KK0048.

\appendix

\section{Green functions from Regge poles}
\label{app:regge}

In the main context, we integrate \eqref{eq:gr} directly along the line slightly above the real line. This is contrast to the analysis in \cite{Dodelson:2023nnr}, where the Green function was obtained from the residues of Regge poles. 
In this appendix, we repeat the analysis but not taking $p = i (\ell + \alpha)$ to be real.

We examine an approximation, where $\ell \in \mathbb{Z}$ is large as $\ell \to \infty$ with keeping $ u =\omega /\ell$ finite.
We consider the case where the equation $\kappa (z)^2  = 0$ has two zeros, whose positions are denoted as $z = z_\pm$ with $z_- \lesssim z_+$. 
We require the ingoing boundary condition at the horizon located at $z \to \infty$. Thus, for $z > z_+$, the solution is given by
\begin{align}
    \psi (z) \sim \frac{1}{\sqrt{\kappa(z)}} e^{i \int_{z_+}^z d z' \kappa (z')}  .
\end{align}
Applying the WKB connection formula in \cite{Berry:1972na,s__c__miller__1953},
the  solution for $z < z_-$ can be found as
\begin{align} \label{eq:psismall}
    \psi (z) \sim \frac{C_+}{\sqrt{\kappa(z)}} e^{i  \int_{0}^z d z' \kappa (z')} + \frac{C_-}{\sqrt{\kappa(z)}} e^{ - i  \int_{0}^z d z' \kappa (z')} 
\end{align}
with
\begin{align}
\begin{aligned}
   & C_+ = \left(\frac{2}{\pi}\right)^{1/2} \left( - \frac{t}{2 e}\right)^{- \frac12 i t} \Gamma \left(\frac12 i t + \frac12\right) \cosh \left(\frac12 t \pi \right) e^{- \frac14 t \pi}   e^{-i  \int_{0}^{z_-} d z' \kappa (z')} , \\
      & C_- = e^{- \frac12 t \pi - \frac12 i \pi}  e^{i  \int_{0}^{z_-} d z' \kappa (z')} .
   \end{aligned}
\end{align}
Here we set
\begin{align} \label{eq:t}
- i \frac{\pi}{2} t =   \int_{z_-}^{z_+}  \kappa (z') d z' \left( \equiv S(z_-,z_+) \right).
\end{align}
For $z < 1/\ell$, the asymptotic solution is written in terms of Hankel functions. Connecting with the wave function for  $z < z_-$, we find \eqref{eq:retardedAdSG}.

Let us examine the Regge poles of the retarded Green function \eqref{eq:retardedAdSG}. 
A series of poles of the retarded Green function are given by
\begin{align}
 - \frac{\pi}{2} t =  i \pi \left( m + \frac{1}{2} \right) - \left( \frac{\pi}{2}\right)^{\frac12} \left( \frac{m + \frac12}{e}\right)^{m+\frac12} \frac{1}{m!} e^{ - \frac{1}{2} i \pi  - i \pi \nu } e^{2i  S_m (0,z_-)} + \mathcal{O} \left (e^{4 i  S_m (0,z_-)} \right)
\end{align}
with $m= 0,1,2,\ldots$. 
The residues are
\begin{align}
\begin{aligned}
    &\text{Res}_{k \to k_m} G_R(\omega , \ell) \\
    &=  \frac{1}{ \nu \Gamma ^2 (\nu)  } \left( \frac{ \omega ^2 - k_m^2}{4}\right)^\nu \frac{\left(2 \pi^3 \right)^{1/2}}{m!} \left( \frac{m + \frac12}{e}\right)^{m + \frac12}e^{  - \frac{1}{2}i \pi \nu } \frac{e^{2 i S_m (0,z_-)}}{\frac{d S_m (z_- , z_+)}{d k}}  + \mathcal{O} (e^{4 i  S_m (0,z_-)} ) .
\end{aligned}
\end{align}
We are interested in the region with $ m \gg 1$.
We may replace sums over $m$ by integrals as
\begin{align} \label{eq:mtok}
    \sum_m
   \simeq \frac{1}{2 \pi} \int_0^\infty d k \frac{dm}{dk} = \frac{1}{ 2 \pi} \int_0^\infty dk \frac{1}{i \pi} \frac{d S (z_- , z_+)}{d k}  .
\end{align}
Applying this, the function \eqref{eq:gr} can be written as
\begin{align} 
\begin{aligned}
    g_R (t , \theta) & \simeq \frac{e^{i \pi \alpha \lfloor \frac{\theta}{\pi} \rfloor - \frac{1}{2}i \pi \nu }}{2^{2 \nu + \alpha} \nu \Gamma (\nu) ^2 \Gamma (\alpha +1) |\sin \theta|^\alpha} \\
 & \quad \times \int _0^\infty d k \int_{\infty + i \delta}^{- \infty + i \delta} d \omega e^{- i \omega t + i k \theta} k^{\alpha}
 (\omega^2 -k^2)^\nu   e^{2 i S (0,z_-)}  .
\end{aligned}
\end{align}
The integral is the same as the one in \eqref{eq:gE2}.

We can study the subleading contributions in $e^{2 i S_m (0,z)}$ for large $m$. The poles of \eqref{eq:retardedAdSG} can be written as
\begin{align}
    \lim_{m \to \infty} \left[t + 2 i \left(m + \frac12\right)\right] = \frac{1}{\pi} \ln \left( 1 + e^{-i \pi \nu + 2 i S (0,z_-)}\right) \, .
\end{align}
The residues are
\begin{align}
\begin{aligned}
    & \lim_{m \to \infty} \text{Res}_{k \to k_m} G_R(\omega , \ell) \\
    &= 
    \frac{4^{-\nu } \left(-1+e^{2 i \pi  \nu }\right) \Gamma (-\nu ) \left(\omega ^2-p^2\right)^{\nu }}{\pi  \Gamma (\nu ) \left(1+e^{i \pi  \nu -2 i S (0 , z_-) }\right)\frac{dt}{d k}}
  \\
    &=  \frac{4^{-\nu }  \left(-1+e^{2 i \pi  \nu }\right) \Gamma (-\nu ) \left(\omega ^2-p^2\right)^{\nu }}{\pi  \Gamma (\nu )\frac{dt}{d k} }\sum_{n=0}^\infty (-1)^{n}e^{-i n \pi \nu + 2 i n S(0 ,z_-)} .
\end{aligned}
\end{align}
The ratios with different $n$ are consistent with those obtained from \eqref{eq:an}.

\section{Boundary condition for AdS gravastar}

 \label{sec:bc-gravastar}
To derive the numerical solution of a scalar field on the gravastar geometry, it is useful to obtain the analytic solution in the dS spacetime.
The metric of the dS spacetime can be expressed as
\begin{align}
    ds^2=-f(r)dt^2+f^{-1}(r)dr^2+r^2d\Omega_{d-1}^2,\;\;f(r)=1-r^2,
\end{align}
where we take the dS length as the unit,  $R_\mathrm{dS}=1$, in this section.
The massive Klein-Gordon equation $(\Box-m^2)\Phi=0$ reduces to
\begin{align}
\label{eq:dS-waveeq}
    \left[\omega^2-l(l+1)\frac{f}{r^2}-m^2 f+f^2\partial_r^2+\left(\frac{2f}{r}+ff'\right)\partial_r\right]\phi_{\omega\ell}(r)=0
\end{align}
by expanding as 
$\phi = e^{-i\omega t}Y_{\ell \vec m}(\Omega)\phi_{\omega\ell}(r)$.
We have
\begin{align}
\label{eq:dS-gen-sol-horizon}
    \phi_{\omega\ell}=(1-y)^{\ell/2}\left\{
    Ay^{i\omega/2}F(1+a-c,1+b-c;2-c;y)
    +By^{-i\omega/2}F(a,b;c;y)
    \right\},
\end{align}
where
\begin{align}
    &a=\frac{d}{4}+\frac{\ell}{2}-\frac{1}{4}\sqrt{d^2-4m^2}+\frac{i\omega}{2},\quad
     b=\frac{d}{4}+\frac{\ell}{2}+\frac{1}{4}\sqrt{d^2-4m^2}+\frac{i\omega}{2},\quad
     c=1+i\omega.
\end{align}
This is a solution that is valid around the cosmological horizon, $y=1-r^2=0$.

For the solution around the center, $y=1-r^2=1$, we can use the following property of the hypergeometric function,
\begin{align}
    F(a,b;c;y)
    &=\Gamma_1(a,b,c)F(a,b;a+b+1-c;1-y)\nonumber\\
    &+\Gamma_2(a,b,c)(1-y)^{c-a-b}F(c-a,c-b;1+c-a-b;1-y),\nonumber\\
    \Gamma_1(a,b,c)&:=\frac{\Gamma(c)\Gamma(c-a-b)}{\Gamma(c-a)\Gamma(c-b)},\;\;
    \Gamma_2(a,b,c):=\frac{\Gamma(c)\Gamma(a+b-c)}{\Gamma(a)\Gamma(b)}.
\end{align}
Substituting the equation into 
the mode with the coefficient $B$ in eq.~\eqref{eq:dS-gen-sol-horizon}, we have
\begin{align}
\label{eq:Bmode-center}
    \phi_{\omega\ell}|_B
    &=By^{i\omega/2}\bigl\{
    (1-y)^{\ell/2}\Gamma_1(a,b,c)F(a,b;a+b+1-c;1-y)\nonumber\\
    &+(1-y)^{-(\ell+1)/2}\Gamma_2(a,b,c)F(c-a,c-b;1+c-a-b;1-y)
    \bigr\}.
\end{align}
The first and second terms are decaying and growing modes at the center, respectively.
Similarly, for the 
mode of $A$, we have
\begin{align}
\label{eq:Amode-center}
    &\phi_{\omega\ell}|_A
    =Ay^{-i\omega/2} \\
    & \times\bigl\{
    (1-y)^{\ell/2}\Gamma_1(1+a-c,1+b-c,2-c)F(1+a-c,1+b-c;1+a+b-c;1-y)\nonumber\\
    & \quad \quad +(1-y)^{-(\ell+1)/2}\Gamma_2(1+a-c,1+b-c,2-c)F(1-a,1-b;1+c-a-b;1-y)
    \bigr\} . \nonumber 
\end{align}
The first and second terms are decaying and growing modes at the center, respectively.
We would like to impose the regularity as the boundary condition at the center.
It requires vanishing of the growing part of $\phi_{\omega\ell}|_A+\phi_{\omega\ell}|_B$ at $y=1$.
The condition for the coefficients reduces to
\begin{align}
\label{eq:regularity-BA}
    \frac{B}{A}
    &=-\frac{\Gamma_2(1+a-c,1+b-c,2-c)}{\Gamma_2(a,b,c)}\nonumber\\
    &=-\frac{\Gamma(a)\Gamma(b)\Gamma(2-c)}{\Gamma(1+a-c)\Gamma(1+b-c)\Gamma(c)}.
\end{align}

Restoring the dS length $R_\mathrm{dS}$, the solution $\psi$ of the wave equation~\eqref{eq:zwaveeq} in the dS geometry that satisfies the regularity condition is given by
\begin{align} \label{eq:psi}
    \psi&=\left(\frac{r}{R_\mathrm{dS}}\right)^{\frac{d-1}{2}+\ell}\Bigl[
    A(1-(r/R_\mathrm{dS})^2)^{\frac{i\omega R_\mathrm{dS}}{2}}{}_2F_1(1+a-c,1+b-c;2-c;1-(r/R_\mathrm{dS})^2)\nonumber\\
    &\qquad +B(1-(r/R_\mathrm{dS})^2)^{\frac{-i\omega R_\mathrm{dS}}{2}}{}_2F_1(a,b;c;1-(r/R_\mathrm{dS})^2)
    \Bigr]\nonumber\\
    &=:\psi_\mathrm{dS},
\end{align}
where
\begin{align} \label{eq:psicoeff}
\begin{aligned}
    a&=\frac d4+\frac \ell2-\frac14 \sqrt{d^2-4m^2R_\mathrm{dS}^2}-\frac{i\omega R_\mathrm{dS}}{2},\\
    b&=\frac d4+\frac \ell2+\frac14 \sqrt{d^2-4m^2R_\mathrm{dS}^2}-\frac{i\omega R_\mathrm{dS}}{2},\\
    c&=1-i\omega R_\mathrm{dS},\\
    A&=\Gamma(1+a-c)\Gamma(1+b-c)\Gamma(c),\\
    B&=-\Gamma(a)\Gamma(b)\Gamma(2-c).
\end{aligned}
\end{align}
For the gravastar with the shell radius $r_0$, the regularity at the center is translated to the boundary condition at $r_0$ by matching the exterior solution to the wave function~\eqref{eq:psi}.
Then the boundary condition is given by
\begin{align}
&\psi(z_0)=\psi_\mathrm{dS}(z_0) , \\
&\left .\frac{d\psi}{dz} \right|_{z=z_0} =
\left .\frac{d\psi_\mathrm{dS}}{dz} \right|_{z=z_0} = \left. \frac{dr}{dz} \frac{d\psi_\mathrm{dS}}{dr} \right|_{z = z_0}=- \left. f(r_0) \frac{d\psi_\mathrm{dS}}{dr} \right|_{z = z_0}
\end{align}
in the tortoise coordinate $z$, where $z_0$ is the shell position.
The numerical solution of $\psi(z)$ in the exterior region $0<z<z_0$ is obtained by integrating the wave equation with this boundary condition at $z=z_0$.


\providecommand{\href}[2]{#2}\begingroup\raggedright\endgroup

\end{document}